\documentclass[10pt,letterpaper]{article}
\usepackage[top=0.85in,left=2.75in,footskip=0.75in,marginparwidth=2in]{geometry}

\usepackage[utf8]{inputenc}

\usepackage{nameref,hyperref}

\usepackage{microtype}
\DisableLigatures[f]{encoding = *, family = * }

\raggedright
\usepackage{changepage}

\usepackage[aboveskip=1pt,labelfont=bf,labelsep=period,singlelinecheck=off]{caption}

\makeatletter
\renewcommand{\@biblabel}[1]{\quad#1.}
\makeatother

\usepackage{lastpage,fancyhdr,graphicx}
\usepackage{epstopdf}
\fancyheadoffset[L]{2.25in}
\fancyfootoffset[L]{2.25in}

\usepackage{color}

\definecolor{Gray}{gray}{.25}

\usepackage{graphicx}

\usepackage{sidecap}
\usepackage{latexsym}
\usepackage{graphics}
\usepackage{float}
\usepackage{multicol}
\usepackage{media9}
\usepackage{multimedia}
\usepackage{adjustbox}
\usepackage{dirtytalk}
\usepackage{amsmath}
\usepackage{amssymb}
\usepackage{multirow}
\usepackage{tabularx}
\usepackage[colorinlistoftodos]{todonotes}
\usepackage{algorithm}
\usepackage[normalem]{ulem}
\usepackage{algpseudocode}
\usepackage{bm}
\def\Var{\text{Var}}
\def\Cov{\text{Cov}}
\def\CovG{\Cov_{\mathrm{G}}}
\usepackage{setspace}

\usepackage{fullpage}
\usepackage[english]{babel}
\usepackage[utf8]{inputenc}

\usepackage{natbib}
\usepackage{siunitx}
\usepackage{wrapfig}
\usepackage[pscoord]{eso-pic}
\usepackage[fulladjust]{marginnote}
\reversemarginpar

\DeclareMathOperator{\Tr}{Tr}
\DeclareMathOperator{\ord}{ord}
\DeclareMathOperator{\sgn}{sgn}
\newcommand{\Ito}{It\^{o}}

\newcommand{\dotprod}{\bm{\cdot}}
\newcommand{\Id}{\mathit{\mathsf{I}}}
\newcommand{\difd}{\mathrm{d}}
\newcommand{\expe}{\mathrm{e}}
\newcommand{\mathi}{\mathrm{i}}
\newcommand{\bExp}{\mathbb{E}}
\DeclareMathOperator{\sinc}{sinc}
\newcommand{\Thmag}{\bar{\Theta}}
\newcommand{\Fmag}{\bar{F}}
\newcommand{\muY}{\mu_Y}
\newcommand{\CY}{C_{Y}}
\newcommand{\celllen}{l}
\newcommand{\colrad}{a}
\newcommand{\flagdisp}[1]{S_{#1}}
\newcommand{\flagdispran}{\delta_{S}}
\newcommand{\bXcol}{\mathbf{X}^{(\mathrm{c})}}
\newcommand{\Thetacol}{\Theta^{(\mathrm{c})}}
\newcommand{\Thetacolgauss}{\Delta\Thetacol_{\mathrm{G}}}

\newcommand{\Thetacolerr}{\Delta\Thetacol_{\mathrm{NG}}}

\newcommand{\Vargauss}{V_{\Delta \Theta,\mathrm{G}}}
\newcommand{\Vargaussp}{\Vargauss^{\prime}}
\newcommand{\Fcol}{\mathbf{F}^{(\mathrm{c})}}
\newcommand{\Torquecol}{T^{(\mathrm{c})}} 
\newcommand{\gamrot}{\gamma_{\mathrm{r}}}
\newcommand{\Vcol}{\mathbf{V}^{(\mathrm{c})}}
\newcommand{\Vflagavgsq}{\Veleff{}^2}
\newcommand{\gamcolr}{\gamrot}
\newcommand{\bW}{\mathbf{W}}
\newcommand{\bx}{\mathbf{x}}
\newcommand{\bzero}{\bm{0}}
\newcommand{\jp}{j^{\prime}}
\newcommand{\jpp}{j^{\prime\prime}}
\newcommand{\tp}{t^{\prime}}
\newcommand{\tpp}{t^{\prime\prime}}
\newcommand{\sigmaThe}{\sigma_{\Theta}}
\newcommand{\varTh}{\delta_{\Theta}}
\newcommand{\Difftrans}{D_{\mathrm{t}}}
\newcommand{\Diffrot}{D_{\mathrm{r}}}
\newcommand{\meanrot}{\Omega^{\ast}_{\mathrm{r}}}
\newcommand{\gamtrans}{\gamma_{\mathrm{t}}}
\newcommand{\Diffroteff}{\Diffrot^{\ast}}
\newcommand{\Difftranseff}{\Difftrans^{\ast}}
\newcommand{\Difftransrot}{\tilde{\Difftrans}}

\DeclareMathOperator{\Ein}{Ein}
\DeclareMathOperator{\Real}{\mathrm{Re}}

\newcommand{\flagpos}[1]{\mathbf{B}_{#1}}
\newcommand{\Fbody}{\Fcol_{\mathrm{b}}}
\newcommand{\Fbodymean}{\langle\Fbody\rangle}

\newcommand{\Torquemean}{\langle \Torquecol \rangle}
\newcommand{\orient}[1]{\hat{\mathbf{e}}_{#1}}
\newcommand{\orientperp}[1]{\orient{#1}^{\perp}}
\newcommand{\centmass}{\mathbf{C}}
\newcommand{\gammamin}{\gamma_-}
\newcommand{\gammamax}{\gamma_+}
\newcommand{\gammaj}{\gamma_{\mathrm{t}1}}
\newcommand{\gammin}{\gamma_{\mathrm{t}2}}
\newcommand{\gmajrot}{\gamma_{\mathrm{tr}1}}
\newcommand{\gminrot}{\gamma_{\mathrm{tr}2}}

\newcommand{\sterr}{\omega}

\newcommand{\bWgrandtherm}{\mathbf{W}^{(\mathrm{T})}}
\newcommand{\bWgrandact}{\mathbf{W}^{(\mathrm{A})}}
\newcommand{\Velrot}{\Omega}
\newcommand{\Veltransbody}{\mathbf{V}^{(\mathrm{c)}}_{\mathrm{b}}}
\newcommand{\Vcolmsq}{\langle|\mathbf{V}^{(\mathrm{c})}|^2\rangle}
\newcommand{\Veleff}{V^{\ast}}

\newcommand{\twoflagvar}{V_{2,2}}
\newcommand{\twoflagmod}{\twoflagvar^{\ast}}
\newcommand{\twoflagthrice}{V_{3,1}}
\newcommand{\threeflagcov}{V_{2,1,1}}
\newcommand{\threeflagcovconj}{\threeflagcov^{\ast}}
\newcommand{\Fmom}[1]{\mu^{(0)}_{F,#1}}
\newcommand{\Fcum}[1]{\kappa^{(0)}_{F,#1}}

\newcommand{\Granddrag}{\mathit{\mathsf{\Gamma}}}
\newcommand{\Granddraginv}{\Granddrag^{-1}}
\newcommand{\Rotmat}[1]{\mathit{\mathsf{R}}_{#1}}
\newcommand{\Rotmatthree}[1]{\tilde{\mathit{\mathsf{R}}}_{#1}}
\newcommand{\Actnoisemat}{\mathit{\mathsf{\Sigma}}}
\newcommand{\bZ}{\mathbf{Z}}
\newcommand{\bs}{\mathbf{s}}
\newcommand{\bmu}{\bm{\mu}}
\newcommand{\Cmat}{\mathit{\mathsf{C}}}
\newcommand{\ufrat}[1]{\vec{\nu}_{#1}}
\newcommand{\ufratperp}[1]{\vec{\nu}_{\perp,#1}}
\newcommand{\wfrat}[1]{\rho_{#1}}
\newcommand{\wfratperp}[1]{\rho_{\perp,#1}}
\begin{document}
\vspace*{0.35in}

% title goes here:
\begin{flushleft}
{\Large
\textbf\newline{Statistical Mobility of Multicellular Colonies of Flagellated Swimming Cells}
}
\newline
% authors go here:
\\
Yonatan L Ashenafi,
Peter R Kramer
\\
\bigskip
\bf{1} Department of Mathematical and Statistical Sciences,
University of Alberta, Edmonton, AB, Canada
\\
\bf{2}  Department of Mathematical Sciences, Rensselaer Polytechnic Institute, Troy, NY, USA
\\
\bigskip
* yashenaf@ualberta.ca

\end{flushleft}

\section*{Abstract}
We study the stochastic hydrodynamics of colonies of flagellated swimming cells,  typified by multicellular choanoflagellates, which can form both rosette and chainlike shapes. The objective is to link cell-scale dynamics to colony-scale dynamics for various colonial morphologies.  Via autoregressive stochastic models for the cycle-averaged flagellar force dynamics and statistical models for demographic cell-to-cell variability in flagellar properties and placement, we derive effective transport properties of the colonies, including cell-to-cell variability.  We provide the most quantitative detail on disclike geometries to model rosettes, but also present formulas for the dynamics of general planar colony morphologies, which includes planar chain-like configurations.

% the * after section prevents numbering
\section{Introduction}
\label{intro}

Various forms of  eukaryotic cells ranging from protozoa to spermatozoa employ flagellar beating to navigate in their fluid environment, for example in search of food, chemicals, and other cells while also avoiding their predators.  Experimental work has investigated vital components of flagella-driven swimming motion such as those concerning fluid rheology \citep{Rafarheology, Matsuirheology}, strength of the propulsive forcing due to flagella \citep{Baylyforceestimate}, the statistics of the orientation of the flagella from its base~\citep{ colonialmotility}, and the ability to sense and react to various environmental cues \citep{aerotax,DeMarcochemotax,Fosterphototaxis}. 
Theoretical models and analyses have made predictions on the flows and mobility mechanisms of the swimming cells, including properties in suspension~\citep{Roperstresslet, BLAKEcilia,Deansuspension,Sparacino_solitary}.  

Some recent experimental studies on swimming microorganisms have turned to multicellular colonies driven by multiple flagella, for example in protozoa~\citep{colonialmotility,GoldsteinGonium} and green algae~\citep{GoldsteinVolvocine}.  The present work is largely motivated by the theoretical analysis in~\cite{colonialmotility} for the observed dynamics of protozoan choanoflagellate colonies. These are colonial eukaryotic organisms which are close relatives of animals, and are being used to understand the pathways to emergence of mutlicelluarity in animals~\citep{BrunetMulticell,King-urmetazoa,KoehlSelective,KoehlCapture}. Choanoflagellate colonies mimic their distant mobile predecessors and choanoflagellate cells are also shown to be quite similar to sponges found in the ocean \citep{leadbeater2015}.  

In~\citet{colonialmotility}, the theoretical connection between the individual cellular behavior and the effective ``aggregate random walker'' model of translational and rotational dynamics for the colony is made through phenomenology and scaling arguments.
We endeavor here to derive more systematically   the colony dynamics beginning from a formulation of specific statistically independent dynamical models for each flagellum together with a rigid-body model for their geometric configuration.   
A key aspect of the swimming performance of colonies is that in both rosette and chain configurations, the flagella of the constituent cells are oriented in various directions and therefore partially or even largely cancel each other other's propulsive force on the colony~\citep{KoehlSelective}.  
Irregularities in the flagellar positioning introduce qualitative changes to the swimming of a colony by mitigating the cancellation of flagellar forces and inducing a net torque causing the colony to rotate~\citep{KoehlSelective,Burkhardt_architecture}. \citet{Roperstresslet} studied feeding fluxes in regular models of colony shapes by consideration of the flows generated by the geometric arrangement of flagellar forces, with brief consideration of irregular flagellar placement.  One of our central objectives  is to  formulate stochastic  models in the flagellar dynamics and statistical models for the flagellar geometry, and to examine in a precise manner how these two sources of randomnness together influence the mobility of the colony.

In Section~\ref{sec:discmodel}, we develop the model of a disclike colony that is constructed in the image of choanoflagellate rosettes. 
The cycle-averaged flagellar forces and geometry are modeled stochastically to account for both observed dynamical variations in time and demographic variations between flagella.  We discuss as well how we estimate the model parameters from the experimental literature.
Next in Section~\ref{summarych1} we summarize the results of our calculations for some key mobility characterizations of the colony:  mean-square speed as well as the effective translational and rotational drift and diffusity.  We present formulas both for an individual colony with its peculiar irregular configuration of flagella as well as for population statistics over an ensemble of colonies with a prescribed demographic distribution of variability in flagellar properties and placement on the colony.
The derivation of these results, which amount to probabilistic calculations and some asymptotic approximations, are explained in Section~\ref{sec:methods}. In Section~\ref{sec:gengeo}, we show how to generalize our characterization of the mobility of a colony of microswimmers of arbitrarily planar geometry in terms of the properties of the individual microswimmers and their geometric arrangement, provided the dynamical noise in the cycle-averaged flagellar dynamics can be taken as relatively small.  We discuss how our conclusions for the simplified model may be expected to apply in more realistic models in Section~\ref{sec:discuss}.

\section{Disclike Colony Model}
\label{sec:discmodel}
In order to obtain analytical relationships between the effective swimming dynamics of the colony in terms of those of the individual cells, we begin with a rather minimal modeling framework.   A simple disclike geometric model for the colony is described in Subsection~\ref{sec:geomod}, followed in Subsection~\ref{sec:cellmod} by the stochastic dynamical model for the propulsion induced by each cell's flagellum.  The resulting dynamical model for the colony is presented in Subsection~\ref{sec:colmod}.  Suitable values for the various biophysical model parameters are discussed in Subsection~\ref{sec:parmod}, which will inform some simplifications in the calculations to follow.

\subsection{Colony Geometry}
\label{sec:geomod}
Our primary geometric model for a colony in a rosette shape is a two-dimensional disc, composed of a fixed number $N$ of swimming cells.  We in particular neglect dynamical changes in cell number due to cell division or other morphology changes, as these take place on a time scale of hours~\citep{Fairclough2010}, much longer than the basic dynamical time scales of seconds to minutes of interest.  The two-dimensionality of our model is chosen simply to avoid the complications of three-dimensional rotational dynamics.   More precisely, we envision the colony as a thin disc in a three-dimensional fluid, whose translational and rotational dynamics are constrained to remain in the plane defined by the disc.  The disclike approximation may actually be reasonable for small \emph{Salpingoeca rosetta} colonies of 4-7 cells; larger colonies exhibit a more three-dimensional structure~\citep{MahadevanChoano}.  But even with a truly flat morphology, we impose a simplification by constraining the colony to only move within the plane defined by its geometry.

In our models, the cells are assumed to be of equal size and symmetrically situated about the disc, with each cell being exposed to the ambient fluid over an arc length $\celllen $ of the disc, as well as along the flat transverse faces of the thin disc.  In particular, this implies the disc representing the colony has a circumference $ N \celllen$ and thus a radius $ \colrad = \frac{N \celllen}{2 \pi}$.  This proportionality between radius and cell number is consistent with a two-dimensional version of the argument in~\citet{SolariHydrodynamics2006}.  In reality, cells in \emph{S. rosettas} colonies do exhibit some size and shape disparities~\citep{Naumannmorphology}, possibly influenced by colony size~\citep{Burkhardt_architecture,MahadevanChoano}, but the available data at this point seems insufficient for a meaningful mathematical model of this cellular variability.   We in particular treat the physical colony in an entirely smooth and symmetric fashion,  neglecting the natural protrusion of each cell with respect to the nominal disc shape of colony, not to mention the microvillar collar~\citep{Fauci_morphology}.  Asymmetries in the flagellar properties and placement will however be represented in our model, as these appear to be more fundamental to the determination of the colony dynamics.  We view these simplifying geometric contrivances, illustrated in Figure~\ref{model schematic1} as still capable of preserving the substance of the interactions between the cellular components in determining the colonial swimming properties.

The flagellum for each cell $i=1,\ldots,N$ is assumed to be within the plane of the colonial disc, with its base attached to the associated arc at a displacement $ \flagdisp{i} \sim U(-\flagdispran,\flagdispran) $ independently and uniformly randomly distributed within an arclength $ \flagdispran \leq \celllen/2$ of the center of the arc.  Each flagellum $ i=1,\ldots,N$ is moreover allowed demographic stochasticity in terms of its mean propulsion force $ F_i^{(0)}$ and its ``relaxed'' angle $ \Theta_i^{(0)}$ which it makes with the normal at its basal point of attachment to the cell.  These are each taken to be independent random variables with  $ \Theta_i^{(0)} \sim N(0,\varTh^2)$,  Gaussian distributed with mean $ 0 $ and standard deviation $ \varTh$. 
The flagellar force $ F_i^{(0)} $ only appears linearly or quadratically in the transport expressions, so we will only characterize it in terms of its first four moments or cumulants.

\iffalse 
\begin{figure}
\centering
    \begin{subfigure}
     \centering
 \includegraphics[scale=0.30]{Colony_Schematic2}\\
  \end{subfigure} 
      \begin{subfigure}
     \centering
     \includegraphics[height=94mm, width=90mm]{Colony_schematic}\\
\end{subfigure} 
\caption{Rendition of a colony of ten cells and a model schematic of eight cells.}
 \label{fig:Model Sch.}
\end{figure}
\fi 

\begin{figure}[H]
{\centering{
\begin{multicols}{2}    \includegraphics[scale=0.29]{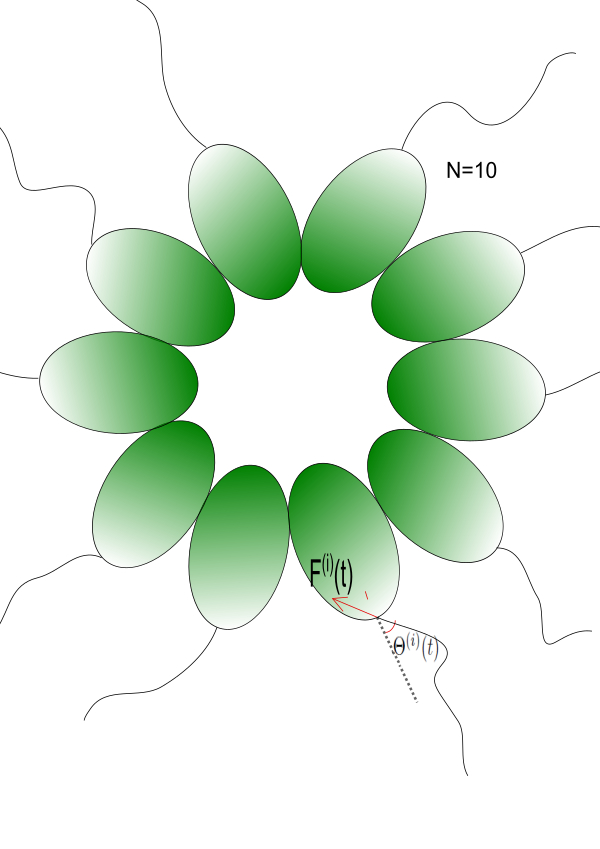}      
   \par
   \includegraphics[height=80mm, width=76mm]{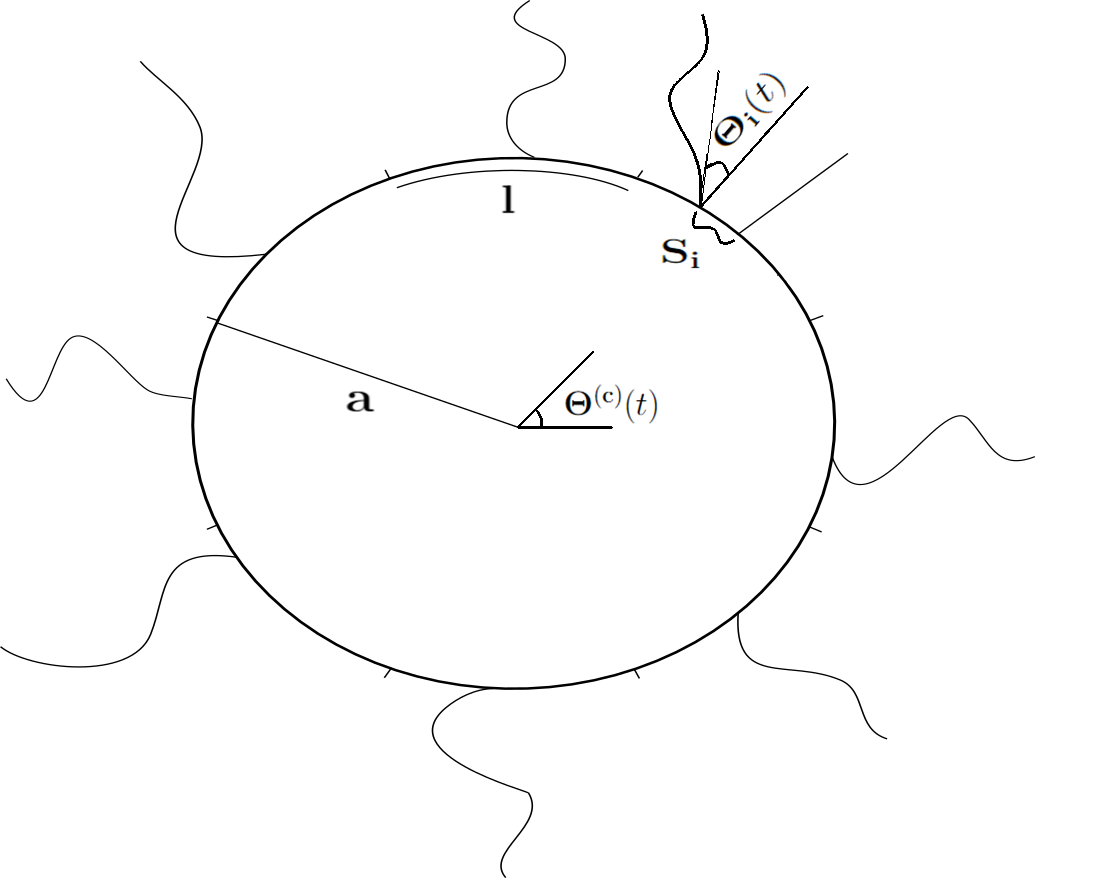}
\end{multicols}}
\caption{Rendition of a colony of ten cells (left) and a model schematic of eight cells (right).  The cell has the shape of a disc of radius $ \colrad$, with each cell represented by an equal arc of length $\celllen$.  The attachment point of the flagellum associated to the $i$th cell is displaced by an arclength $ \flagdisp{i}$ from the center of its arc.  The direction of the flagellar force is dynamically directed along a direction $ \Theta_i (t)$ with respect to the normal.  The colony's dynamical orientation is denoted by an angle $ \Theta^{(c)} (t) $ with respect to a fixed reference direction which we take to be along the rightward horizontal.   As usual, all quantities associated to rotational measures are taken to be positive in the counterclockwise direction.     
}
\label{model schematic1}}
\end{figure}

\subsection{Cell Propulsive Forcing Model}
\label{sec:cellmod}

Experimental and theoretical studies have characterized the approximately periodic dynamics of flagellar beating, including the coupling between flagellar dynamics in various cells and multicellular organisms~\citep{BrumleyVolvox,JoannyFlow}. For at least choanoflagellates, however, the flagella appear to be sufficiently distant and irregularly positioned to make correlations between their dynamics unobservable \citep{colonialmotility}.  In our model, we will therefore neglect correlations between the flagellar beating dynamics.  The detailed representation of the periodic beating of each flagellum $ i=1,\ldots,N$ does not then appear to be necessary, and we represent the flagellar dynamics instead by a locally cycle-averaged force $ F_i (t)$ and cycle-averaged orientation $ \Theta_i (t)$ of the flagellum with respect to the normal at its basal attachment to the cell.  We treat these variables as continuous in time; they could be thought of as windowed time averages over a period width centered at the current time $ t $.  Since flagellar beating is known to have effectively random disturbances due to the noise in the dynein activity driving them as well as environmental effects, we model the dynamics of the cycle-averaged flagellar force magnitude and orientation for cell $i $ via autoregressive stochastic processes with relaxed states $F_i^{(0)}$ and $\Theta_i^{(0)}$, dynamical variances $ \sigma_F^2$ and  $\sigma_{\Theta}^2$, and damping parameters (or inverse correlation times)
$\gamma_F$ and $\gamma_\Theta$ respectively.
The SDEs for these processes are given as:
\begin{equation}   \label{force magnitude sde 1}
\begin{aligned}
dF_i(t)=&-\gamma_F(F_i(t)-F_i^{(0)})dt+\sqrt{2\sigma_F^2\gamma_F} \, \difd W_i^F(t)
\end{aligned} 
\end{equation}
\begin{equation}   \label{force direction sde 1}
\begin{aligned}
d\Theta_i(t)=&-\gamma_\Theta(\Theta_i(t)-\Theta_i^{(0)})dt+\sqrt{2\sigma_\Theta^2\gamma_\Theta} \, \difd W_i^\Theta(t)
\end{aligned} 
\end{equation}
Here $\{W_i^F\}_{i=1}^N$ and $\{W_i^\Theta\}_{i=1}^N$ stand for independent standard Wiener processes.  Note that we do not associate the noise of the flagellar dynamics to temperature, as the noise from active driving processes is apparently dominant.  Our model does not account for a possible jumpy change in the swimming direction of a cell, as seen for fast swimming \emph{S. rosetta} cells in~\citet{Sparacino_solitary}, but it is slow swimmers that form \emph{S. rosetta} colonies~\citep{dayelmorphogenesis}.
 In our analytical and numerical computations, we take the initial conditions for the flagella to be independently distributed according to their stationary distributions, so $ F_i(0) $ and $ \Theta_i(0)$, are normally distributed with mean $ F_i^{(0)}$ and standard deviation $ \sigma_F $ (respectively, $ \Theta_i^{(0)}$ and $ \sigma_{\Theta} $).

\subsection{Colony Dynamics Model}
\label{sec:colmod}

We assume the colony is sufficiently isolated that we can neglect external interaction or hydrodynamic forces.  The Reynolds number for colonial microswimmers  is approximately $10^{-4}-10^{-3}$ 
so we may use low Reynolds number hydrodynamics based on the linear Stokes equation to represent the hydrodynamic drag on the colony motion. Using the drag formulas from~\citet[Ch. 3]{kimmicrohyd} for a spheroid in the oblate limit,  we take for, respectively, the translational and rotational drag coefficient:
\begin{equation} 
\begin{aligned}
\gamtrans=\frac{32}{3}a\eta%\text{Pa.sec.}
, \hspace{0.6 cm}  \gamrot=\frac{32}{3}a^3\eta%\text{Pa.sec.} 
\end{aligned}
\end{equation}
where $\eta$ is the dynamic viscosity of the fluid. 

Balancing the active and drag forces and torque, and incorporating thermal noise, we have the following overdamped Langevin equations for the
center of mass ($\bXcol$) and angular orientation ($\Theta^{(c)}$):
\begin{equation} \label{Colony position sde 1}
\begin{aligned} d\bXcol (t)=\frac{\mathbf{F}^{(c)}(t)}{\gamtrans} dt+\sqrt{2\Difftrans}d\bW^{x,c}(t)
\end{aligned}
\end{equation}
\begin{equation} \label{Colony orientation sde 1}
\begin{aligned}
d\Theta^{(c)}(t)=\frac{T^{(c)}(t)}{\gamrot}dt+\sqrt{2\Diffrot}dW^{\Theta,c}(t).
\end{aligned}
\end{equation}
Here $ \Difftrans = k_B T/\gamtrans$ and $ \Diffrot = k_B T/\gamrot $ are, respectively, the thermally induced translational and rotational diffusivity, 
$k_B$ is the Boltzmann constant and T is temperature, and $ \bW^{X,c}$ and $ \bW^{\Theta,c}$ are independent Wiener processes representing thermal noise.  The aggregate active force on the colony is obtained by adding up the vectorial contributions from each flagellum:
\begin{equation} \label{Agg. active force}
\begin{aligned}
 \mathbf{F}^{(c)}=\sum_{i=1}^{N} -F_i(t)\begin{pmatrix}\cos(2\pi[\frac{i-\frac{1}{2}+\frac{S_i}{l}}{N}]+\Theta_i(t)+\Theta^{(c)}(t))\\\sin(2\pi[\frac{i-\frac{1}{2}+\frac{S_i}{l}}{N}]+\Theta_i(t)+\Theta^{(c)}(t))\end{pmatrix}. 
\end{aligned}
\end{equation}
 \normalsize 
The aggregate active torque is similarly obtained by adding up the torque contributions from each flagellum:
\begin{equation} \label{Agg. active torque}
\begin{aligned}
  T^{(c)}=-\sum_{i=1}^N F_i(t)a\sin \Theta_i(t)
\end{aligned}
\end{equation}
\normalsize
To obtain the representation~\eqref{Agg. active force} for the total force on the colony, we have defined the colony orientation angle $ \Thetacol (t) $ to be the angle with respect to direction of the horizontal made by the radius from the center of the colony to the beginning of the arc defining cell $1$, with all angular orientations taken in the usual counterclockwise sense (see Figure~\ref{model schematic1}).  Then the center of the arc associated to cell $i $ will, at time $ t $, be situated at angular position $ \Thetacol (t)+ \frac{2\pi}{N} \left(i-\frac{1}{2}\right)$, and the associated flagellum base is at angular position 
$ \Thetacol (t)+ \frac{2\pi}{N} \left(i-\frac{1}{2}+ \frac{\flagdisp{i}}{\celllen}\right)$.  The force applied by flagellum $i$ will be directed in the opposite direction (into the cell body), after correction for the current cycle-averaged displacement $ \Theta_i (t)$ of the flagellar orientation from the normal direction to the cell.

Because the environment in our model is spatially homogenous and isotropic, without loss, we will take the initial configuration of the colony to satisfy $ \bXcol (t=0)=\bzero$ and $ \Thetacol (t=0) = 0$.

\subsection{Parameter Estimation}
\label{sec:parmod}

Estimates for the various biophysical parameters in our model are presented in Table~\ref{Tab:Parameter table}.  We next explain how they were obtained from the experimental literature.

\begin{table}[H]   
\caption{Estimates of model parameters}
\begin{adjustbox}{width=\columnwidth,center} 
\centering  
\begin{tabular}{ |c|c| }   
 \hline
 Parameter &  Estimated value 
 \\ 
  \hline Number of cells (N)   & 2-10    \\ 
 \hline
 Arclength of cell exposure ($\celllen$) & $ 2 \pi \,\si{\micro\metre}$
 \\ \hline Flagellar force ($F^{(0)}_i$) & $2 \pm1\, \si{\pico\newton}$ 
 \\ \hline Relaxed flagellar orientation with respect to normal ($\Theta^{(0)}_i$) & $\pm 0.02 $  \\ 
  \hline Damping parameter of force orientation ($\gamma_\Theta$)  & $10\, \si{\sec}^{-1}$  \\ 
 \hline Damping parameter of force magnitude ($\gamma_F$) & $10 \,\si{\sec}^{-1}$   \\ 
  \hline Dynamical variance of force orientation ($\sigma_\Theta^2$)  &  $0.002$ \\ 
   \hline Dynamical variance of force magnitude ($\sigma_F^2$)  & $0.1 \, \si{\pico\newton}^2 $    \\
\hline Temperature(T) & 300 K  \\ \hline Dynamic viscosity ($\eta$) & $\SI{0.01}{\gram/\cm \sec}$ \\
\hline 
\end{tabular}
\end{adjustbox}
\label{Tab:Parameter table}
\end{table}

\subsubsection{Estimation of Geometric Parameters}
Images in~\citet{dayelmorphogenesis,colonialmotility}) indicate the individual cells in an \emph{S. Rosetta} colony have a diameter on the order of $ \SI{5}{\micro\metre}$, and~\citet{Faircloughintercellularbridge} report intercellular bridges between cells have lengths of about $\SI{0.15}{\micro\metre}$.  The experimental observations in~\citet{colonialmotility}, which are the target of our modeling efforts, report  colonial projected areas of $20-250\si{\micro\metre^2}$.  If we assume the colony cross sections to be approximately circular, this would correspond to circumferences of $ 10-60\si{\micro\metre}$.  This would correspond to about $N\sim 2-10$ cells under our two-dimensional disc model if we partition the circumference of the colony into segments for each cell and their intercellular bridges to their neighbors, and approximate the arc length occupied by a cell by its diameter.      \citet{MahadevanChoano} find in fact \emph{S. rosetta} to proceed through an approximately two-dimensional phase of growth of 4-7 cells to a three-dimensional phase of growth from 8-12 cells.  \citet{KoehlCapture} report 
$2-13$ cells in \emph{Salpingoeca helianthica} colonies.  We note that~\citet{MahadevanChoano} report shrinking cell sizes as the colony increases, but the reported data does not seem to permit a clean scaling representation.  One could of course allow the exposed cell arclength $l$ to decrease as desired with the number of cells $N$; our analysis is only affected in terms of the dependence of the asymptotic errors on the colony size.  We take $ \celllen = 2 \pi\, \si{\micro\metre}$ for modeling convenience since it is roughly consistent with observations.  
\subsubsection{Estimation of Flagellar Force Parameters}
\begin{figure}[H]	  
{\centering{
\includegraphics[scale=0.7]{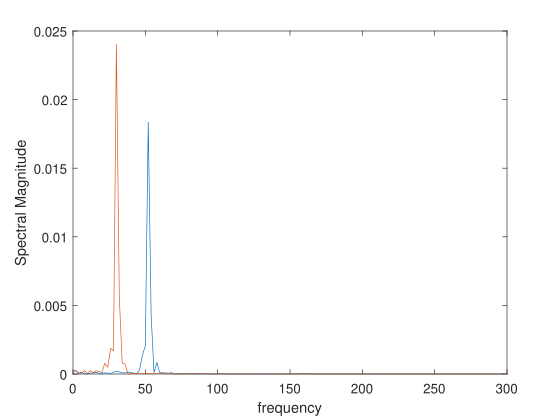}}
\caption{The spectral distributions of the time series for the orientation of two flagellaa in \citet[Fig. 2a. Fig. 2b]{colonialmotility} 
}
\label{Spectrum}}
\end{figure}

The typical flagellar force magnitude $ F_i^{(0)}$ is taken from~\citet{Roperstresslet}. To estimate the dynamical variance of the flagellar force, we assume the force is proportional to the frequency~\citep{JoannyFlow} and consider the periodogram estimates of the frequency spectrum (Figure~\ref{Spectrum}) using the time series from Figures 2a and 2b in~\cite{colonialmotility}.  
Gaussian fits to each resulting frequency distribution yielded estimates for the ratio of the frequency spread (standard deviation) to the mean, which we took as an estimate for the ratio of the dynamical standard deviation of the force to its mean.  The peak frequency of the two flagella differ by about a factor of $2$, suggesting a rather broad demographic distribution of the flagellar force, consistent with the spread of flagellar frequencies reported in \citet[Fig. 2e]{colonialmotility}.  By contrast, the polydispersivity of beating frequencies in \emph{Volvox carteri}, whose partial metachronal synchronization is out of the scope of our study, is observed to be only about 10\%, or 3\% when controlled for spatial location~\citep{BrumleyVolvox}.

\begin{figure}[H]	  
\centering{
\includegraphics[scale=0.8]{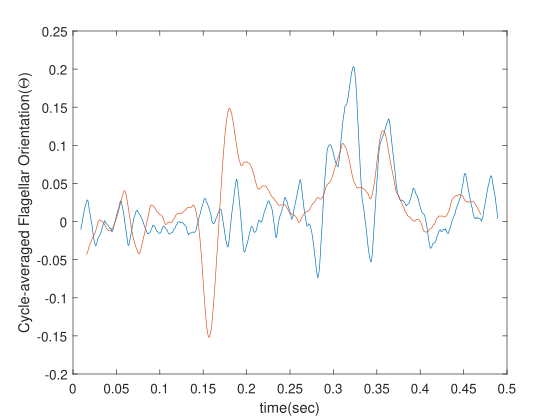}}
\caption{The cycle-averaged smoothing
of the time series for the flagellar orientation $ \Theta $ reported in Figures 2a (blue) and 2b (red) in (\citet{colonialmotility}
\label{Processed images}}
\end{figure}

For estimating the cycle-averaged statistics of the flagellar orientation, we applied a moving average filter over a window equal to one period as estimated by the frequency peak
to the  sample time series of the orientation of two flagella in Figures 2a and 2b of  \cite{colonialmotility}, obtaining the plots in Figure~\ref{Processed images}.
The average values were both found to be $ 0.02$, so we used that value as a typical magnitude (with either sign) of the relaxed or dynamical mean $ \Theta_i^{(0)}$ of the flagellar orientation with respect to the normal.
The dynamical variance $ \sigma_{\Theta}^2$ of the flagellar orientation was then estimated as the variance of these cycle-averaged time courses of the flagellar orientation.  

The damping parameter $ \gamma_{\Theta} $ for the cycle-averaged force orientation is easily inferred from the decay times of the orientation correlation presented in~\citet[Fig. 2f]{colonialmotility}.  This decay time was seen to have a broad variation across flagella, but we neglect the demographic stochasticity of this parameter in our model.  We make the plausible choice $ \gamma_F=\gamma_{\Theta}$ since we are not aware of time series data on the flagellar force.

\section{Statistics of Mobility} \label{summarych1}
We present here a summary of the effective statistics of motion of the disclike colony model described in Section~\ref{sec:discmodel}.  In Subsection~\ref{sec:rotsum} we report the effective rotational drift and diffusivity of the colony, while in Subsection~\ref{sec:transsum} we report the mean-square speed and translational diffusivity of the colony.  Our model has two sorts of randomness:  1) dynamical stochasticity arising from the noisy driving terms on the flagellar force and orientation in Subsection~\ref{sec:cellmod}, and 2) demographic stochasticity, meaning random variations between cells of the static features $ F_i^{(0)}$, $ \Theta_i^{(0)}$, and $ \flagdisp{i}$ of their associated flagella as defined in Subsection~\ref{sec:geomod}.  For each mobility descriptor, we first report the formulas for a particular colony as a function of its peculiar values of the static flagellar features, but averaged over the dynamical noise.
We  use $ \langle \cdot \rangle $ to denote such averages over the dynamical stochasticity, with a fixed realization of the static flagellar properties.  After the mobility quantifier formula for an individual colony is presented, we then express its average  and variance with respect to the demographic distribution of flagellar properties described in Subsection~\ref{sec:geomod}.  In this summary section, we reserve the notation $ \bExp$ and $ \Var$  for this demographic average and variance.  The demographic moments of the relaxed flagellar force will be denoted $ \Fmom{k} \equiv \bExp\left[ (F^{(0)})^k \right]$, and their corresponding cumulants by  $ \Fcum{k}$.

The derivation of these results will follow in Section~\ref{sec:methods}.  We provide some visualization of the formulas and comparison with Monte Carlo simulations; the method for conducting the simulations is discussed in Appendix~\ref{sim_settings}.   We assume the flagella are initialized with their statistically stationary distribution for their force and orientation.  If they are not, the formulas to follow only apply at times $ t \gg \gamma_F^{-1}, \gamma_{\Theta}^{-1} $ over which the flagella relax to their stationary distribution. The diffusivity formulas are inherently long-time so unaffected by the initial conditions.

\subsection{Rotational Mobility Statistics}
\label{sec:rotsum}
\subsubsection{Rotational Drift}: 
\label{sec:rotdriftsum}
\begin{equation} \label{eff rotational drift_summary}
\begin{aligned}
\meanrot  \equiv \frac{\difd \langle \Thetacol (t) \rangle}{\difd t} 
=-\frac{a}{\gamrot} e^{-\frac{1}{2}\sigma_\Theta ^2} \sum_{i=1}^N F_i^{(0)}\sin(\Theta_i^{(0)})
\end{aligned}
\end{equation}
The rotational drift is driven by the natural relaxed torque diminished by the factor $ \expe^{-\frac{1}{2} \sigma_\Theta^2} $ arising from the dynamical variation of the angular orientation of the flagella of the cells,  while being resisted by rotational drag due to the fluid environment.

\begin{figure}[H]   
\centering
\includegraphics[height=80mm,width=120mm]{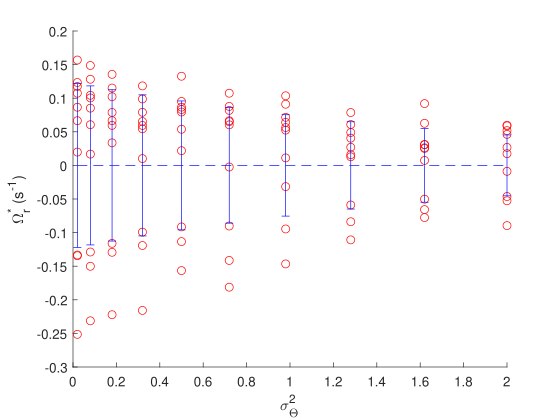}
\caption{
Comparison of theoretical (blue) and Monte Carlo simulation (red) results  of rotational drift over a long time ($t=10000 \si{\second}$)  for a colony of $N=10$ cells.   The analytical result has zero mean with demographic standard deviation from (\ref{eff rotational drift_sd_summary}) indicated with error bars.  Simulation results are shown separately for 1 simulation of each of 10 colony samples at each value of  $\sigma_{\Theta}^2$, with the flagellar displacements  modeled as $S_i\sim U(\frac{-l}{2},\frac{l}{2})$. The rest of the parameters are specified in Appendix~\ref{sim_settings} and Table \ref{Tab:Parameter table}.}
\label{rot drift sigma fig}
\end{figure}

\begin{figure}[H]   
\centering
\includegraphics[height=80mm,width=120mm]{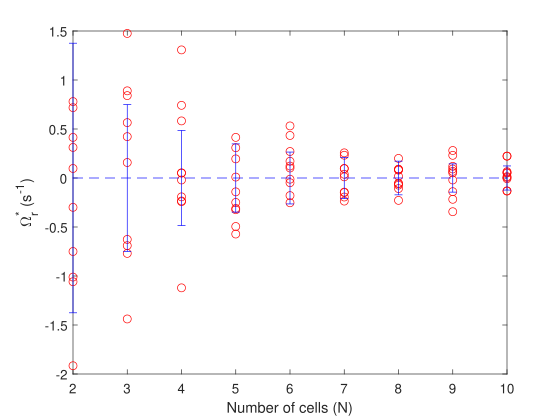}
\caption{
Comparison of theoretical (blue) and Monte Carlo simulation (red) results  of rotational drift over a long time ($t=10000 \si{\second}$)  for colonies of various sizes.   The analytical result has zero mean with demographic standard deviation from Eq.~\eqref{eff rotational drift_sd_summary} indicated with error bars.  Simulation results are shown separately for 1 simulation of each of 10 colony samples at each colony size, with the flagellar displacements  modeled as $S_i\sim U(\frac{-l}{2},\frac{l}{2})$. The rest of the parameters are specified in Appendix~\ref{sim_settings} and Table \ref{Tab:Parameter table}. }
\label{long time rot drift N}
\end{figure}

Because we assume $ \Theta_i^{(0)} $ is symmetrically distributed about $0 $, the demographic average of the rotational rate of the colony is zero. Individual colonies will break the symmetry and have some rotational drift.
We can characterize the typical magnitude of the rotational drift by its demographic standard deviation:
\begin{equation} \label{eff rotational drift_sd_summary}
    \sigma_{\meanrot} = \sqrt{\Var[\meanrot]}
    = \frac{a}{\gamcolr}e^{-\frac{1}{2}\sigma_\Theta ^2} \sqrt{N\Fmom{2}
    \left(\frac{1 - \expe^{-2\varTh^2}}{2}\right)}
\end{equation}
We see that the magnitude of the rotational drift of a colony increases monotonically with an increase in the demographic variability $ \varTh$ of the flagellar basal attachment angles, with linear proportionality at small demographic variations.  A comparison of this demographic spread of rotational drift with Monte Carlo simulations at various dynamical variances $ \sigmaThe^2$ is presented in Figure~\ref{rot drift sigma fig}.  As illustrated again by comparison with Monte Carlo simulations in Figure~\ref{long time rot drift N}, the magnitude of the rotational drift will tend to decrease with colony size as $N^{-3/2}$ since the rotational drag coefficient $ \gamcolr$ scales cubically with the number of cells $N$ while the radius $ \colrad$ scales linearly.

\subsubsection{Rotational Diffusion}
\label{sec:rotdiffsum}
\begin{equation} \label{eff rotational diffusion_summary} 
\begin{aligned}
\Diffroteff &\equiv \lim\limits_{t\rightarrow \infty}\frac{(\Theta^{(c)}-\meanrot t)^2}{2t}=\Diffrot \\
& \qquad \qquad + 
\frac{1}{2\gamrot^2}
\Big(\sum_{i=1}^N\big[{(aF_i^{(0)})^2}e^{-\sigma_\Theta^2}\gamma_\Theta^{-1}(-\mathrm{Ein} (-\sigma_\Theta^2)+\cos(2\Theta_i^{(0)})\mathrm{Ein} (\sigma_\Theta^2))\\& \qquad \qquad 
+a^2\sigma_F^2\gamma_F^{-1}\big({}_1 F_1[1,\frac{\gamma_F}{\gamma_\theta}+1,-\sigma_\Theta^2]-\cos(2\Theta_i^{(0)})e^{-2\sigma_\Theta^2}{}_1 F_1[1,\frac{\gamma_F}{\gamma_\theta}+1,\sigma_\Theta^2]\big)\big]\Big)
\end{aligned}
\end{equation}
$\mathrm{Ein} (z) \equiv \int_0^z (1-\expe^{-z^{\prime}})/z^{\prime} \, \difd z^{\prime}$ is the complementary exponential integral and ${}_1 F_1$ is Kummer's confluent hypergeometric function \citep{abramowitz64:handbook}.

Taking the mean and variance of this rotational diffusion formula over the demographic distribution of flagella, we obtain:
\begin{equation}  \label{demog avg long time variance of colony orientation}
\begin{aligned}
\bExp\left[\Diffroteff\right]&= \Diffrot + 
 \frac{N\colrad^2}{2\gamrot^2}\left(
 \frac{\expe^{-\sigma_\Theta^2}}{\gamma_{\Theta}}\Fmom{2}\Big(-\mathrm{Ein} (-\sigma_\Theta^2)+\expe^{-2 \varTh^2}\mathrm{Ein} (\sigma_\Theta^2)\Big)\right.\\
 &\left.\qquad \qquad +\frac{\sigma_F^2}{\gamma_F} ({}_1 F_1[1,\frac{\gamma_F}{\gamma_\theta}+1,-\sigma_\Theta^2]-\expe^{-2 \varTh^2-2\sigma_\Theta^2}{}_1 F_1[1,\frac{\gamma_F}{\gamma_\theta}+1,\sigma_\Theta^2])\Big)\right), \\
\Var\left[\Diffroteff\right] &= 
\frac{N\colrad^4}{4\gamrot^4} \\
&  \times \left(\frac{\expe^{-2\sigma_\Theta^2}}{\gamma_\theta^2}\Fmom{4}\Big(\mathrm{Ein}^2 (-\sigma_\Theta^2)-2\Ein (-\sigmaThe^2) \Ein (\sigmaThe^2) \expe^{-2 \varTh^2} + \frac{1+\expe^{-8\varTh^2}}{2} \Ein^2(\sigmaThe^2)\Big)\right.\\
& \qquad \qquad  - \frac{\expe^{-2\sigma_\Theta^2}}{\gamma_\theta^2}(\Fmom{2})^2\Big(-\mathrm{Ein} (-\sigma_\Theta^2)+ \expe^{-2 \varTh^2} \Ein(\sigmaThe^2)\Big)^2 \\
& \qquad \qquad \left.+\frac{\sigma_F^4 \expe^{-4\sigma_\Theta^2}}{2\gamma_F^2} (1 - \expe^{-4 \varTh^2})^2 ({}_1 F_1^2[1,\frac{\gamma_F}{\gamma_\theta}+1,\sigma_\Theta^2])\right), 
\end{aligned}
\end{equation}

\begin{figure}[H]   
\centering
\includegraphics[height=80mm,width=120mm]{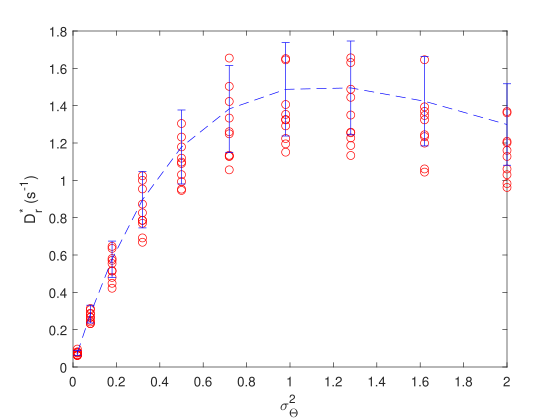}
\caption{Comparison of theoretical (blue) and Monte Carlo simulation (red) results  of rotational diffusivity over a long time ($t=10000 \si{\second}$)  for a colony of $N=10$ cells.   The analytical result is plotted as error bars centered at the mean with width equal to one standard deviation using the formulas from~\eqref{demog avg long time variance of colony orientation}.  The parameters and simulations are the same as in Figure~\ref{rot drift sigma fig}.}
\label{rot diff sigma fig}
\end{figure}

\begin{figure}[H]   
\centering
\includegraphics[height=80mm,width=120mm]{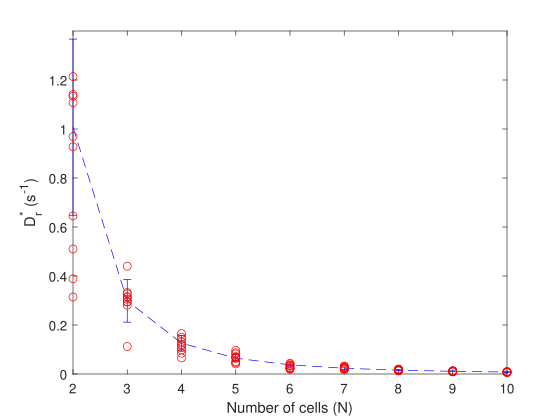}
\caption{
Comparison of theoretical (blue) and Monte Carlo simulation (red) results  of rotational diffusivity over a long time ($t=10000 \si{\second}$)  for colonies of various sizes.   The analytical result is plotted as error bars centered at the mean with width equal to one standard deviation using the formulas from~\eqref{eq:rotdiffmeansmall}, which are visually indistinguishable from the more precise expressions in Eq.~\eqref{demog avg long time variance of colony orientation}. 
The parameters and simulations are the same as in Figure~\ref{long time rot drift N}.}
\label{long time rot diff N}
\end{figure}

These expressions can be expressed more simply for the relevant regime of small flagellar angular displacements (dropping the fourth order terms in the asymptotic expansion):
\begin{equation} \label{eff rotational diffusion_summary_small}
\begin{aligned}
\Diffroteff &= \Diffrot +\frac{a^2}{\gamrot^2} \left(\sum_{i=1}^N\left[(F_i^{(0)})^2 \frac{\sigma_\Theta^2}{\gamma_{\Theta}}
+\sigma_F^2\left(
 \frac{\left(\Theta_i^{(0)}\right)^2}{\gamma_F} + \frac{\sigma_{\Theta}^2}{\gamma_F+\gamma_{\Theta}} \right)\right]\right) 
\end{aligned}
\end{equation}
with demographic mean and standard deviation:
    \begin{align}
\bExp[\Diffroteff] &= \Diffrot + 
\frac{Na^2}{\gamrot^2} \Big(\bExp[(F^{(0)})^2] \frac{\sigma_\Theta^2}{\gamma_{\Theta}}
+\sigma_F^2\left(
 \frac{\varTh^2}{\gamma_F} + \frac{\sigma_{\Theta}^2 }{\gamma_F+\gamma_{\Theta}} \right)\Big), \label{eq:rotdiffmeansmall} 
\\
\Var[\Diffroteff] &= 
\frac{Na^4}{\gamrot^4} \Big(\Var[(F^{(0)})^2] \frac{\sigma_\Theta^4}{\gamma_\Theta^2}
+\frac{2\sigma_F^4
 \varTh^4}{\gamma_F^2}\big)\Big) \nonumber
    \end{align}
These formulas are illustrated as a function of dynamical variance $ \sigmaThe^2$ in comparison with Monte Carlo simulations in Figure~\ref{rot diff sigma fig}.  Using the biophysical parameter values from Table~\ref{Tab:Parameter table}, we verify that the contribution from rotational diffusion due to the active flagellar driving is significantly stronger than the thermal one.  This can be seen in the comparison of the formulas for various colony sizes against Monte Carlo simulations in Figure~\ref{long time rot diff N}.  Since the rotational drag is  proportional to the cube of the colony size the long time rotational diffusion enhancement will decrease with colony size as  $\frac{1}{N^3}$, and its demographic standard deviation as $ \frac{1}{N^{7/2}}$.  As can be expected on physical grounds, the active contribution to the rotational diffusion relies  on dynamical torque fluctuations from the flagella induced both by orientational fluctuations and force fluctuations along a quenched angle deflected from the normal by the cell's orientation within the colony. 

On comparing Figures~\ref{rot drift sigma fig} and~\ref{rot diff sigma fig}, we see that the rotational drift is typically small compared to the rotational diffusivity over most of the potential range of dynamical angular variances plotted.  But when the flagellar orientations are small deviations from the normal, as it appears to be for choanoflagellates (Table~\ref{Tab:Parameter table}), the typical rotational drift and diffusivity for a colony are comparable (Figures~\ref{long time rot drift N} and~\ref{long time rot diff N}).
Indeed, taking the relaxed orientations as symmetrically distributed about the normal with small demographic standard deviation $ \varTh \ll 1$ and neglecting the thermal contribution $ \Diffrot$ to the rotational diffusivity, the mean rotation drift rate~\eqref{eff rotational drift_summary} should scale as $ \meanrot \sim \sqrt{N} \Fmag \colrad \varTh/\gamrot$, while the effective rotational diffusivity should scale as $ \Diffroteff \sim N  \colrad^2 /\gamrot^2 [(\Fmag^2 \sigmaThe^2/\gamma_{\Theta} + \sigma_F^2 (\varTh^2/\gamma_F+\sigmaThe^2/(\gamma_F+\gamma_{\Theta})] $, where $ \Fmag$ denotes a typical magnitude of the force $ F_i^{(0)}$ of a flagellum.  The ratio of the effective rotational drift and effective rotational diffusivity is thus
\begin{equation}
    \frac{\meanrot}{\Diffroteff} \sim N^{-1/2} \left(\frac{\gamrot}{\Fmag \colrad}\right)
    \frac{\varTh}{\sigmaThe^2 \gamma_{\Theta}^{-1} + (\sigma_F/\Fmag)^2 (\varTh^2\gamma_F^{-1}+\sigmaThe^2(\gamma_F+\gamma_{\Theta})^{-1})} \label{eq:rotratiofull}
\end{equation}
Taking $ \sigmaThe^2$ and $ \varTh^2$ as comparable, and $\sigma_F/\Fmag$ as small, consistent with Table~\ref{Tab:Parameter table}, we can simplify this to:
\begin{equation}
    \frac{\meanrot}{\Diffroteff} \sim N^{-1/2} \frac{\gamrot\gamma_{\Theta}}{\Fmag \colrad} \frac{\varTh}{\sigmaThe^2}.
 \label{eq:rotratio}
\end{equation}
We notice the ratio of the physical parameters can be viewed as the ratio of the relaxation rate $ \gamma_{\Theta}$ of the flagellar orientation and the nominal angular frequency scale $ \Fmag \colrad/\gamrot$ a flagellum would impart to the colony, if its force were applied normally to the colony surface (which it is not).  From the numerical parameters in Table~\ref{Tab:Parameter table}:
\begin{equation}
    \frac{\gamrot\gamma_{\Theta}}{\Fmag \colrad} \sim 0.05 N^2, \qquad \frac{\varTh}{\sigmaThe^2} = 10 \label{eq:ratioest}
\end{equation}
so $ \meanrot/\Diffroteff \sim 0.5 N^{3/2}$.  Thus,  for the biophysical parameters in Table~\ref{Tab:Parameter table}, the effective rotational drift and diffusivity are comparable for small colonies but the effective rotational diffusivity is reduced at a faster $ N^{-3}$ rate relative to the $N^{-3/2}$ dependence of the rotational drift.  

The relation between the effective rotational drift and diffusivity can also be expressed somewhat more naturally, under just the assumption that the relaxed flagellar orientations $ \Theta_j^{(0)}$ are symmetrically distributed about the normal,
\begin{equation}
    \frac{\Diffroteff} {\meanrot{}^2}\sim  \frac{\sigmaThe^2}{\varTh^2 \gamma_{\Theta}} + \frac{\sigma_F^2}{\gamma_F\Fmag^2}\left(1+\frac{\sigmaThe^2}{\varTh^2}\right),
    \label{eq:rotsqdiff}
\end{equation} 
These relative sizes of rotational drift and diffusivity have substantial impact on the translational diffusivity as will be seen in Subsection~\ref{sec:transdiff}.

\subsection{Translational Mobility Statistics}
\label{sec:transsum}
Because the environment is isotropic, any initial mean velocity of the colony will dissipate on the time scale of rotational diffusion, and therefore the colony does not exhibit sustained translational drift.

\subsubsection{Mean-Square Speed}
\label{sec:msqspeed}
We therefore look at the instantaneous colony velocity induced by active flagellar forces, $ \Vcol \equiv \frac{\mathbf{F}^{(c)}}{\gamtrans}$.  The thermal Brownian motion component to the colony velocity is not well-defined in our model~\eqref{Colony position sde 1} which coarse-grains inertial dynamics, but on physical grounds would have mean-square value $ \frac{2k_B T}{m} $ where $m$ is the mass of the colony.  We will neglect this thermal component in the results we are about to present, and verify afterward that its physical contribution would be subdominant for the biophysical parameter values in Table~\ref{Tab:Parameter table}. 

An individual colony's mean-square speed (from active flagellar-driven contributions) is:
\begin{align}
\Vcolmsq &=
\gamtrans^{-2}
\left[N \sigma_F^2 + e^{-\sigma_{\Theta}^2}\left|\sum_{j=1}^N F_j^{(0)} 
\expe^{\mathi \Theta_j^{(0)}
+\frac{2\pi\mathi}{N}\left(j+\frac{\flagdisp{j}}{\celllen}\right)}    \right|^2
    + (1-\expe^{-\sigma_{\Theta}^2}) \sum_{j=1}^N\left(F_j^{(0)}\right)^2 \right] 
    \label{eq:msqspeedsum}
\end{align}
The squared modulus of the complex sum is simply the squared norm of the total force on the colony when each flagellum $j$ is exerting its mean force $ F_j^{(0)}$ and its relaxed orientation $ \Theta_j^{(0)} $.  The low variance $ \sigmaThe^2$ of the angular orientation of the flagellar forcing implies a substantial impact of the asymmetrical contributions of forcing from the cells on the instantaneous velocity of the colony. 

We remark that the mean-square displacement of the colony adjusts from an initial ballistic phase to a second ballistic phase over time scales between those of the flagellar dynamics and the colony dynamics:
\begin{align}
&\bExp|\bXcol (t+\tau)-\bXcol (t)|^2 \sim 4 \Difftrans \tau + \Vflagavgsq \tau^2 \text{ for }  \gamma_F^{-1},\gamma_{\Theta}^{-1} \ll \tau \ll \meanrot{}^{-1},\Diffroteff{}^{-1}, \nonumber \\
&\Vflagavgsq \sim 
\gamtrans^{-2} e^{-\sigma_{\Theta}^2}\left|\sum_{j=1}^N F_j^{(0)} 
\expe^{\mathi \Theta_j^{(0)}
+\frac{2\pi\mathi}{N}\left(j+\frac{\flagdisp{j}}{\celllen}\right)}    \right|^2 
= \Vcolmsq - \frac{N \sigma_F^2}{\gamtrans^2}-
\frac{(1-\expe^{-\sigma_{\Theta}^2})}{\gamtrans^2} \sum_{j=1}^N\left(F_j^{(0)}\right)^2\label{eq:coarsemsq}
\end{align}
where the expression for the mean-square velocity inferred over these intermediate time scales is approximated through second order in the dynamical flagellar fluctuations.  We note this mean-square velocity, which might be more relevant for comparison to experiments, differs from the instantaneous mean-square velocity~\eqref{eq:msqspeedsum} by terms second order in the dynamical flagellar fluctuations.  It is simply the square of the colony speed $\Veleff$ induced by each flagellum $j$ applying its mean force $F_j^{(0)}$ about its mean orientation $ \Theta_j^{(0)} $ relative to the normal from the attachment point, multiplied by $ \expe^{-\sigmaThe^2}$ to account for the averaging over dynamical flagellar fluctuations.  We continue our discussion in terms of the instantaneous mean-square velocity~\eqref{eq:msqspeedsum}; rather similar conclusions with slight adjustments apply for the mean-square velocity observed over longer intervals in Eq.~\eqref{eq:coarsemsq}.

The demographic average and standard deviation, derived in Appendix~\ref{app:demovar}, are:
\begin{align}
&\bExp[\Vcolmsq] 
= N\gamtrans^{-2} \left(\sigma_F^2+ \Fmom{2}
-\expe^{-\varTh^2-\sigmaThe^2} \sinc (\flagdispran/\colrad) (\Fmom{1})^2 \right) \label{eq:vmsqmean}\\ 
&     \Var (\Vcolmsq) 
 = N^2 \expe^{-2\sigmaThe^2}\gamtrans^{-4}  \left(\Fmom{2}-\expe^{-\varTh^2} \sinc^2 (\flagdispran/\colrad) (\Fmom{1})^2 )\right)^2 \label{eq:vmsqvar} \\
     & \qquad \qquad +4 N\gamtrans^{-4}(1- \expe^{-\sigmaThe^2} \expe^{-\varTh^2}  \sinc^2(\flagdispran/\colrad))  \Fcum{3} \Fmom{1} \nonumber\\
     & \qquad \qquad 
     +N \gamtrans^{-4} \Fcum{4} +
     N \gamtrans^{-4} (2- \expe^{-2\sigmaThe^2}- \expe^{-2\sigmaThe^2-4\varTh^2}
    \sinc^2(2\flagdispran/\colrad))  (\Fmom{2})^2 \nonumber\\
& \qquad \qquad -2N\gamtrans^{-4}(1-4\expe^{-\sigmaThe^2-\varTh^2}\sinc^2(\flagdispran/\colrad)+3\expe^{-2\sigmaThe^2-2 \varTh^2}
    \sinc^4(\flagdispran/\colrad))  (\Fmom{1})^4 
    \nonumber\\
& \qquad \qquad    +4N \gamtrans^{-4}\sinc^2(\flagdispran/\colrad)
    (-2\expe^{-\sigmaThe^2-\varTh^2}+ \expe^{-2\sigmaThe^2-3\varTh^2}\sinc(2\flagdispran/\colrad)+\expe^{-2\sigmaThe^2-\varTh^2})(\Fmom{1})^2  \Fmom{2}.    \nonumber
\end{align}
 The representation of the variance fully in terms of cumulants, by the substitution $ \Fmom{2} = \Fcum{2}+(\Fmom{1})^2 $ and $ \Fmom{1} = \Fcum{1}$ leads to a more cumbersome expression.
We present the demographic statistics for the mean-square speed rather than the root-mean-square speed because they have a less complicated analytic expression.  We can simplify the expressions somewhat by neglecting the demographic and dynamical fluctuation in flagellar orientation relative to that in the flagellar force ($ \varTh, \sigmaThe \downarrow 0$) to obtain: 
    \begin{align}
\bExp[\Vcolmsq] 
&= N \gamtrans^{-2} \left(\sigma_F^2 + \Var[F^{(0)}]+\left(1-\sinc\left(\frac{\flagdispran}{\colrad}\right)\right)(\bExp[F^{(0)}])^2\right) \label{demographic_speed_disp_force_summary} \\ 
     \Var (\Vcolmsq) 
     & = N^2 \gamtrans^{-4} \left[\Var[F^{(0)}]+\left(1-\sinc^2\left(\frac{\flagdispran}{\colrad}\right)\right)(\bExp[F^{(0)}])^2\right]^2\nonumber\\
  & \qquad \qquad +4N \gamtrans^{-4}(1- \sinc^2(\flagdispran/\colrad)) \Fcum{3} \Fmom{1} \nonumber\\
     & \qquad \qquad 
     +N \gamtrans^{-4}\Fcum{4} + N\gamtrans^{-4}(1- 
    \sinc^2(2\flagdispran/\colrad)) (\Fmom{2})^2 \nonumber\\
& \qquad \qquad 
-2N\gamtrans^{-4}(1-4\sinc^2(\flagdispran/\colrad)+3
    \sinc^4(\flagdispran/\colrad))(\Fmom{1})^4 
    \nonumber\\
& \qquad \qquad    +4N \gamtrans^{-4}\sinc^2(\flagdispran/\colrad)
    (-1+\sinc(2\flagdispran/\colrad)) (\Fmom{1})^2  \Fmom{2}    \nonumber    
\end{align}
Further neglecting the demographic fluctuation in flagellar placement, so that only variations in flagellar force magnitude are taken into account, we take $ \flagdispran/\colrad, \varTh, \sigmaThe \downarrow 0$ to obtain:
    \begin{align}
\bExp[\Vcolmsq] 
&= N \gamtrans^{-2} \left(\sigma_F^2 + \Var[F^{(0)}]\right) \label{demographic_speed_only_force_summary} \\ 
     \Var (\Vcolmsq) 
     & = \gamtrans^{-4} \left[N^2 (\Var[F^{(0)}])^2 + N \Fcum{4}\right]. \nonumber
\end{align}

\begin{figure}[H]	 
{\centering{
\includegraphics[height=80mm,width=130mm]{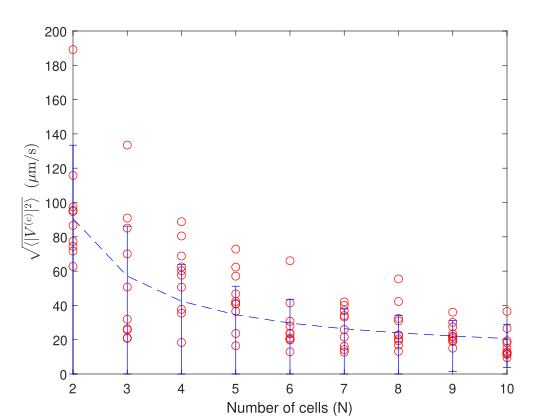}}
\caption{Comparison of theoretical (blue) and simulated (red) results of colony root mean square speed plotted against population size for a colony with $S_i\sim U(\frac{-l}{2},\frac{l}{2})$ for each cell.
The colony population size ranges from 2 to 10 cells, with 10 independently sampled colonies for each cell size. The theoretical values are plotted as $\sqrt{\bExp[\Vcolmsq]}$ with error bars extending from $\sqrt{\bExp[\Vcolmsq] \pm \sqrt{\Var(\Vcolmsq)}} $, using the formulas in Eq.~\eqref{demographic_speed_disp_force_summary}. 
For most cell sizes, in fact the theoretical standard deviation of the mean-square speed exceeds its mean, in which case the error bars are terminated below at $0$ speed.
The rest of the parameters are specified in Appendix~\ref{sim_settings} and Table \ref{Tab:Parameter table}.
}
\label{speedsize}}
\end{figure}
  
The approximation used in the theoretical formulas~\eqref{demographic_speed_disp_force_summary} seem to be well satisfied for the biophysical parameter values in Table~\ref{Tab:Parameter table}, and indeed are shown in Figure~\ref{speedsize} to agree well with Monte Carlo simulations.
The even simpler formula~\eqref{demographic_speed_only_force_summary} for the statistics of the root-mean-square velocity, which neglects variability in flagellar placement, is not accurate for small colony sizes and underpredicts within about 20\% of the more precise formula~\eqref{demographic_speed_disp_force_summary} for $ N \geq 6$ for our parameter choices.  
To check on the neglect of the thermal contribution to the colony root-mean-square speed, we take a single cell to be a sphere of diameter $\SI{5}{\micro\meter}$, the diameter of a cell~\citep{dayelmorphogenesis,colonialmotility}, and have mass density on the order of $ 1 \si{\gram/\centi\meter^3}$.  Then we estimate the mass of a single cell as $M \approx 60 \si{\pico\gram}$, and thus the root-mean-square thermal speed scale of a single cell as $ \sqrt{\frac{2k_B T}{M}} \sim \SI{10}{\micro\metre/\second}$.  The thermal speed of the colony will decrease with size, and we confirm from Figure~\ref{speedsize} that the colony speed contribution from active flagellar forces is about an order of magnitude greater than that of thermal effects.

On comparison with the experimental data for the choanoflagellate \emph{S. rosetta} from \citet[Fig. 4]{colonialmotility}, the speeds from our theoretical model are an order unity multiple too large, but this may be attributed to an underrepresentation of the theoretical drag by our treatment of the swimming cells as wedges rather than osculating discs or spheres of the observed diameter.  More importantly, our theoretical model predicts a general $N^{-1/2}$ decrease of the root-mean-square-speed with colony size (as $ \gamtrans \propto N$), while~\citet{colonialmotility} observe a slow increase of speed with colony size. This was attributed, in a three-dimensional model, to the number of flagella scaling as the square of the colony radius while the drag scales only linearly.  This effect would be lost in our two-dimensional disclike model, where the number of flagella scales only linearly with colony radius.  The observed trend of a speed increase with colony size is still a bit mysterious because it would require all the flagella in the three-dimensional model to coordinate their orientation to a common direction, which would seem a bit awkward for the cells situated away from the rear of the direction of motion.  Our model treats each flagellum as autonomous and indifferent to the direction of colony propulsion so would have no such coordination effect.  One other feature observed in~\citet[Fig. 4]{colonialmotility} is the wide variability in the speed of colonies of a certain size, with the standard deviation being about half the average swimming speed.  Figure~\ref{speedsize} shows similar behavior in our model, with greater variability at smaller colony sizes.  For large colony sizes, our theoretical results~\eqref{eq:vmsqmean} and~\eqref{eq:vmsqvar} give
\begin{equation*}
 \frac{\sqrt{\Var (\Vcolmsq)}}{\bExp[\Vcolmsq]} 
\sim  \frac{\expe^{-\sigmaThe^2} \Fmom{2}}
{\sigma_F^2+ \Fmom{2}} \text{ as } N \rightarrow \infty,
\end{equation*}
which should be around $1$ when the demographic variability in the flagellar force $\Var[F^{(0)}]$ is substantially greater than the dynamical variance in flagellar force $ \sigma_F^2$ and the dynamical orientation variance $ \sigmaThe^2$ is small, as is the case for the biophysically relevant parameter values inferred in Table~\ref{Tab:Parameter table}. 

\citet{KoehlCapture} compare the swimming speeds of unicellular and multicellular forms of \emph{S. helianthica}, another choanoflagellate with a similar morphology to \emph{S. rosetta}, but presumably different biophysical parameters.  One of their  observations is the swimming speed of multicellular colonies (of up to $N=13 $ cells) is comparable to that of unicellular colonies, and that the standard deviation of their speeds are no more than about a third of the mean.  Our model does not even qualitatively explain these features.  Among various reasons for the disagreement, one could be that our model is for an isolated swimming colony while the experiments generally involve suspensions.

\subsubsection{Translational Diffusivity}
\label{sec:transdiff}
The translational diffusivity of the colony becomes unwieldy to calculate for general noise parameters, so we
will specialize to the case where the dynamical fluctuations $ \sigmaThe$ of flagellar orientation are small, and the magnitude $ \sigma_F $ of dynamical force variations  is small relative to the magnitude of the relaxed flagellar forces $ F_j^{(0)}$.
Our results are presented  perturbatively through second order in these parameters being assumed small.     Note we are allowing substantial demographic variations in flagellar force magnitudes $ F_j^{(0)} $ and relaxed orientations $ \Theta_j^{(0)} $, and are stipulating only that the \emph{dynamical} variations in flagellar force magnitude and orientation are relatively small.  This is important for the generalization of our calculation to other geometries in Section~\ref{sec:gengeo}.

We also assume
that the effective rotational drift and diffusivity of the colony, including the active contributions from the flagella, occur slowly relative to the flagellar fluctuation time scale:
\begin{align}
\meanrot, \Diffroteff \ll  \gamma_F, \gamma_{\Theta}. \label{eq:rottimesep}
\end{align}
The colony force autocorrelation function will then decay on a longer time scale than the time scale of adjustment of the individual flagellar forces.  This in particular requires the thermal rotational diffusion rate $\Diffrot$ to be small compared to $ \gamma_F$ and $ \gamma_{\Theta} $, which amounts to the well-satisfied relation $0.4 N^{-3} \si{\sec}^{-1} \ll 10 \si{\sec}^{-1}$ for the biophysical parameters from Table~\ref{Tab:Parameter table}.   Figure~\ref{long time rot diff N} show that the inequality~\eqref{eq:rottimesep}, when including active contributions, is still satisfied by about a factor of 10, with a wider ratio for larger colonies, for the same specified biophysical parameter values.  

The above assumptions are fundamental to our approximate calculation, and a further mild technical assumption allows us to avoid tedious consideration of uninteresting cases:
\begin{itemize}
   \item The time scales of relaxation of flagellar force and orientation are comparable: 
   \begin{equation}
\gamma_F/\gamma_{\Theta} \sim \ord(1). \label{eq:gamone}
   \end{equation}
\end{itemize}
Indeed, we find this condition satisfied by the parameters we inferred from experimental observations in Table~\ref{Tab:Parameter table}.

Under the assumptions described above, we obtain the following approximation for the translational diffusivity:

\begin{align} 
 \lim\limits_{t\rightarrow \infty}\frac{\langle \bXcol (t) \otimes \bXcol (t)\rangle}{2t} &= \Difftranseff \Id , \nonumber\\
 \Difftranseff &\equiv 
\Difftrans + \frac{\Vflagavgsq}{2}
 \frac{\Diffroteff}{\Diffroteff{}^2+\meanrot{}^2} + \Difftransrot \label{eff translational diffusion_summary}
\end{align}
where $ \Vflagavgsq$ is the mean-square velocity coarse-grained over the flagellar time scale~\eqref{eq:coarsemsq}, and
\begin{align*}
&\Difftransrot \equiv   
 \frac{\colrad \meanrot}{2\gamtrans^2 \gamrot (\meanrot{}^2+\Diffroteff{}^2)}\left[\frac{2\sigma_F^2}{\gamma_F}\sum_{j,\jp=1}^N F_j^{(0)} \sin \Theta_{\jp}^{(0)} 
    \cos\left(\Theta_j^{(0)}-\Theta_{\jp}^{(0)}
    +\frac{2\pi}{N}\left(j-\jp+\frac{\flagdisp{j}-\flagdisp{\jp}}{\celllen}\right)\right)\right.\nonumber \\
    & \left. \qquad + \frac{\sigma_{\Theta}^2}{\gamma_{\Theta}} \sum_{j,\jp=1}^N 
    F_j^{(0)} F_{\jp}^{(0)} (F_{\jp}^{(0)} \cos (\Theta_{\jp}^{(0)}) -F_{j}^{(0)} \cos (\Theta_{j}^{(0)}))  
    \sin\left(\Theta_j^{(0)}-\Theta_{\jp}^{(0)}
    +\frac{2\pi}{N}\left(j-\jp+\frac{\flagdisp{j}-\flagdisp{\jp}}{\celllen}\right)\right)\right]
\end{align*}
Here $\meanrot$ is the effective rotational drift coefficient and $\Diffroteff$ is the effective rotational diffusion coefficient given in (\ref{eff rotational drift_summary}) and (\ref{eff rotational diffusion_summary}) respectively.  As shown in Subsection~\ref{sub:transstat}, the error in our calculation is essentially fourth order in the small parameters $ \sigmaThe$ and $ \sigma_F$.

The first two terms of the expression in Eq.~\eqref{eff translational diffusion_summary} are exactly the translational diffusion for a colony with thermal diffusivity $ \Difftrans$, moving at a constant speed $ \Veleff$ along an orientation with constant rotational drift $ \meanrot$ and constant rotational diffusion $ \Diffroteff$.  
As we shall shortly explain, the complicated terms in  $ \Difftransrot$ are included for completeness and consistency with the results presented for general geometries in Subsection~\ref{sub:gensum}, but should typically be considerably smaller than the simpler terms presented in Eq.~\eqref{eff translational diffusion_summary} for disclike geometries.

Due to the nonpolynomial dependence of the translational diffusion on the effective rotational statistics, a rigorous computation of the demographic mean is complicated.  We therefore content ourselves with an estimate for the demographic mean by averaging separately the numerator and denominator, neglecting demographic correlations between the colony speed and rotational characteristics, and dropping the small contribution from $ \Difftransrot $:
\begin{equation}
 \label{demographic eff translational diffusion_summary}
   \bExp[ \Difftranseff] \approx  
   \bExp[\Vflagavgsq]
 \frac{\bExp[\Diffroteff]}{2(\Var[\Diffroteff] + (\bExp[\Diffroteff])^2+\Var[\meanrot])}
+ \Difftrans.
\end{equation}
All component demographic averages have been reported above, other than $ \bExp[\Vflagavgsq]$ which is clearly representable as a linear combination of $ \bExp[\Vcolmsq]$ and $ \bExp[(F_j^{(0)})^2]$.  We do not report a demographic standard deviation because it is cumbersome, and in any case is simply a compounding of the demographic fluctuations from the mean-square speed~\eqref{eq:vmsqvar} and the the rate of velocity decorrelation induced by the effective rotational drift~\eqref{eff rotational drift_sd_summary} and rotational diffusion~\eqref{eq:rotdiffmeansmall} of the colony.  

\begin{figure}[H]	 
{\centering{
\includegraphics[height=75mm,width=130mm]{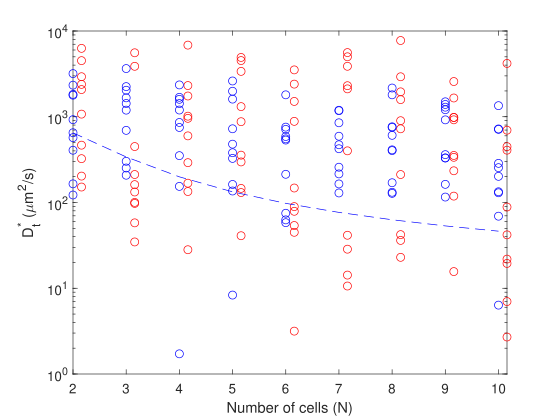}
}
\caption{
Comparison of theoretical (blue) and simulated (red) results of the effective  translational diffusivity plotted against colony size. For each colony size, we generate 10 colonies with flagellar displacements modeled as $S_i\sim U(\frac{-l}{2},\frac{l}{2})$ for each cell. For each of these colonies, the theoretical diffusivity computed from~\eqref{eff translational diffusion_summary} using the peculiar flagellar parameters of that colony is plotted as a blue circle, while the demographic mean formula~\eqref{demographic eff translational diffusion_summary} is plotted as a blue dashed line.  The contribution from terms in $ \Difftransrot $ are neglected in the theoretical calculation.  The red circles, offset horizontally for clarity, represent the translational diffusivity estimated from simulations.
The rest of the parameters are specified using Section~\ref{sim_settings} and Table~\ref{Tab:Parameter table}. 
\label{diffusion_size}} }
\end{figure}

As noted in Subsection~\ref{sec:msqspeed}, the mean-square colony speed scales inversely with colony size $ N$.  

The second factor in the expressions~\eqref{eff translational diffusion_summary} and~\eqref{demographic eff translational diffusion_summary} is essentially a rotational decorrelation time scale, with this rotational time being determined by rotational diffusivity when it dominates rotational drift, and by a shorter time scale when rotational drift is stronger.  The reason for the reduction in the translational diffusivity by rotational drift is that the steady rotational drift causes (in absence of rotational diffusion) a perfect cancellation of sustained translational motion.  
From Figures~\ref{long time rot drift N} and~\ref{long time rot diff N}, we see the typical rotational drift is somewhat larger than the rotational diffusivity using the biophysical parameters from Table~\ref{Tab:Parameter table}, noting moreover from the discussion in Subsection~\ref{sec:rotdriftsum} and Subsection~\ref{sec:rotdiffsum} that rotational drift decays more slowly ($\sim N^{-3/2}$) with colony size than rotational diffusion ($\sim N^{-3}$).  Thus, when the rotational drift dominates in Eq.~\eqref{demographic eff translational diffusion_summary}, the rotational decorrelation time scale (second factor) is not substantially varying with $ N$ and the demographically averaged translational diffusivity should decrease inversely with colony size $N$.

We show in Figure~\ref{diffusion_size} a plot of  the theoretical formulas for translational diffusivity compared against simulations.  The actual magnitude is presumably an order of magnitude too large because of our disc model's underestimate of the translational drag.  Focusing on the qualitative features, we see the theoretical predictions based on the properties of specific colonies (blue circles) agree roughly with the simulated results (red circles), while the approximation~\eqref{demographic eff translational diffusion_summary} for the demographic average of the translational diffusivity (dashed line) does not seem so accurate.  In particular, the approximation~\eqref{demographic eff translational diffusion_summary} for the demographic average predicts an order of magnitude decrease in the theoretical translational diffusivity as the colony size increases from $N=2$ to $10$, while neither the simulations nor the distribution of theoretical predictions for specific colonies show evidence of a decrease with colony size.  We also observe the demographic variability over two orders of magnitude for the theoretical translational diffusivity for a given colony size, while the simulations show a somewhat more pronounced variability, with a few colonies tending to show considerably higher translational diffusivity than theoretically predicted.  The strong variabilities in the theoretical translational diffusivity, and therefore the inadequacy of the averaging approximation in~\eqref{demographic eff translational diffusion_summary}, can be traced to the variability in the rotational drift contribution $ \meanrot{}^2$, which becomes typically more dominant in the denominator as the colony size increases. Recall from Subsection~\ref{sec:rotdriftsum} that the rotational drift has mean zero due to symmetry but a substantial demographic variation.  For colonies with a low rotational drift, the rotational decorrelation time is anomalously increased and the theoretical translational diffusivity well above the demographic mean.

Referring to Eq.~\eqref{eq:rotratiofull}, we see that
the ratio of the more complex expression proportional to $ \meanrot$ to the expression proportional to $\Diffroteff$ in Eq.~\eqref{eff translational diffusion_summary} is no larger than
\begin{align}
\frac{2\Difftransrot (\Diffroteff{}^2+\meanrot{}^2)} 
{\Diffroteff \Vflagavgsq}
&\sim    O\left(\frac{\meanrot}{\Diffroteff}\frac{\colrad \Fmag}{\gamrot}\left[\sigmaThe^2\gamma_{\Theta}^{-1} + (\sigma_F^2/\Fmag^2)\gamma_F^{-1}\varTh\right]\right) \sim 
    O\left(N^{-1/2}\varTh \frac{\sigmaThe^2\gamma_{\Theta}^{-1} + (\sigma_F^2/\Fmag^2)\gamma_F^{-1}\varTh}{\sigmaThe^2\gamma_{\Theta}^{-1} + (\sigma_F^2/\Fmag^2)\gamma_F^{-1} \varTh^2}\right)  \label{eq:sincosrat} \\
    & \sim \begin{cases}
    O(N^{-1/2}\varTh^2) & \text{ if } (\sigma_F^2/\Fmag^2)\gamma_F^{-1} \varTh \lesssim \sigmaThe^2\gamma_{\Theta}^{-1}, \\
    O\left(N^{-1/2}\varTh^2 \frac{ (\sigma_F^2/\Fmag^2)\gamma_F^{-1}}{\sigmaThe^2\gamma_{\Theta}^{-1}}\right) & \text{ if } (\sigma_F^2/\Fmag^2)\gamma_F^{-1} \varTh^2 \ll
    \sigmaThe^2 \gamma_{\Theta}^{-1} \ll  (\sigma_F^2/\Fmag^2)\gamma_F^{-1} \varTh \\
    O(N^{-1/2}) & \text { if } (\sigma_F^2/\Fmag^2)\gamma_F^{-1} \varTh^2 \gg \sigmaThe^2 \gamma_{\Theta}^{-1}.
    \end{cases} \nonumber
\end{align}

This is small when the colony size is large, and particularly small if $ \varTh$ is small and the relative strength of force magnitude fluctuations is not much larger that those of the force orientation fluctuations.
 
For typical situations within the paradigm of small stochastic variations in the flagellar forces and orientation variations from the normal to the surface, $ \Difftransrot$ should not contribute substantially relative to the simple terms in~\eqref{eff translational diffusion_summary}.  Indeed, for the biophysical parameters listed in Table~\ref{Tab:Parameter table}, the ratio in Eq.~\eqref{eq:sincosrat} is approximately $ 0.04 N^{-1/2}$.  This correction term is not sufficiently potent to rectify the discrepancies seen in Figure~\ref{diffusion_size}.

Note the terms $ \Difftransrot$  arise from correlations between individual flagellar dynamics and the rotation of the colony as a whole.  It appears to be related to the wobbling of swimming choanoflagellate cells observed in the computational simulations of~\citet{Fauci_morphology}, though we'd expect the wobbling to be reduced for colonies due to the increased rotational drag.  The argument for the smallness of $ \Difftransrot$ relies largely on the disclike colony configuration, with nearly normal flagella, and does not apply to general colony configurations.  Indeed, the rotational drift should become even stronger relative to the rotational diffusivity in less symmetric colonies.  Thus the terms which generalize $ \Difftransrot $ to more general nearly planar colony configurations are retained in Eq.~\eqref{eq:transdiffgen}.

\section{Methods}
\label{sec:methods}
In this section we will present the detailed derivations for the transport statistics reported in Section~\ref{summarych1} for the disc colony described in Section~\ref{sec:colmod}.   Rotational statistics are computed in Subsection~\ref{sub:rotstat}, and translational statistics in Subsection~\ref{sub:transstat}.  
All statistical averaging operations here, denoted by angle brackets $ \langle \cdot \rangle$, are over dynamical noise only, for fixed colony parameters.  Thus, in contrast to Section~\ref{summarych1}, we will use $ \Cov$ and $ \Var$ to denote covariance and variance of random variables with respect to the dynamical noise.  The subsequent demographic averages are all straightforward except for the mean-square speed; the demographic statistic for that quantity is derived in detail in Appendix~\ref{app:demovar}.

\subsection{Rotational Statistics}
\label{sub:rotstat}
Under the model described in Subsection~\ref{sec:cellmod}, the statistics for the flagellar force magnitude $ F_i (t)$ and orientation $ \Theta_i (t) $ are all independent, statistically stationary Gaussian random processes with means $ \langle F_i \rangle = F_i^{(0)} $ and $\langle \Theta_i\rangle =\Theta_i^{(0)} $ and correlation functions

\begin{equation}  \label{autocorrelation}
\Cov(\Theta_i(t),\Theta_i(t'))=(\sigma_\Theta^2)e^{-\gamma_\Theta|t-t'|}, \qquad \qquad
\Cov(F_i(t),F_i(t'))=(\sigma_F^2)e^{-\gamma_F|t-t'|}
\end{equation}
In particular, the correlation time of $\Theta_i$ is $\frac{1}{\gamma_\Theta}$ and for $F_i$ is $\frac{1}{\gamma_F}$.

We proceed next to compute the first and second order statistics of the total torque~\eqref{Agg. active torque} on the colony.  Beginning with the mean, we use 

the independence of $F_i$ and $\Theta_i$ and Eq.~\eqref{eq:trigavg} to write:
 
\begin{equation} \label{mean torque}
\begin{aligned}
\langle T^{(c)}(t)\rangle =-\sum_{i=1}^N \langle F_i(t)\rangle a\langle \sin(\Theta_i(t))\rangle =-\sum_{i=1}^N F_i^{(0)}a\sin(\Theta_i^{(0)})e^{-\frac{1}{2}\sigma_\Theta ^2}
\end{aligned}
\end{equation}
This mean torque induces the mean rotational velocity for a given colony (see Eq.~\eqref{Colony orientation sde 1}):
\begin{equation*}
    \meanrot \equiv \lim_{t \rightarrow \infty} \frac{\langle \Thetacol (t) - \Thetacol (0) \rangle}{t} = -\frac{a}{\gamcolr}e^{-\frac{1}{2}\sigma_\Theta ^2} \sum_{i=1}^N F_i^{(0)}\sin(\Theta_i^{(0)})
\end{equation*}

To compute the variance of the colony orientation, we require the autocorrelation function of the colony torque.  We begin by using independence of all component stochastic processes to write:
\begin{equation} 
\begin{aligned}
\Cov[T^{(c)}(t),T^{(c)}(t')]&=\Cov[-\sum_{i=1}^N F_i(t)a\sin(\Theta_i(t)),-\sum_{j=1}^N F_j(t')a\sin(\Theta_j(t'))]
\\&=\sum_{j=1}^N \Cov [F_j(t)a\sin(\Theta_j(t)),F_j(\tp)a\sin(\Theta_j(\tp))] \\
&=  \sum_{j=1}^N \left[
 \langle (F_j (t) a \sin \Theta_j (t))(F_j (\tp) a \sin \Theta_j (\tp)) \rangle -  \langle F_j (t) a \sin \Theta_j (t) \rangle
 \langle F_j (\tp) a \sin \Theta_j (\tp) \rangle\right]
   \\
&   = a^2 \sum_{j=1}^N \langle F_j (t) F_j (\tp) \rangle 
  \frac{1}{2} \langle \cos (\Theta_j (t)-\Theta_j (\tp)) - \cos (\Theta_j (t) +\Theta_j (\tp))\rangle\\
  & \qquad \qquad - a^2 \sum_{j=1}^N (F_j^{(0)})^2 \sin^2 \Theta_j^{(0)} \expe^{- \sigma_{\Theta}^2}  
\end{aligned} 
\end{equation}

Next, we compute each summand using the formula~\eqref{eq:trigavg} for the average of the sinusoid of a Gaussian random variable:
\begin{equation*}
\begin{aligned}
&\Cov [F_j(t)a\sin(\Theta_j(t)),F_j(\tp)a\sin(\Theta_j(\tp))]  \\
  &  \qquad \qquad  = \frac{1}{2} a^2 \left(\sigma_F^2 \expe^{-\gamma_F |t-\tp|}+(F_j^{(0)})^2\right) \left(\expe^{-\frac{1}{2} \Var[\Theta_j (t)-\Theta_j (\tp)]} - \cos (2\Theta_j^{(0)})\expe^{-\frac{1}{2} \Var[ \Theta_j (t)+\Theta_j (\tp)]}\right)
  \\
  & \qquad \qquad \qquad \qquad - a^2 (F_j^{(0)})^2 \sin^2 \Theta_j^{(0)} \expe^{- \sigma_{\Theta}^2}  \\
& \qquad \qquad =\frac{a^2}{2} \left((\sigma_F^2)e^{-\gamma_F|t-t'|}+(F_j^{(0)})^2\right)(e^{[-\sigma_\Theta^2+\sigma_\Theta^2e^{-\gamma_\Theta|t-t'|}]}-\cos(2\Theta_j^{(0)})e^{[-\sigma_\Theta^2-\sigma_\Theta^2e^{-\gamma_\Theta|t-t'|}]})\\
&   \qquad \qquad \qquad \qquad  -a^2(F_j^{(0)})^2\frac{1}{2}(1-\cos(2\Theta_i^{(0)})))e^{-\sigma_\Theta^2}
\end{aligned} 
\end{equation*}

We thereby obtain
\begin{equation} \label{autocorrelation of torque}
\begin{aligned}
&\Cov[T^{(c)}(t),T^{(c)}(t')] \\
& \qquad \sum_{j=1}^N \frac{a^2}{2} ((\sigma_F^2)e^{-\gamma_F|t-t'|}+(F_j^{(0)})^2) (e^{[-\sigma_\Theta^2+\sigma_\Theta^2e^{-\gamma_\Theta|t-t'|}]}-\cos(2\Theta_j^{(0)})e^{[-\sigma_\Theta^2-\sigma_\Theta^2e^{-\gamma_\Theta|t-t'|}]})\\
   &  \qquad \qquad -\frac{a^2}{2}(F_j^{(0)})^2(1-\cos(2\Theta_j^{(0)})))e^{-\sigma_\Theta^2}
\end{aligned}
\end{equation}

We can then express the evolution of the variance of the colony orientation by noting the independence of the active torque and thermally driven rotational diffusion:
\begin{equation}\label{eq:thetafromT} 
\begin{aligned}
\Var(\Theta^{(c)}(t))&= \Var(\frac{1}{\gamrot^2}\int_0^t T^{(c)}(t')dt')+\Var(\sqrt{2\Diffrot}W^{\Theta,c}(t))\\
&=\left(\frac{1}{\gamrot}\right)^2\int_0^t\int_0^t \Cov(T^{(c)}(t''),T^{(c)}(t'))dt''dt'+2\Diffrot t
\end{aligned}
\end{equation}
We are primarily interested in the long-time statistical behavior of the orientation.  To this end, we observe that $ \Cov(T^{(c)}(t''),T^{(c)}(t'))$ is a function only of $ |\tpp-\tp|$, and that for integrable functions $ g$,
a change of variables to $u=t''-t'$ and $u^{\prime}=t''+t'$ gives:
\begin{equation}
\begin{aligned}
\int\limits_{0}^{t}\int\limits_{0}^{t}g(|t''-t'|)dt'dt''&=
\frac{1}{2}\int\limits_{-t}^{t}\int\limits_{|u|}^{2t-|u|}g(|u|)\difd u^{\prime} \, \difd u 
=2\int\limits_{0}^{t}(t-u)g(u)\, \difd u \sim 2t \int_0^{\infty} g(u) \, \difd u \text{ as } t \rightarrow \infty. \label{eq:kuboint}
\end{aligned}
\end{equation}

To evaluate the time integral, we use the following formulas, which involve a change of integration variable to $ u=\expe^{-kt}$
\begin{equation*}
\begin{aligned}
    \int_0^{\infty} (\expe^{c\expe^{-kt}}-1)  \,  \difd t &= \int_0^1 \frac{\expe^{cu}-1}{ku} \, \difd u = -k^{-1} \Ein (-c) \text{ for } c,k \in \mathbb{R}, k >0, \\
    \int_0^{\infty}
    \expe^{-bt+c\expe^{-kt}} \,  \difd t &=k^{-1} \int_0^1 u^{b/k-1} \expe^{cu} \, \difd u = \frac{1}{b} {}_1 F_1 (b/k,b/k+1,c) \\& = \frac{1}{b} \expe^{c} {}_1F_1 (1,b/k+1,-c)
     \text{ for } b,c,k \in \mathbb{R}, b,k >0
\end{aligned}
\end{equation*}
where we have used integral representations and identities for the entire exponential integral $ \Ein $  and the confluent hypergeometric function of the first kind ${}_1 F_1  $.  Applying the asymptotic integral formula~\eqref{eq:kuboint} with these integral identities to the integral of the active torque correlation function~\eqref{autocorrelation of torque} yields the result~\eqref{eff rotational diffusion_summary}.

\subsection{Translational Statistics Derivations}
\label{sub:transstat}
To compute the translational mobility of the colony, we next examine the force on the colony~\eqref{Agg. active force}. We will use complex notation for expressing vector calculations, with $\mathi$ in these calculations representing the imaginary unit.
\iffalse Again, terms requiring noisy fluctuations from both the magnitude and direction of the cell's propulsive forcing will be neglected.\fi
\begin{align} \label{mean colony forcing 1}
 \mathbf{F}^{(c)} (t)&=\sum_{j=1}^{N} -F_j(t)\begin{pmatrix}\cos\left(2\pi\left[\frac{j-\frac{1}{2}+\frac{S_j}{l}}{N}\right]+\Theta_j(t)+\Theta^{(c)}(t)\right)\\\sin\left(2\pi\left[\frac{j-\frac{1}{2}+\frac{S_j}{l}}{N}\right]+\Theta_j(t)+\Theta^{(c)}(t)\right)\end{pmatrix} \nonumber \\
&\sim -\sum_{j=1}^N F_j(t)e^{\mathi\left[2\pi\left(\frac{j-\frac{1}{2}+\frac{S_j}{l}}{N}\right)+\Theta_j(t)+\Theta^{(c)}(t)\right]}
\end{align}

While individual colonies will have asymmetrically arranged flagella, the isotropic enviroment will ensure that the colony orientation $ \Thetacol (t)$ will eventually become uniformly distributed, implying an eventual zero mean to the total force vector on the colony.  In more detail, as the calculations in~\citet{Ashenafi_thesis} show, the mean force on the colony relaxes on a time scale $ \max (\gamma_F^{-1},\gamma_\Theta^{-1}) $ from the initial flagellar forces and orientations to a flagellar-averaged form, which then rotates with drift $ \meanrot $~\eqref{eff rotational drift_summary} and dissipates at a rate $ \Diffroteff$ due to rotational diffusion.  The flagellar-averaged root-mean-square force will be recovered in the calculation of the colony force covariance to follow.  The translational drift of the colony is proportional to the colony force, and thus will eventually tend to zero due to the randomization of its direction.

The homogeneity and isotropy of the environment implies the long-time dynamics will also be statistically isotropic, so  we have
\begin{equation}
    \Cov(\Fcol (t),\Fcol (\tp)) \sim \frac{1}{2}\langle \Fcol (t) \cdot \Fcol (\tp) \rangle \Id \text{ for } t, \tp \rightarrow \infty.
    \label{eq:fcoliso}
\end{equation}
where $ \Id$ denotes the identity matrix.  Using the complex representation of the vectorial colony force~\eqref{mean colony forcing 1}, we have:
\begin{equation*}
   \langle |\Fcol (t)|^2\rangle = \sum_{j=1}^N \sum_{\jp=1}^N 
  \expe^{\frac{2\pi \mathi}{N}\left(j-\jp+\frac{\flagdisp{j} - \flagdisp{\jp}}{\celllen}\right)} \langle F_j (t) F_{\jp} (t) \expe^{\mathi (\Theta_j (t)-\Theta_{\jp} (t))}\rangle.  
\end{equation*}
Now the flagellar forces and orientations are assumed independently distributed with respect to their Gaussian stationary distributions. Thus, we can evaluate the average over the orientation fluctuations using the formula~\eqref{eq:jointmgf} for the average of the exponential of a Gaussian random variable. 
\begin{equation*}
\begin{aligned}
   \langle |\Fcol (t)|^2\rangle &= \sum_{j=1}^N \sum_{\jp=1}^N 
  \expe^{\frac{2\pi \mathi}{N}\left(j-\jp+\frac{\flagdisp{j} - \flagdisp{\jp}}{\celllen}\right)} \langle F_j (t) F_{\jp} (t)\rangle \langle \expe^{\mathi (\Theta_j (t)-\Theta_{\jp} (t))}\rangle
  \\& \qquad \qquad = 
  \sum_{j=1}^N (\sigma_F^2 + (F_j^{(0)})^2) 
+    \sum_{j=1}^N \sum_{\substack{\jp=1 \\ \jp \neq j}}^N 
  \expe^{\frac{2\pi \mathi}{N}\left(j-\jp+\frac{\flagdisp{j} - \flagdisp{\jp}}{\celllen}\right)} F_j^{(0)} F_{\jp}^{(0)} 
  \expe^{\mathi (\Theta_j^{(0)}-\Theta_{\jp}^{(0)})} \expe^{-\sigma_{\Theta}^2} \\
 &= N \sigma_F^2 +\left[e^{-\sigma_{\Theta}^2}\left|\sum_{j=1}^N F_j^{(0)} 
\expe^{\mathi \Theta_j^{(0)}
+\frac{2\pi\mathi}{N}\left(j+\frac{\flagdisp{j}}{\celllen}\right)}    \right|^2
    + (1-\expe^{-\sigma_{\Theta}^2}) \sum_{j=1}^N\left(F_j^{(0)}\right)^2 \right]  \label{eq:fcolrms}
 \end{aligned}
\end{equation*}
Using the proportionality of the colony velocity (neglecting the thermal contribution) to the colony force via the translational drag, we obtain the instantaneous mean-square velocity~\eqref{mean colony forcing 1} for a colony.  

To obtain the effective translational diffusivity of the colony requires integrating the autocorrelation function of the colony velocity, which in turn requires us to consider the colony force correlation function at unequal times $ t,\tp$: 

\begin{equation}
    \langle \Fcol (t) \cdot \Fcol (\tp) \rangle = 
    \Real \sum_{j=1}^N \sum_{\jp=1}^N 
  \expe^{\frac{2\pi \mathi}{N}\left(j-\jp+\frac{\flagdisp{j} - \flagdisp{\jp}}{\celllen}\right)} \langle F_j (t) F_{\jp} (\tp) \expe^{\mathi (\Theta_j (t)+\Thetacol (t)-\Theta_{\jp} (\tp)-\Thetacol (\tp))}\rangle
  \label{eq:forcecorrstart}
  \end{equation}
Here the calculation in general is complicated relative to our previous ones due to the correlations between the colony rotation $ \Thetacol (\tp)-\Thetacol(t)$ and the flagellar forces and orientations.  Moreover, the calculation of the average is complicated by the fact that $ \Thetacol (t) $ is non-Gaussian.
We proceed by assuming the dynamical force fluctuation standard deviation $ \sigma_F $ is small relative to  the magnitude of the flagellar forces $ F_j^{(0)} $, which we denote by $ \Fmag$, and the dynamical orientation standard deviation $ \sigmaThe$ is also small. Our attempts to split off higher terms by Taylor expansion led, when substituted into the integral~\eqref{eq:xfromf}, to pessimistic error bounds that compete with significant terms.  We thus take a different approximation route by treating $ \Thetacol (t)-\Thetacol (\tp), \Theta_j(t), \Theta_{\jp} (\tp), F_j (t), F_(\jp) (\tp)$ as jointly Gaussian random variables.  We use the notation $ \langle \cdot \rangle_{\mathrm{G}}$ to denote an average computed under this approximation.
The assumption here is on the change in colony orientation $\Thetacol (t)-\Thetacol (\tp)$, which is non-Gaussian because the colony torque~\eqref{Agg. active torque} is not Gaussian.  But they are Gaussian through first order in the flagellar force and orientation fluctuations, which we are assuming to  be small.  Moreover, by the functional central limit theorem, $ \Thetacol (t) -\Thetacol (\tp) $ can be treated as approximately Gaussian when $ |t-\tp|\gg \max (\gamma_F^{-1},\gamma_{\Theta}^{-1})$, the correlation time of the colony torque (see Eq.~\eqref{autocorrelation of torque}).  Actually, this does not tell us that $ \Thetacol (t) - \Thetacol (\tp) $ is \emph{jointly} Gaussian with the other random variables, but as we will see, they are weakly correlated so this concern should not be of much import.   

Proceeding with this plausible jointly Gaussian approximation, we have from the formulas in Appendix~\ref{app:gaussavg}:
\begin{align}
&    \langle F_j (t) F_{\jp} (\tp) \expe^{\mathi (\Theta_j (t)+\Thetacol (t)-\Theta_{\jp} (\tp)-\Thetacol (\tp))}\rangle_G  = \expe^{\mathi\left(\Theta_j^{(0)}-\Theta_{\jp}^{(0)}+\langle\Thetacol (t)-\Thetacol (\tp)\rangle\right)}
\label{eq:fgstart} \\
& \qquad \qquad = \left( F_j^{(0)} - \mathi \Cov(F_j (t), \Thetacol(\tp)-\Thetacol(t))\right) \left( F_{\jp}^{(0)} - \mathi \Cov(F_{\jp} (\tp), \Thetacol(\tp)-\Thetacol(t)\right) \nonumber \\
& \qquad \qquad \qquad \qquad \cdot 
\expe^{-\frac{1}{2} (2\sigmaThe^2 + \Var (\Thetacol (\tp)-\Thetacol (t)))-2\Cov (\Theta_j (t)-\Theta_{\jp}(\tp), \Thetacol (\tp)-\Thetacol(t))}  
\text{ for } j \neq \jp \nonumber
\end{align}

We next express the colony rotation in terms of its mean (first term) and fluctuating behavior:  
\begin{align*}
&\Thetacol (\tp)-\Thetacol(t) =
\meanrot (\tp-t) 
- \int_t^{\tp} \frac{\colrad}{\gamrot}\sum_{k=1}^{N} [F_k (\tpp)  \sin (\Theta_k (\tpp)) 
- F_k^{(0)} \sin (\Theta_k^{(0)}) \expe^{-\sigmaThe^2/2}]\, \difd \tpp \\
& \qquad \qquad \qquad \qquad +\sqrt{2\Diffrot} (W^{\Theta,c}(\tp)-W^{\Theta,c}(t)) \nonumber \\
\end{align*}
and then decompose this expression into a Gaussian component and a smaller non-Gaussian component:  
\begin{align}
&\Thetacol (\tp)-\Thetacol(t) =
\Thetacolgauss (\tp,t) + \Thetacolerr (\tp,t), 
\label{eq:gausapprox} \\
&\Thetacolgauss (\tp,t) 
= \meanrot (\tp-t) +\sqrt{2\Diffrot} (W^{\Theta,c}(\tp)-W^{\Theta,c}(t)) \\
& \qquad \qquad \qquad \qquad 
- \frac{\colrad}{\gamrot} \int_t^{\tp} \sum_{k=1}^N \left[
F_k^{(0)} \cos (\Theta_k^{(0)}) (\Theta_k(\tpp)-\Theta_k^{(0)})+(F_k(\tpp)
-F_k^{(0)}) \sin(\Theta_k^{(0)}) \expe^{-\sigmaThe^2/2}\right] 
\, \difd \tpp \nonumber\nonumber \\
&\Thetacolerr (\tp,t) = -  \sum_{k=1}^N\int_t^{\tp} \frac{\colrad}{\gamrot} F_k (\tpp)
\left(
\sin \Theta_k (\tpp) - \sin (\Theta_k^{(0)}) \expe^{-\sigmaThe^2/2}-\cos (\Theta_k^{(0)}) (\Theta_k(\tpp)-\Theta_k^{(0)})
\right)
\, \difd \tpp \nonumber \\
& \qquad \qquad \qquad \qquad - \sum_{k=1}^N \int_t^{\tp}\frac{\colrad}{\gamrot}
(F_k(\tpp)-F_k^{(0)}) \cos (\Theta_k^{(0)}) (\Theta_k(\tpp)-\Theta_k^{(0)}) \,\difd \tpp \nonumber
\end{align}

The covariances in Eq.~\eqref{eq:fgstart} can now be seen to be clearly dominated by just the covariance with the Gaussian component so we compute:
\begin{align}
&    \Cov(F_j (t), \Thetacol(\tp)-\Thetacol(t)) =  \Cov (F_j (t), \Thetacolgauss (\tp,t))  +\Cov (F_j (t), \Thetacolerr (\tp,t)) \nonumber \\ &\qquad \qquad =
    - \frac{\colrad}{\gamrot} \int_{t}^{\tp} \left[\sigma_F^2  \sin \Theta_j^{(0)} + O (\sigma_F^2 \sigmaThe^4)\right] \expe^{-\sigmaThe^2/2} \expe^{-\gamma_F|\tpp-t|} \, \difd \tpp \nonumber \\
    &\qquad \qquad = \frac{\colrad \sigma_F^2 \sin \Theta_j^{(0)}}{\gamrot} \frac{\sgn(\tp-t)(\expe^{-\gamma_F|t-\tp|}-1)}{\gamma_F} \left(1 + O(\sigmaThe^2)\right), 
    \label{eq:covfth}\\
& \Cov (\Theta_j (t)-\Theta_{\jp}(\tp), \Thetacol (\tp)-\Thetacol(t)) \nonumber \\
& \qquad \qquad = \Cov (\Theta_j (t)-\Theta_{\jp}(\tp), \Thetacolgauss(\tp,t)) +   \Cov (\Theta_j (t)-\Theta_{\jp}(\tp), \Thetacolerr(\tp,t)) \nonumber \\
& \qquad \qquad = -\int_{t}^{\tp} \colrad \sigmaThe^2 
 \frac{F_j^{(0)} \cos (\Theta_j^{(0)}) \expe^{-\gamma_{\Theta} |t-\tpp|}- F_{\jp}^{(0)} \cos (\Theta_{\jp}^{(0)}) \expe^{-\gamma_{\Theta} |\tp-\tpp|}}{\gamrot} \left(1+O(\sigmaThe^2)\right) \nonumber \\
& \qquad \qquad = - \frac{\colrad \sigmaThe^2 \left(F_j^{(0)} \cos (\Theta_j^{(0)}) - F_{\jp}^{(0)} \cos (\Theta_{\jp}^{(0)})\right)}{\gamrot} \frac{\sgn(\tp-t)(1-\expe^{-\gamma_{\Theta} |t-\tp|})}{\gamma_{\Theta}} \left(1+ O(\sigmaThe^2)\right)
\label{eq:covthth}
\end{align}
and similarly
\begin{equation}
     \Cov(F_{\jp} (\tp), \Thetacol(\tp)-\Thetacol(t)) 
       = \frac{\colrad \sigma_F^2 \sin \Theta_{\jp}^{(0)}}{\gamrot} \frac{\sgn(\tp-t)(\expe^{-\gamma_F|t-\tp|}-1)}{\gamma_F} \left(1 + O(\sigmaThe^2)\right) \label{eq:covfthp}
\end{equation}
We estimate the variance of the colony rotation in terms of its long term-asymptotics using a generalization of Eq.~\eqref{eq:thetafromT}:
\begin{align*}
    \Var (\Thetacol(\tp)-\Thetacol(t)) &= 
\frac{2}{\gamrot^2}\int_0^{|\tp-t|} (|\tp-t|-u)\Cov(T^{(c)}(t),T^{(c)}(t+u))\, \difd u+2\Diffrot |\tp-t| \\
&= 2 \Diffroteff |\tp-t| + \Vargauss (|\tp-t|)
\end{align*}
with the error term estimated via Taylor expansion of the torque correlation function~\eqref{autocorrelation of torque} with respect to the small parameters $ \sigmaThe^2$ and $ \sigma_F^2/\Fmag^2$
\begin{align}
    \Vargauss (\tau) &\equiv
- \frac{2\tau}{\gamrot^2}
\int_{\tau}^{\infty} \Cov(T^{(c)}(t),T^{(c)}(t+u)) \, \difd u
-\frac{2}{\gamrot^2} \int_0^{\tau} u \Cov(T^{(c)}(t),T^{(c)}(t+u)) \, \difd u \nonumber\\ 
&\sim O\left(\frac{N\colrad^2}{\gamrot^2 \gammamin} 
    (\sigma_F^2 + \Fmag^2 \sigmaThe^2 )\left[\tau \expe^{-\gammamin \tau} + \gammamin^{-1}
    \right]\right) \sim O\left(\frac{\Diffroteff-\Diffrot}{\gammamin} \left[\tau \expe^{-\gammamin \tau} + \gammamin^{-1}
    \right]\right) \nonumber \\
\Vargaussp (\tau) &\sim O\left(\frac{N\colrad^2}{\gamrot^2 \gammamin} 
    (\sigma_F^2 + \Fmag^2 \sigmaThe^2 )\left[(1+\gammamin \tau)\expe^{-\gammamin \tau} 
    \right]\right)\sim O(\left(\frac{\Diffroteff-\Diffrot}{\gammamin} \left[(1+\gammamin \tau)\expe^{-\gammamin \tau} 
    \right]\right) \label{eq:vargaussest}
\end{align}

Substituting Eqs.~\eqref{eq:covfth}, \eqref{eq:covthth}, and~\eqref{eq:covfthp} into Eq.~\eqref{eq:fgstart}, we obtain 
\begin{equation}
\begin{aligned}
&\langle F_j (t) F_{\jp} (\tp) \expe^{\mathi (\Theta_j (t)+\Thetacol (t)-\Theta_{\jp} (\tp)-\Thetacol (\tp))}\rangle_G \\
&=    \left(F_j^{(0)} + \mathi \frac{\colrad \sigma_F^2}{\gamrot \gamma_F}  \sin(\Theta_j^{(0)}) \sgn(\tp-t) (1 - \expe^{-\gamma_F|\tp-t|}) (1+O(\sigmaThe^2)) \right) \\
& \qquad \qquad \cdot
 \left(F_{\jp}^{(0)} + \mathi \frac{\colrad \sigma_F^2}{\gamrot \gamma_F}  \sin(\Theta_{\jp}^{(0)}) \sgn(\tp-t) (1 - \expe^{-\gamma_F|\tp-t|}) (1+O(\sigmaThe^2))\right) \\
 & \qquad \qquad \cdot
e^{\mathi[\Theta_j^{(0)}-\Theta_{\jp}^{(0)}]-\sigma_\Theta^2-\mathi  \meanrot (\tp-t)-(\Diffroteff-\Diffrot)|\tp-t|-\frac{1}{2}\Vargauss (|\tp-t|)} \\
& \qquad \qquad \cdot\expe^{-\frac{a\sigmaThe^2}{\gamrot \gamma_{\Theta}} (F_j^{(0)} \cos (\Theta_j^{(0)})-F_{\jp}^{(0)} \cos (\Theta_{\jp}^{(0)})) \sgn(\tp-t)(1- \expe^{-\gamma_{\Theta} (\tp-t)})(1+O(\sigmaThe^2))} \text{ for } j \neq \jp.
\end{aligned}
\label{eq:fjjp}
\end{equation}
A similar calculation gives:
\begin{equation}
\begin{aligned}
&\langle F_j (t) F_{\jp} (\tp) \expe^{\mathi (\Theta_j (t)+\Thetacol (t)-\Theta_{\jp} (\tp)-\Thetacol (\tp))}\rangle_G
e^{-\sigmaThe^2 (1-\expe^{-\gamma_{\Theta}(\tp-t)})-\mathi  \meanrot (\tp-t)-(\Diffroteff-\Diffrot)|\tp-t|-\frac{1}{2}\Vargauss |\tp-t|} \\ 
& \qquad \cdot  
 \left(\left(F_j^{(0)}+ \mathi\frac{\colrad \sigma_F^2}{\gamrot \gamma_F} \sin(\Theta_j^{(0)})\sgn(\tp-t)(1 - \expe^{-\gamma_F(\tp-t)})(1+O(\sigmaThe^2)) \right)^2
 + \sigma_F^2 \expe^{-\gamma_F (\tp-t)}\right) 
\text{ for } j = \jp\\
\end{aligned}
\label{eq:fjj}
\end{equation}
The expressions in~\eqref{eq:fjj} and~\eqref{eq:fjjp} can be simplified when the time lag $ |\tp-t|$ is large compared to the time scale of flagellar force and orientation relaxation:  

\begin{align}  \label{Col_force_autocorr}
&\langle F_j (t) F_{\jp} (\tp) \expe^{\mathi (\Theta_j (t)+\Thetacol (t)-\Theta_{\jp} (\tp)-\Thetacol (\tp))}\rangle_G
=
e^{-\sigma_{\Theta}^2-\mathi\meanrot(\tp-t)-\Diffroteff |\tp-t|-\frac{1}{2}\Vargauss (|\tp-t|)}  \\
& \qquad  \cdot \left(F_j^{(0)} +\mathi
\frac{\colrad \sigma_F^2}{\gamrot \gamma_F} \sin \Theta_j^{(0)}
\sgn(\tp-t)(1+O(\sigmaThe^2))\right)
\left(F_{\jp}^{(0)}+\mathi
\frac{\colrad \sigma_F^2}{\gamrot  \gamma_F} \sin \Theta_{\jp}^{(0)}\sgn(\tp-t)(1+O(\sigmaThe^2))\right) \nonumber \\
& \qquad   \cdot \expe^{\mathi (\Theta_j^{(0)}-\Theta_{\jp}^{(0)})}
\expe^{-\frac{a\sigma_{\Theta}^2}{\gamrot\gamma_{\Theta}} \sgn(\tp-t)(F_j^{(0)} \cos \Theta_j^{(0)} -F_{\jp}^{(0)}\cos \Theta_{\jp}^{(0)})(1+O(\sigmaThe^2))}\nonumber \\
& \qquad \qquad \qquad \qquad 
\text{ for } |\tp-t|\gg \gammamin^{-1}.\nonumber
\end{align}
 
We note for what follows that the error in using this expression also for short time lags can be estimated as: 
\begin{equation}
\label{eq:shorterr}
    O\left( \left(1+\frac{\colrad\Fmag} {\gamrot\gammamin}\right) \left[(\sigma_F/\Fmag)^2+\sigmaThe^2\right]
    \right).
\end{equation}

Continuing with the Gaussian approximation and decorating averages with a subscript G to denote this approximation, we examine the mean-square displacement over observation intervals large compared to the flagellar dynamics but short compared to the colony dynamics:
\begin{equation*}
\bExp|\bXcol (t+\tau)-\bXcol (t)|^2 = \Big(\frac{1}{\gamtrans}\Big)^2\int_t^{t+\tau}\int_t^{t+\tau} \Fcol (\tp)\cdot \Fcol (\tpp))  d\tp \, d\tpp+4\Difftrans \tau.
\end{equation*}
Provided $ \tau $ is short compared to the time scale on which the colony as a whole rotates, the force correlation function in the integrand simply adjusts from its zero lag value to a different constant value inferred from substituting Eq.~\eqref{Col_force_autocorr} into Eq.~\eqref{eq:forcecorrstart}.   Approximating the integrand by the second constant and noting the error in this approximation from Eq.~\eqref{eq:shorterr}, we obtain:
\begin{align}
&\bExp|\bXcol (t+\tau)-\bXcol (t)|^2 \sim 4 \Difftrans \tau \label{eq:msqdint}\\
& \qquad +
\frac{e^{-\sigma_{\Theta}^2}\tau^2}{2\gamtrans^2}
 \Real \sum_{j=1}^N\sum_{\jp=1}^N \left[
\left(F_j^{(0)} +\mathi
\frac{\colrad \sigma_F^2}{\gamrot \gamma_F} \sin \Theta_j^{(0)}(1+O(\sigmaThe^2))\right)
\left(F_{\jp}^{(0)}+\mathi
\frac{\colrad \sigma_F^2}{\gamrot  \gamma_F} \sin \Theta_{\jp}^{(0)}(1+O(\sigmaThe^2))\right) \right.\nonumber \\
& \qquad \qquad \cdot \expe^{\mathi (\Theta_j^{(0)}-\Theta_{\jp}^{(0)})
+\frac{2\pi\mathi}{N}\left(j-\jp+\frac{\flagdisp{j}-\flagdisp{\jp}}{\celllen}\right)}
\expe^{-\frac{a\sigma_{\Theta}^2}{\gamrot\gamma_{\Theta}} (F_j^{(0)} \cos \Theta_j^{(0)} -F_{\jp}^{(0)}\cos \Theta_{\jp}^{(0)})(1+O(\sigmaThe^2))}  \nonumber \\
& \qquad \qquad +  O\left( \left(1+\frac{\colrad\Fmag} {\gamrot\gammamin}\right) \left[(\sigma_F/\Fmag)^2+\sigmaThe^2\right]
    \gamtrans^{-2}\gammamin^{-1} \tau
    \right)  \text{ for }    \gammamin^{-1} \ll \tau \ll \meanrot{}^{-1},\Diffroteff{}^{-1} 
    \nonumber 
\end{align}
The term proportional to $ \tau^2$ term corresponds to a second ballistic phase, whose coefficient gives the mean-square speed that would be estimated from observations taken at a resolution longer than that of the flagellar dynamics.  Expanding this expression with respect to $ \sigma_F$ and $ \sigmaThe$, the complicated terms at second order all cancel out in the double sum, and we arrive at the value reported in Eq.~\eqref{eq:coarsemsq}, with a relative error
\begin{equation*}
    \sim O\left( \left(1+\frac{\colrad\Fmag} {\gamrot\gammamin}\right)^2 \left[(\sigma_F/\Fmag)^2+\sigmaThe^2\right]^2\right)
\end{equation*}
Indeed, the expression in Eq.~\eqref{eq:coarsemsq} would follow by a simple calculation starting from Eq.~\eqref{mean colony forcing 1} and freezing the colony rotation.  Note the ballistic term should dominate the error term in Eq.~\eqref{eq:msqdint} for $ \tau \gg \gammamin^{-1}$.

Finally, we can compute the long-time covariance of the displacement of the colony from integrating the velocity autocorrelation function, which is directly proportional to the force autocorrelation function, and adding the translational diffusion component:
 
\begin{equation}
\begin{aligned}
\CovG (\bXcol (t),\bXcol (t))  &=\Big(\frac{1}{\gamtrans}\Big)^2\int_0^t\int_0^t \CovG (\Fcol (\tp), \Fcol (\tpp))  dt'dt''+2\Difftrans \Id t \end{aligned}
\label{eq:xfromf}
\end{equation}

We use now the expressions~\eqref{eq:fcoliso}, ~\eqref{eq:forcecorrstart}, and~\eqref{Col_force_autocorr} for the covariance of the colony forcing at large time lag,

We note that the expression~\eqref{Col_force_autocorr} exhibits complex conjugation symmetry under the joint interchange $ t \leftrightarrow \tp$, $ j \leftrightarrow \jp$, which in turn makes the force autocorrelation function~\eqref{eq:forcecorrstart} depend only on the absolute value of the time lag $ |t-\tp| $, as it must.  Thus, we use again Eq.~\eqref{eq:kuboint} to get:  
\begin{align}
&\lim_{t\rightarrow \infty} \frac{\CovG (\bXcol (t),\bXcol (t))}{2t} \label{eq:dtcomp}\\
& =\Difftrans \Id + 
\frac{1}{2 \gamtrans^2} \Id
\int_0^{\infty} \Real \sum_{j=1}^N \sum_{\jp=1}^N 
  \expe^{\frac{2\pi \mathi}{N}\left(j-\jp+\frac{\flagdisp{j} - \flagdisp{\jp}}{\celllen}\right)}  \langle F_j (0) F_{\jp} (u) \expe^{\mathi (\Theta_j (0)+\Thetacol (0)-\Theta_{\jp} (u)-\Thetacol (u)}\rangle_G \, \difd u\nonumber \\
 &= \Difftrans \Id + \Real \Id \frac{e^{-\sigma_{\Theta}^2}}{2\gamtrans^2}
 \sum_{j=1}^N\sum_{\jp=1}^N \left[
\left(F_j^{(0)} +\mathi
\frac{\colrad \sigma_F^2}{\gamrot \gamma_F} \sin \Theta_j^{(0)}(1+O(\sigmaThe^2))\right)
\left(F_{\jp}^{(0)}+\mathi
\frac{\colrad \sigma_F^2}{\gamrot  \gamma_F} \sin \Theta_{\jp}^{(0)}(1+O(\sigmaThe^2))\right) \right.\nonumber \\
& \qquad \qquad \cdot \expe^{\mathi (\Theta_j^{(0)}-\Theta_{\jp}^{(0)})
+\frac{2\pi\mathi}{N}\left(j-\jp+\frac{\flagdisp{j}-\flagdisp{\jp}}{\celllen}\right)}
\expe^{-\frac{a\sigma_{\Theta}^2}{\gamrot\gamma_{\Theta}} (F_j^{(0)} \cos \Theta_j^{(0)} -F_{\jp}^{(0)}\cos \Theta_{\jp}^{(0)})(1+O(\sigmaThe^2))}  \nonumber \\
& \qquad \qquad\cdot
\int_0^{\infty} 
e^{-\mathi\meanrot u-\Diffroteff u-\frac{1}{2}\Vargauss (u)} \, 
  \, \difd u\nonumber\\
  & \left.\qquad \qquad \qquad \qquad + O\left( 
  \int_0^{\gammamin^{-1}} \left(1+\frac{\colrad\Fmag}{\gamrot\gammamin}\right) \left[(\sigma_F/\Fmag)^2+\sigmaThe^2\right]\, \difd u\right) 
 \right] \nonumber
\end{align}
Integrating by parts in the main integral, noting from Eq.~\eqref{eq:vargaussest} that $ \Vargauss (u)$ is subdominant to $ \Diffroteff u$ for large $u$ and $ \Vargauss (0)=0$:
\begin{equation*}
\int_0^{\infty} 
e^{-\mathi\meanrot u-\Diffroteff u-\frac{1}{2}\Vargauss (u)} \, \difd u
= \frac{1}{\mathi \meanrot + \Diffroteff}
\left[1-\frac{1}{2}\int_0^{\infty} \Vargaussp (u)
e^{-\mathi\meanrot u-\Diffroteff u-\Vargauss (u)} \, \difd u\right].
\end{equation*}
Applying the estimate on $ \Vargaussp (u)$ from Eq.~\eqref{eq:vargaussest}, we have:
\begin{equation*}
  \int_0^{\infty} 
e^{-\mathi\meanrot u-\Diffroteff u-\Vargauss (u)} \, \difd u
= \frac{1}{\mathi \meanrot + \Diffroteff}
\left[1+ O\left(\frac{\Diffroteff-\Diffrot}{\gammamin}\right)\right]. 
\end{equation*}
For the correction term in Eq.~\eqref{eq:dtcomp}, we estimate:
\begin{align*}
&  \int_0^{\gammamin^{-1}} \left(1+\frac{\colrad\Fmag}{\gamrot\gammamin}\right) \left[(\sigma_F/\Fmag)^2+\sigmaThe^2\right]
\, \difd u
  \\
  &\sim O\left( 
 \left(1+\frac{\colrad\Fmag}{\gamrot\gammamin}\right) \left[(\sigma_F/\Fmag)^2+\sigmaThe^2\right] \gammamin^{-1}
\right)
\end{align*}
Next computing the real part of the resulting complex expression in Eq.~\eqref{eq:dtcomp}, we obtain:
\begin{align}
&\lim_{t\rightarrow \infty} \frac{\CovG (\bXcol (t),\bXcol (t))}{2t} =\Difftrans \Id + 
\frac{e^{-\sigma_{\Theta}^2}}{2\gamtrans^2 (\meanrot{}^2+\Diffroteff{}^2)} \Id \label{eq:dthyp}\\
&\qquad\qquad\cdot \left[\sum_{j,\jp=1}^N F_j^{(0)} \left(F_{\jp}^{(0)} (\Diffroteff +O(\sterr)))+ \frac{2\colrad \sigma_F^2 \sin \Theta_{\jp}^{(0)} (\meanrot+O(\sterr))}{\gamrot \gamma_F}\right) \right.\nonumber \\
& \qquad \qquad \qquad \qquad \cdot   \cosh\left(\frac{a\sigma_{\Theta}^2}{\gamrot\gamma_{\Theta}} (F_{j}^{(0)} \cos (\Theta_j^{(0)}) -F_{\jp}^{(0)} \cos (\Theta_{\jp}^{(0)}))\right)  \nonumber \\
   & \qquad \qquad \qquad \qquad \cdot
    \cos\left(\Theta_j^{(0)}-\Theta_{\jp}^{(0)}
    +\frac{2\pi}{N}\left(j-\jp+\frac{\flagdisp{j}-\flagdisp{\jp}}{\celllen}\right)\right)\nonumber \\
    &\qquad \qquad +  \sum_{j,\jp=1}^N 
    F_j^{(0)} \left(F_{\jp}^{(0)} (\meanrot + O(\sterr)) - 
 \frac{2\colrad \sigma_F^2 \sin \Theta_{\jp}^{(0)} (\Diffroteff+O(\sterr))}{\gamrot \gamma_F}\right)  \nonumber \\
& \qquad \qquad \qquad \qquad \cdot     \sinh\left(\frac{a\sigma_{\Theta}^2}{\gamrot\gamma_{\Theta}} (F_{\jp}^{(0)} \cos (\Theta_{\jp}^{(0)}) -F_{j}^{(0)} \cos (\Theta_{j}^{(0)}))\right)  \nonumber \\
   & \qquad \qquad \qquad \qquad \left.\cdot
    \sin\left(\Theta_j^{(0)}-\Theta_{\jp}^{(0)}
    +\frac{2\pi}{N}\left(j-\jp+\frac{\flagdisp{j}-\flagdisp{\jp}}{\celllen}\right)\right)\right]\nonumber \\
& \qquad \qquad \qquad \cdot 
\left[1+ O\left(\frac{\Diffroteff-\Diffrot}{\gammamin}\right)+O\left(\frac{\colrad \Fmag ((\sigma_F/\Fmag)^2+\sigmaThe^2)\sigmaThe^2}{\gamrot \gammamin}\right)+O\left(\frac{\colrad^2 \sigma_F^4}{\gamrot^2\gamma_F^2\Fmag^2}\right)\right]
\end{align}
where
\begin{equation*}
    \sterr \equiv  
 \left(1+\frac{\colrad\Fmag}{\gamrot\gammamin}\right) \left[(\sigma_F/\Fmag)^2+\sigmaThe^2\right] 
 \frac{\meanrot{}^2+\Diffroteff{}^2}{\gammamin},
\end{equation*}
a quantity with dimensions of rate, denotes the error arising from the integration over times $ \lesssim \gammamin^{-1}$. 

In $ \sterr$ we have separated out a nondimensional factor of $ \frac{\Fmag \colrad}{\gamrot \gammamin}$ which from Eq.~\eqref{eq:ratioest}, is $ \sim 20/N^2 $ and thus order unity for the biophysical parameters in Table 1 and reasonable colony sizes.  But to allow for situations in which this factor may be large and thus somewhat counteract the asymptotically small factors of $ \sigmaThe$ and $ \sigma_F/\Fmag$, we recast errors involving this factor in terms of the effective rotational diffusion via the expresssion~\eqref{eff rotational diffusion_summary} Taylor expanded with respect to these small factors:

\begin{align}
\label{eq:estaf}
    \frac{a\sigma_{\Theta}^2}{\gamrot\gamma_{\Theta}} F_{j}^{(0)}  &\sim O(N^{-1/2}(\Diffroteff-\Diffrot)^{1/2} \gammamax^{1/2} \gammamin^{-1} \sigma_{\Theta}),
    \\
    \frac{\colrad \sigma_F^2}{F_j^{(0)} \gamrot \gamma_F}
    &\sim O(N^{-1/2}(\Diffroteff-\Diffrot)^{1/2} \gammamax^{1/2} \gammamin^{-1}(\sigma_{F}/\Fmag)), \nonumber \\
    \frac{\colrad \Fmag}{\gamrot\gammamin}
  \left[(\sigma_F/\Fmag)^2+\sigmaThe^2\right]&\sim O\left(N^{-1/2}(\Diffroteff-\Diffrot)^{1/2}\gammamax^{1/2} \gammamin^{-1}(\sigmaThe + (\sigma_F/\Fmag))\right). \nonumber
\end{align} 

Applying now the assumption~\eqref{eq:gamone} that $ \gammamax/\gammamin$ is order unity, we see that
the squares of the first two expressions in Eq.~\eqref{eq:estaf} are  small compared to the  $ O((\Diffroteff-\Diffrot)\gammamin^{-1})$ integration error, and the functions of them in Eq.~\eqref{eq:dthyp} should be self-consistently represented by their first order Taylor approximations.  
The last expression implies that
\begin{equation*}
\sterr  \sim O\left(
 \left(\left[(\sigma_F/\Fmag)^2+\sigmaThe^2\right] 
 + \left(\frac{\Diffroteff-\Diffrot}{N \gammamin}\right)^{1/2}\left[(\sigma_F/\Fmag)+\sigmaThe\right] \right)
 \frac{\meanrot{}^2+\Diffroteff{}^2}{\gammamin}\right)
\end{equation*}
With these simplifications and reductions we arrive at:
\begin{align} \label{eq:difftranslong}
&\lim_{t\rightarrow \infty} \frac{\CovG (\bXcol (t),\bXcol (t))}{2t} =\Difftrans \Id   + \frac{e^{-\sigma_{\Theta}^2}}{2\gamtrans^2 (\meanrot{}^2+\Diffroteff{}^2)} \Id \\
&\qquad\qquad\cdot \left\{\left[\sum_{j,\jp=1}^N F_j^{(0)} \left(F_{\jp}^{(0)} (\Diffroteff +O(\sterr)))+ \frac{2\colrad \sigma_F^2 \sin \Theta_{\jp}^{(0)} (\meanrot+O(\sterr))}{\gamrot \gamma_F}\right) \right.\right. \nonumber \\
   & \qquad \qquad \qquad \qquad \cdot
    \cos\left(\Theta_j^{(0)}-\Theta_{\jp}^{(0)}
    +\frac{2\pi}{N}\left(j-\jp+\frac{\flagdisp{j}-\flagdisp{\jp}}{\celllen}\right)\right)\nonumber \\
    &\qquad \qquad + \frac{a (\meanrot + O (\sterr)) \sigma_{\Theta}^2}{\gamrot\gamma_{\Theta}} \sum_{j,\jp=1}^N 
    F_j^{(0)} F_{\jp}^{(0)} (F_{\jp}^{(0)} \cos (\Theta_{\jp}^{(0)}) -F_{j}^{(0)} \cos (\Theta_{j}^{(0)}))  \nonumber \\
   & \qquad \qquad \qquad \qquad \left.\cdot
    \sin\left(\Theta_j^{(0)}-\Theta_{\jp}^{(0)}
    +\frac{2\pi}{N}\left(j-\jp+\frac{\flagdisp{j}-\flagdisp{\jp}}{\celllen}\right)\right)\right]\nonumber \\
& \qquad \qquad \qquad \cdot 
\left[1+ O\left(\frac{\Diffroteff-\Diffrot}{\gammamin}\right)+O\left( \left(\frac{\Diffroteff-\Diffrot}{N\gammamin}\right)^{1/2}\left[(\sigma_F/\Fmag)+\sigmaThe\right]\sigmaThe^2\right)\right]
\end{align}
  
The error terms are all formally second order in the small parameters $ \sigma_F/\Fmag$ and $ \sigmaThe$, but we've also noted when these small parameters are therefore better expressed in terms of the effective rotational diffusivity, which Eq.~\eqref{eff rotational diffusion_summary_small} shows is itself a second order quantity in these small parameters.  
We now explain why we retain some second order terms in explicit form.  First of all, we note that $ \sterr$ has, beyond the second order factors of smallness in dynamical flagellar variation, an additional factor of $ (\meanrot{}^2+\Diffroteff{}^2)/\gammamin$ which is small compared to the larger of $ \meanrot $ and $ \Diffroteff$ under the self-consistency assumption~\eqref{eq:rottimesep} that the effective colony dynamics are slow relative to the flagellar dynamics.  Secondly, the explicitly presented second order terms are proportional to $ \meanrot$, which we have seen in Subsection~\ref{sec:rotsum} to typically be larger than $ \Diffroteff$, particularly for larger colonies.   So even though their ratio to $ \meanrot$ is $ O(N^{-1} (\Diffroteff-\Diffrot)^{1/2} ((\sigma_F/\Fmag)+\sigmaThe))$ and thus negligible to the level of our approximation, their ratio to $ \Diffroteff$ won't be as small. 

In the summary discussion in Subsection~\ref{sec:transdiff}, we report the translational diffusivity based on the expression~\eqref{eq:difftranslong}, dropping the error terms.  As discussed at the beginning of this calcluation, we did not attempt to quantify the error in computing averages as if the various quantities are jointly Gaussian.  Informal considerations would suggest these errors would arise at fourth order, which would be negligible with respect to the results of our more precise estimates within the Gaussian approximation framework.

\section{Generalized Colony Geometries}
\label{sec:gengeo}

The disclike model from Section~\ref{sec:discmodel} may be an appropriate simplified model for a rosette colony, but
choanoflagettes are known also to form chain colonies~\citep{dayelmorphogenesis}.  We show in this section how much of our analysis of the mobility of disclike colonies can be extended to more general thin two-dimensional colony shapes.   We do continue to consider the colonies as rigid, which is reasonable even for 
choanoflagellate chain colonies  due to their intercellular bridges~\citep{Roperstresslet}. The flagellar force model remains of the same Langevin type as in Subsection~\ref{sec:cellmod}; we only generalize in Subsection~\ref{sec:genmod} the representation of the relaxed locations and orientations of the flagella and the hydrodynamic resistance of the colony. Within this generalized model, we present mobility statistics for the colony in Subsection~\ref{sub:gensum} under the general assumption of small dynamical fluctuations of the flagellar force magnitudes and angles.  We do not present demographic mobility statistics, as these would be too complex to develop in a general geometric setting.  In Subsection~\ref{sec:genderive}, we sketch how these results can be derived using the methods of Section~\ref{sec:methods}.

\subsection{Model}
\label{sec:genmod}

We first appeal to low Reynolds number hydrodynamics  to  define the  resistance supermatrix~\citep[Sec. 5.4] {jh:lrhsa} for the joint translation and rotational dynamics of the rigid two-dimensional colony within its own plane:
\begin{equation}
    \Granddrag = \begin{pmatrix} 
    \gammaj & 0 & \gmajrot \cr
    0 & \gammin & \gminrot \cr
    \gmajrot & \gminrot & \gamrot 
    \end{pmatrix} \label{eq:granddrag}
\end{equation}
We have chosen a body frame
coordinate system with the Cartesian axes aligned with the principal axes of the translational component of the friction matrix in the upper left $2 \times 2$ block.  If the translational friction coefficients $ \gammaj \leq \gammin $ along the two axes are unequal, we take the convention of choosing the first Cartesian axis to lie along the direction with a lower friction coefficient. If the colony had a thin ellipsoidal shape, this would be equivalent to choosing the body frame coordinates based on the major and minor axes of the elliptical cross section.  The origin of the body frame is chosen to be the center of mass.  Then $ \gamrot$ is the rotational drag coefficient, and $ \gmajrot$, $\gminrot $ are the frictional coupling coefficients between translational motion along the principal body axes and rotation.  

\begin{figure} [H]
\centering
\includegraphics[height=60mm,width=110mm]{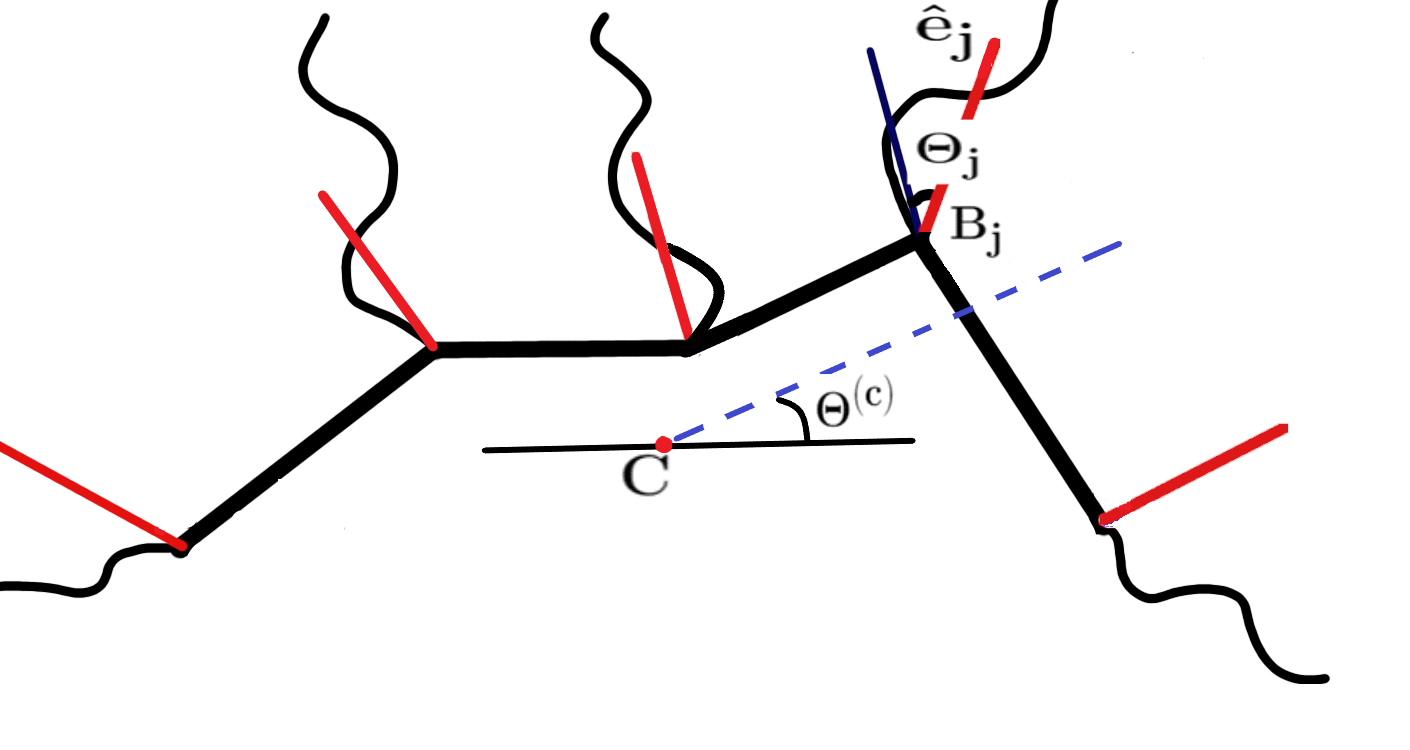}\\
\caption{Model Schematic: A colony of five cells arranged in a chain with   
with colony orientation $\Theta^{(c)}$ describing the rotation, about the center of mass at $ \centmass$, of the major axis of the hydrodynamic friction matrix from the horizontal
$\flagpos{j} $ is the location of the juncture of flagellum $j$ with the colony surface, and the relaxed orientation of its force is $ \orient{j}$. 
$\Theta_j$ is the angle between the momentary flagellar force orientation relative to its relaxed orientation.  
}
\label{chain colony model}
\end{figure}

For each of the $ j=1,\ldots,N$ flagella, we specify the two-dimensional position $ \flagpos{j} $ of the base of each flagellum,  a unit vector $ \orient{j}$ describing the mean direction of force applied by that flagellum, and a force magnitude $ F_j^{(0)} $ of that flagellum. A schematic for a chain colony shape is given in Figure~\ref{chain colony model}. 
For the disc colony model from Subsection~\ref{sec:colmod}, the body frame orientation is arbitrary,
\begin{equation}
\flagpos{j}=    a\begin{pmatrix}\cos(2\pi[\frac{j-\frac{1}{2}+\frac{S_j}{l}}{N}])\\\sin(2\pi[\frac{j-\frac{1}{2}+\frac{S_j}{l}}{N}]\end{pmatrix}, \orient{j} = 
\begin{pmatrix}-\cos\left(2\pi[\frac{j-\frac{1}{2}+\frac{S_j}{l}}{N}]+\Theta_j^{(0)}\right)\\-\sin\left(2\pi[\frac{j-\frac{1}{2}+\frac{S_j}{l}}{N}]+\Theta_j^{(0)}\right)\end{pmatrix},
\label{eq:discspec}
\end{equation}

The  force summation rules of the generalized colony in the body frame are: 
\begin{equation}
\begin{aligned}
\label{eq:chainft}
\Fbody &=  \sum_{j=1}^N F_j (t) \Rotmat{\Theta_j (t)} \orient{j}, \\
\Torquecol &= \sum_{j=1}^N F_j (t)
(\flagpos{j} - \centmass) \cdot\Rotmat{\Theta_j (t)}  \orientperp{j}
\end{aligned}
\end{equation}
where $ \Theta_j (t)$ is now the force orientation for flagellum $ j$ measured with respect to its natural orientation $ \orient{j} $.  For the disclike case considered in the previous sections, we introduced an angle $ \Theta_j^{(0)}$ for the angle between the flagellum's natural orientation $\orient{j}$ and the surface normal.  For an asymmetric colony, there is no natural reference angular position so there is no point in introducing such a demographic parameter; it is embedded in $ \orient{j}$ (as can be seen for the disclike case in Eq.~\eqref{eq:discspec}).  Thus, we carry over the flagellar dynamic models from Subsection~\ref{sec:cellmod} by simply setting $ \Theta_j^{(0)}\equiv 0$ in Eq.~\eqref{force direction sde 1} for the generalized geometry model under current discussion. 

\subsection{Colony Transport Statistics}
\label{sub:gensum}
In this subsection, without recurrent comment, all results are presented only to second order in the force fluctuation relative to its mean value $ \sigma_F/\Fmag$ and angular fluctuation magnitude $ \sigmaThe $, both assumed small.  We express the results with $ \sim$ to indicate they are asymptotic approximations in this sense. We find it useful to represent some results using the rotation matrix 
\begin{equation*}
    \Rotmat{\theta} \equiv \begin{pmatrix}
    \cos \theta & - \sin \theta \cr
    \sin \theta & \cos \theta  
    \end{pmatrix}.
\end{equation*}
and the perpendicular direction $ \orientperp{j}=\Rotmat{-\pi/2} \orient{j}$ to the orientation of flagellum $j$, directed to give the appropriate sign for the torque in Eq.~\eqref{eq:chainft}.   The colony transport statistics are naturally expressed in terms of 
sums of quantities associated with the  geometric orientation of each flagellum $j$ in relation to the colony body:
\begin{itemize}
\item the vectors (with units of time per mass) describing the ratio of the colony velocity induced per force directed along the flagellum's relaxed orientation $ \orient{j}$:
\begin{equation*}
    \ufrat{j} \equiv \left[\Granddraginv \begin{pmatrix} \orient{j} \cr (\flagpos{j} - \centmass)\cdot \orientperp{j}  \end{pmatrix} \right]_{\mathbf{x}}
\end{equation*}
and per unit force directed transversely (along $ \orientperp{j} $) to the flagellum's relaxed orientation:
\begin{equation*}
\ufratperp{j} \equiv
    \left[\Granddraginv
    \begin{pmatrix} \orientperp{j} \cr -(\flagpos{j} - \centmass)\cdot \orient{j}
    \end{pmatrix}\right]_{\bx}
\end{equation*}
 with $ [\cdot]_{\bx}$ denoting the projection of the three-dimensional supermatrix onto the translational components (first two indices).  
\item the scalar (with units of time per mass per length) describing the ratio of the the colony angular velocity induced per unit force directed along the flagellum's relaxed orientation $ \orient{j}$:
\begin{equation*}
    \wfrat{j} \equiv \left[\Granddraginv \begin{pmatrix} \orient{j} \cr (\flagpos{j} - \centmass)\cdot \orientperp{j}  \end{pmatrix} \right]_{\theta}
\end{equation*}
and per unit force directed transversely (along $ \orientperp{j} $) to the flagellum's relaxed orientation:
\begin{equation*}
\wfratperp{j} \equiv
    \left[\Granddraginv
    \begin{pmatrix} \orientperp{j} \cr -(\flagpos{j} - \centmass)\cdot \orient{j}
    \end{pmatrix}\right]_{\theta}
\end{equation*}
\end{itemize}
The results naturally are not dependent on simultaneous sign reversal of $ \ufratperp{j}$ and $ \wfratperp{j}$, corresponding to choosing the opposite ``transverse'' direction.  For the disc configuration developed in Section~\ref{sec:discmodel}, we'd have $ \ufrat{j} = \gamtrans^{-1} \orient{j}$, $ \ufratperp{j} = \gamtrans^{-1} \orientperp{j} $, $ \wfrat{j} = -\colrad \gamrot^{-1} \sin \Theta_j^{(0)}$, and $ \wfratperp{j} = \colrad \gamrot^{-1} \cos \Theta_j^{(0)}$.

The fundamental transport statistics for a colony under our flagellar forcing model can then be expressed in terms of the geometric configuration: 
\begin{itemize}
    \item \textbf{Rotational Drift}:
\begin{equation}
    \meanrot =  \sum_{j=1}^N F_j^{(0)} \expe^{-\frac{1}{2} \sigmaThe^2} \wfrat{j} \label{eq:rotdriftgen}
\end{equation}
where $ [\cdot]_{\theta} $ the angular (third index) component of the three-dimensional supervector in the brackets.
    \item \textbf{Effective Rotational Diffusion}:
    \begin{align}
    \Diffroteff &\sim \Diffrot +
        \sigma_F^2 \gamma_F^{-1} \sum_{j=1}^N 
        \wfrat{j}^2 + \sigmaThe^2 \gamma_{\Theta}^{-1} \sum_{j=1}^N (F_j^{(0)})^2    \wfratperp{j}^2
    \label{eq:rotdiffgen}
\end{align}
The first term is the thermal contribution
\begin{equation*}
\Diffrot =     \frac{k_B T}{\gamrot - \gmajrot^2/\gammaj - \gminrot^2/\gammin} 
\end{equation*}while the second term arises from the active flagellar force fluctuations  and the third term from the orientational fluctuations in the active flagellar forces.
    \item \textbf{Mean-Square Instantaneous Speed}:
    \begin{align}
        \Vcolmsq&\sim \expe^{-\sigmaThe^2}
         \left|\sum_{j=1}^N F_{j}^{(0)}\ufrat{j}\right|^2 
         +   \sigma_F^2 \sum_{j=1}^N \left|\ufrat{j}\right|^2     +   
         \sigmaThe^2 \sum_{j=1}^N  (F_j^{(0)})^2
    \left|\ufratperp{j}\right|^2
    \label{eq:msqspeedgen}
    \end{align}
The first term is the squared speed when all fluctuations are neglected, multiplied by $ \expe^{-\sigmaThe^2}$ to account for the mitigation in the mean flagellar force applied due to angular fluctuations.  The second and third term are contributions measuring the magnitudes of velocities induced by rapid dynamical flagellar fluctuations, and we show next that they disappear when viewed over a time scale long compared to these fluctuations.   
    \item \textbf{Mean-Square Speed Over Coarse Time Intervals}
The mean-square displacement exhibits a ballistic regime over time scales long compared to the flagellar time scales but short compared to the rotational dynamics of colony:    
\begin{align}
\bExp|\bXcol (t+\tau)-\bXcol (t)|^2 &\sim 4 \Difftrans \tau + \Vflagavgsq \tau^2 \text{ for }  \gamma_F^{-1},\gamma_{\Theta}^{-1} \ll \tau \ll \meanrot{}^{-1},\Diffroteff{}^{-1}, \nonumber \\
\Vflagavgsq &\sim \expe^{-\sigmaThe^2}
         \left|\sum_{j=1}^N F_{j}^{(0)}\ufrat{j}\right|^2 \label{eq:coarsespeedgen}
\end{align}
    \item \textbf{Effective Translational Diffusion}:  The colony will have no sustained drift due to statistical isotropy, so its mean position will remain bounded.  
The long-time diffusivity matrix will similarly be a simple multiple of the identity:
\begin{equation*}
    \lim_{t\rightarrow \infty} \frac{\Cov (\bXcol (t),\bXcol (t))}{2t} = \Difftranseff \Id
\end{equation*}
with 
\begin{align}
&\Difftranseff \sim \Difftrans +  \frac{\Diffroteff \Vflagavgsq}{2(\Diffroteff{}^2+\meanrot{}^2)}
  \label{eq:transdiffgen}
 - \frac{\meanrot k_B T}{(\meanrot{}^2+\Diffroteff{}^2)}
\frac{\sigma_F^2}{\gamma_F}  \sum_{j=1}^N F_j^{(0)} \ufrat{j}\cdot \left[ \Granddraginv\right]_{\bx,\theta} \\
& \qquad   - \frac{\meanrot}{(\meanrot{}^2+\Diffroteff{}^2)}
\frac{\sigma_F^2}{\gamma_F}  \sum_{j=1}^N F_j^{(0)} \ufrat{j} \cdot \sum_{\jp=1}^{N} 
\wfrat{\jp} \ufrat{\jp} - \frac{\meanrot}{(\meanrot{}^2+\Diffroteff{}^2)}
\frac{\sigmaThe^2}{\gamma_\Theta} \sum_{j=1}^N F_j^{(0)} \ufrat{j}  \cdot \sum_{\jp=1}^{N} (F_{\jp}^{(0)})^2\wfratperp{\jp}
\ufratperp{\jp} \nonumber
\nonumber
\end{align}
\end{itemize}
where $ \Difftrans$ is the thermal diffusivity, $\Vflagavgsq$ is the mean-square velocity observed on scales long compared to flagellar time scales~\eqref{eq:coarsespeedgen}, and $ [\Granddraginv]_{\bx,\theta}$ is the vector describing the translational-rotational coupling of the inverse resistance matrix in the body frame of the colony.
This translational diffusivity result requires the additional assumption~\eqref{eq:rottimesep} that the effective rotational diffusivity is small compared with the flagellar fluctuation rates together with the assumption~\eqref{eq:gamone} that the relaxation rates of flagellar force and orientation are comparable.

The first two terms in Eq.~\eqref{eq:transdiffgen} correspond to an isotropic particle moving at constant speed $ \Vflagavgsq$ in a direction governed by rotation at rate $ \meanrot$ together with rotational diffusion $ \Diffroteff$.  
For the disc-like case, our argument for the subdominance of the terms beyond the first two in Eq.~\eqref{eq:transdiffgen} relied on relationships derived in Subsection~\ref{sec:rotdiffsum} between the effective rotational drift and diffusivity for a nearly symmetric disc arrangement.  These cannot be expected to hold for general geometries, and those formally second order terms might be significant when the rotational velocity $ \meanrot$ is substantially greater than the effective rotational diffusivity $ \Diffroteff$ of the colony.

All these results can be verified to agree with the results obtained for the disc-like model when specialized to that geometry.
Computing demographic averages of these quantities analytically is considerably more challenging than for the disc-like model from Section~\ref{sec:geomod}.  For example, demographic variability across colonies will  affect, in a correlated way, the resistance matrix $ \Granddrag$ in~\eqref{eq:granddrag} and other geometric features of the colony which are multiplied together in all mobility expressions.  But these transport quantities are easily computed (without the need for running long Monte Carlo simulations) for any given realization of a colony's physical makeup, and thus we can easily generate a statistical distribution of the mobility statistics for a specified demographic colony distribution model by direct sampling, just as we did for translational diffusivity in the disclike colony model (blue circles in Figure~\ref{diffusion_size}).

\subsection{Justification of Generalized Geometry Formulas}
\label{sec:genderive}
The derivation of the formulas for the effective translation and rotation of colonies with a general configuration follows the same mathematical procedure as for the disc shape from Section~\ref{sec:methods}, just with more notationally complex versions of the formulas.  We will only discuss some key components of the argument and add details only where they are novel to the generalized colony geometry setting.  All calculations are only accurate to second order in $ \sigma_F/\Fmag $ and $ \Thmag$; we do not report formal error estimates here.

We first compute
the first and second order statistics of the total body force and torque on the colony:
\begin{itemize}
    \item the mean total force on the colony, in the body frame:
    \begin{equation}
        \Fbodymean  = 
        \sum_{j=1}^N F_j^{(0)} \orient{j} \expe^{-\frac{1}{2} \sigma_{\Theta}^2}
    \end{equation}
     \item the mean torque on the colony:
    \begin{equation}
        \Torquemean  =\sum_{j=1}^N F_j^{(0)} \expe^{-\frac{1}{2} \sigma_{\Theta}^2}(\flagpos{j} - \centmass) \cdot\orientperp{j}
    \end{equation}
    \item the covariance matrix of the total force on the colony, in the body frame
    \begin{equation*}
        \Cov(\Fbody,\Fbody) = \sigma_F^2 \sum_{j=1}^N \orient{j} \otimes \orient{j} + \sigmaThe^2 \sum_{j=1}^N (F_j^{(0)})^2 
        \orientperp{j} \otimes \orientperp{j}
    \end{equation*}
    \item the variance of the torque on the colony:
    \begin{equation*}
        \Var (\Torquecol) \sim \sum_{j=1}^N  (F_j^{(0)})^2 \sigma_{\Theta}^2
        \left|(\flagpos{j} - \centmass) \dotprod \orient{j}\right|^2 +\sigma_F^2 \left|(\flagpos{j} - \centmass) \cdot \orientperp{j}\right|^2
    \end{equation*}
    \item the covariance of the torque and vectorial colony force in the body frame
    \begin{equation}
        \Cov(\Torquecol,\Fbody) = 
\sum_{j=1}^N \sigma_F^2\left[(\flagpos{j} - \centmass)\cdot \orientperp{j}\right] \orient{j}
        - (F_j^{(0)})^2 \sigmaThe^2 \left[(\flagpos{j} - \centmass) \cdot \orient{j}\right] \orientperp{j})
        \label{eq:covtf}
    \end{equation}
\end{itemize}
These calculations are quite similar to those described in Subsection~\ref{sub:rotstat},  averaging over the stationary distribution of the flagellar variables $ \{\Theta_j (t)\}_{j=1}^N$, $ \{F_j (t)\}_{j=1}^N$.    The main complication is the need to introduce a more general friction matrix~\eqref{eq:granddrag}, which moreover couples rotational and translational motion in general~\citep[Sec. 5.4]{jh:lrhsa}.  As a consequence, we need to account for correlations between rotations and translations in computing the effective rotational diffusivity.  The covariance~\eqref{eq:covtf} was obtained by the following calculation together with an expansion with respect to small $ \sigmaThe$:
\begin{align*}
        \Cov(\Torquecol,\Fbody) &= 
        \sum_{j=1}^N \sigma_F^2(\flagpos{j} - \centmass)\cdot \langle \Rotmat{\Theta_j (t)} \orientperp{j}\otimes \Rotmat{\Theta_j (t)}\orient{j} \rangle \\
        & \qquad \qquad + (F_j^{(0)})^2 (\flagpos{j} - \centmass) \cdot\Cov(\Rotmat{\Theta_j(t)} \orientperp{j},\Rotmat{\Theta_j (t)} \orient{j}), \\
        \langle \Rotmat{\Theta_j (t)} \orientperp{j} \otimes \Rotmat{\Theta_j (t)}\orient{j} \rangle
        &= \langle \cos^2 \Theta_j (t) \rangle \orientperp{j}\otimes \orient{j}
        - \langle \sin^2 \Theta_j (t) \rangle (\orient{j}\otimes \orientperp{j}) \\
& \qquad \qquad        + \langle \sin \Theta_j (t) \cos \Theta_j (t) \rangle (\orient{j} \otimes \orient{j}-\orientperp{j}\otimes\orientperp{j})
        \\
        & = \frac{1+\langle \cos 2\Theta_j (t)\rangle}{2} 
        \orientperp{j}\otimes \orient{j}
        - \frac{1 - \langle \cos 2 \Theta_j (t)\rangle}{2}  \orient{j}\otimes \orientperp{j} 
        + 0(\orient{j} \otimes \orient{j}-\orientperp{j}\otimes\orientperp{j})
        \\
       & = \frac{1 + \expe^{-2 \sigmaThe^2}}{2} \orientperp{j}\otimes \orient{j}
        - \frac{1 - \expe^{-2 \sigmaThe^2}}{2}
        \orient{j}\otimes \orientperp{j} \\
        \Cov(\Rotmat{\Theta_j(t)} \orientperp{j},\Rotmat{\Theta_j (t)} \orient{j}) &= \langle \Rotmat{\Theta_j (t)} \orientperp{j} \otimes \Rotmat{\Theta_j (t)}\orient{j} \rangle - (\langle \Rotmat{\Theta_j(t)}\rangle \orientperp{j}) \otimes  (\langle \Rotmat{\Theta_j(t)}\rangle \orient{j}) \\
        &=\frac{1 + \expe^{-2 \sigmaThe^2}}{2} \orientperp{j}\otimes \orient{j}
        - \frac{1 - \expe^{-2 \sigmaThe^2}}{2}
        (\orient{j}\otimes \orientperp{j})
        - (\expe^{-\sigmaThe^2/2} \orientperp{j})\otimes(\expe^{-\sigmaThe^2/2} \orient{j}) \\
        &= \frac{(1-\expe^{-\sigmaThe^2})^2}{2} \orientperp{j} \otimes \orient{j}
        - \frac{1 - \expe^{-2 \sigmaThe^2}}{2}
        (\orient{j}\otimes \orientperp{j})
\end{align*}
Similarly, to compute the covariance matrix of the body force (which contributes to the enhanced rotational diffusion through the rotational-translational coupling)
\begin{align*}
     \Cov(\Fbody,\Fbody) &= \sum_{j=1}^N \sigma_F^2 \langle (\Rotmat{\Theta_j (t)} \orient{j}) \otimes (\Rotmat{\Theta_j (t)} \orient{j})\rangle  + \sum_{j=1}^N (F_j^{(0)})^2 \Cov (\Rotmat{\Theta_j (t)} \orient{j}, \Rotmat{\Theta_j (t)} \orient{j}), \\
   \langle \Rotmat{\Theta_j (t)} \orient{j} \otimes \Rotmat{\Theta_j (t)}\orient{j} \rangle
        &= \langle \cos^2 \Theta_j (t) \rangle \orient{j}\otimes \orient{j}
        + \langle \sin^2 \Theta_j (t) \rangle (\orientperp{j}\otimes \orientperp{j}) \\
    & \qquad \qquad     - \langle \sin \Theta_j (t) \cos \Theta_j (t) \rangle (\orient{j}\otimes \orientperp{j} + \orientperp{j}\otimes \orient{j}) \\
        &= \frac{1+\langle \cos 2\Theta_j (t)\rangle}{2} 
        \orient{j}\otimes \orient{j}
        + \frac{1 - \langle \cos 2 \Theta_j (t)\rangle}{2}  \orientperp{j}\otimes \orientperp{j} + 0 (\orient{j}\otimes \orientperp{j} + \orientperp{j}\otimes \orient{j}) \\
       &= \frac{1 + \expe^{-2 \sigmaThe^2}}{2} \orient{j}\otimes \orient{j}
        + \frac{1 - \expe^{-2 \sigmaThe^2}}{2}
        \orientperp{j}\otimes \orientperp{j}, \\
         \Cov(\Rotmat{\Theta_j(t)} \orient{j},\Rotmat{\Theta_j (t)} \orient{j}) &= \langle \Rotmat{\Theta_j (t)} \orient{j} \otimes \Rotmat{\Theta_j (t)}\orient{j} \rangle - (\langle \Rotmat{\Theta_j(t)}\rangle \orient{j}) \otimes  (\langle \Rotmat{\Theta_j(t)}\rangle \orient{j}) \\
        &=\frac{1 + \expe^{-2 \sigmaThe^2}}{2} \orient{j}\otimes \orient{j}
        + \frac{1 - \expe^{-2 \sigmaThe^2}}{2}
        \orientperp{j}\otimes \orientperp{j}
        - (\expe^{-\sigmaThe^2/2} \orient{j})\otimes(\expe^{-\sigmaThe^2/2} \orient{j}) \\
        &= \frac{(1-\expe^{-\sigmaThe^2})^2}{2} \orient{j} \otimes \orient{j}
        + \frac{1 - \expe^{-2 \sigmaThe^2}}{2}
        \orientperp{j}\otimes \orientperp{j}
\end{align*}

The instantaneous mean-square speed~\eqref{eq:msqspeedgen} follows from the second-order statistics
via the relation:
\begin{equation}
    \Vcol (t) = \Rotmat{\Thetacol (t)}\left[\Granddraginv 
    \begin{pmatrix}
    \Fbody (t) \cr \Torquecol (t)
    \end{pmatrix}    \right]_{\mathbf{x}} \label{eq:vcolinst}
\end{equation}
which implies:
\begin{equation*}
    \Vcolmsq = \left|\left[\Granddraginv 
    \begin{pmatrix}
    \Fbodymean \cr \Torquemean 
    \end{pmatrix}    \right]_{\mathbf{x}}\right|^2
    +\Tr\left[
    \Granddraginv \begin{pmatrix} 
    \Cov(\Fbody,\Fbody) & \Cov (\Fbody,\Torquecol) \cr
    \Cov(\Torquecol,\Fbody) & \Var(\Torquecol)
    \end{pmatrix}
\Granddraginv    
    \right]_{\bx},
\end{equation*}
The matrix appearing here takes the form of a sum of dyads from each flagellum which allows the somewhat simplified representation in Eq.~\eqref{eq:msqspeedgen}.

To compute the long-run drift and diffusion, we turn to the equations of microhydrodynamics which balance the active forces and torques, drag force and torque, and the Brownian forces and torques, enhanced by the active flagellar noise.  The flagellar force fluctuations (which contribute to the terms proportional to $ \sigma_F^2$) decay exponentially with time scale $ \gamma_F^{-1}$, and the  force orientation fluctuations decay exponentially with time scale $ \gamma_{\Theta}^{-1}$.  By standard results on white-noise approximations of Ornstein-Uhlenbeck noise~\citep[Sec. 8.1]{cwg:hsm}, the square of the coefficient of the white noise should be chosen as twice the product of the variance and correlation time of the Ornstein-Uhlenbeck process to have an equivalent effect on slower variables.  Indeed the derivation of this result relies on the formula~\eqref{eq:kuboint}.  Applying this coarse-graining in conjunction with the single-time statistics of the colony force and torque listed above, the microhydrodynamics equations for the colony in its body frame reads~\citep{jfb:sd} 
\begin{align}
\label{eq:microhydro}
\begin{pmatrix}
 \Fbodymean \cr \Torquemean 
    \end{pmatrix}
+
    \sqrt{2\Actnoisemat}
    \frac{\difd \bWgrandact}{\difd t}
+ \sqrt{2 k_B T \Granddrag} \frac{\difd \bWgrandtherm (t)}{\difd t}
&= \Granddrag \begin{pmatrix}
    \Veltransbody \cr \Velrot
\end{pmatrix},
\end{align}
We have represented generic three-dimensional white noise representing thermal fluctuations by the formal expression $ \frac{\difd \bWgrandtherm (t)}{\difd t}$ to facilitate the re-expression of the microhydrodynamic equations in the language of stochastic differential equations for the colony position and orientation in the lab frame below.  We have another independent copy of white noise $ \frac{\difd \bWgrandtherm (t)}{\difd t}$ representing the coarse-graining of the active flagellar fluctuations on the colony's body force and torque; the active noise matrix is given by
\begin{align*}
    \Actnoisemat 
    &= 
    \sigma_F^2 \gamma_F^{-1} \sum_{j=1}^N 
    \begin{pmatrix} \orient{j} \cr (\flagpos{j} - \centmass)\cdot \orientperp{j}
    \end{pmatrix}
    \begin{pmatrix}
     \orient{j} & (\flagpos{j} - \centmass)\cdot \orientperp{j}
    \end{pmatrix} \\
& \qquad \qquad + \sigmaThe^2 \gamma_{\Theta}^{-1} \sum_{j=1}^N (F_j^{(0)})^2    
\begin{pmatrix} \orientperp{j} \cr -(\flagpos{j} - \centmass)\cdot \orient{j}
    \end{pmatrix}
    \begin{pmatrix}
     \orientperp{j} & -(\flagpos{j} - \centmass)\cdot \orient{j}
    \end{pmatrix} 
\end{align*}
The vectors comprising the first and second dyadic sums can be readily understood as the impact of a force fluctuation along the natural orientation of a flagellum, respectively an orientation fluctuation at the natural force of a flagellum, on the supervector of colony force and torque. 

We now re-express Eq.~\eqref{eq:microhydro} as a stochastic differential system for the colony configuration variables: 
\begin{align}
\begin{pmatrix}
\difd \bXcol (t) \cr \difd \Thetacol (t)
\end{pmatrix}
&= \Rotmatthree{\Thetacol (t)}
\begin{pmatrix}
\Veltransbody \, \difd t \cr \Velrot \, \difd t
\end{pmatrix} \label{eq:supercol}\\
& = \Rotmatthree{\Thetacol (t)} 
\left[\Granddraginv 
\begin{pmatrix}
\Fbodymean \cr \Torquemean
\end{pmatrix}
\, \difd t
+ \sqrt{2 k_B T \Granddraginv} \, \circ \difd \bWgrandtherm (t)
+ \Granddraginv \sqrt{2\Actnoisemat}
    \circ \difd \bWgrandact (t)
\right]. \nonumber
\end{align}
where
\begin{equation*}
    \Rotmatthree{\theta} \equiv \begin{pmatrix}
    \cos \theta & - \sin \theta & 0 \cr
    \sin \theta & \cos \theta & 0 \cr
    0 & 0 & 1 
    \end{pmatrix}.
\end{equation*}
is the augmented rotation matrix which maps the colony configuration supervector from the body frame to the lab frame.  Note that here we must distinguish the Stratonovich interpretation of the stochastic differentials~\citep[Sec. 4.4]{gardiner}, which we have denoted by $ \circ \difd \bW $, since the colony rotation $ \Thetacol (t)$ now appears as a variable coefficient.  The Stratonovich interpretation is appropriate because of the usual  reasons of taking short-correlation-time limits of physical processes.
The colony orientation can be solved first to give:
\begin{equation}
    \Thetacol (t) = \Thetacol (0) + \left[\Granddraginv 
\begin{pmatrix}
\Fbodymean \cr \Torquemean
\end{pmatrix}
t
+ \sqrt{2 k_B T \Granddraginv} \, \bWgrandtherm (t)
+ \Granddraginv \sqrt{2\Actnoisemat}
    \bWgrandact (t)\right]_{\theta}. \label{eq:thetacolsolve}
\end{equation}
From here, the rotational statistics~\eqref{eq:rotdriftgen} and~\eqref{eq:rotdiffgen} follow, using the form of $ \Actnoisemat$ as a sum of dyads to simplify the representation of the effective rotational diffusivity $ \Diffroteff = k_B T \Granddraginv + \Granddraginv \Actnoisemat \Granddraginv$.

A detailed justification of the lack of long-time translational drift can be developed by a direct generalization of the calculation for the disc colony in Subsection~\ref{sub:transstat}.   We begin by rewriting the translational component of Eq.~\eqref{eq:supercol} in \Ito\ form, which produces a drift correction from the correlated fluctuations in the colony rotation and flagellar fluctuations~\citep[Sec. 4.4]{gardiner}:
\begin{align*}
\difd \bXcol (t) 
& = \Rotmat{\Thetacol (t)} 
\left[\Granddraginv 
\begin{pmatrix}
\Fbodymean \cr \Torquemean
\end{pmatrix} 
\, \difd t
+ \sqrt{2 k_B T \Granddraginv} \,  \difd \bWgrandtherm (t)
+ \Granddraginv \sqrt{2\Actnoisemat}
    \difd \bWgrandact (t)
\right]_{\mathbf{x}} \\
& \qquad \qquad + 
\Rotmat{\Thetacol (t)+\pi/2} 
\left[\sqrt{2 k_B T \Granddraginv} \difd \langle \bWgrandtherm (t),\Thetacol (t)\rangle
+
\frac{1}{2}\Granddraginv \sqrt{2\Actnoisemat} \difd \langle \bWgrandact (t), \Thetacol (t) \rangle\right]_{\mathbf{x}}.
\end{align*}
Here we used the fact that $ \partial \Rotmat{\theta}/\partial \theta = \Rotmat{\theta+\pi/2}$ and next compute the quadratic variations:
\begin{align*}
    \difd \langle \bWgrandact (t), \Thetacol (t) \rangle
    &=  \left[\Granddraginv \sqrt{2\Actnoisemat}\right]^{T}_{\theta} \, \difd t, \\
    \difd \langle \bWgrandtherm (t),\Thetacol (t) \rangle &= 
    \left[\sqrt{2 k_B T \Granddraginv}\right]_{\theta}^{T} \, \difd t
\end{align*}
The notation $ [\cdot]_{\theta}^T$ applied to $3\times 3$ matrices means to take the third row (corresponding to the $ \theta$ component) and transpose it to a 3-dimensional column vector.
Effecting these evaluations in the drift correction term, we obtain:
\begin{align}
\difd \bXcol (t) 
& = \Rotmat{\Thetacol (t)} 
\left[\Granddraginv 
\begin{pmatrix}
\Fbodymean \cr \Torquemean
\end{pmatrix} \, \difd t
+ \sqrt{2 k_B T \Granddraginv} \,  \difd \bWgrandtherm (t)
+ \Granddraginv \sqrt{2\Actnoisemat}
    \difd \bWgrandact (t)
\right]_{\mathbf{x}} \nonumber \\
& \qquad \qquad +  2\Rotmat{\Thetacol (t) +\pi/2} \left[k_B T\Granddraginv+\Granddraginv \Actnoisemat \Granddraginv\right]_{\bx,\theta}
\, \difd t  \label{eq:afterito}
\end{align}
The notation $ [\cdot]_{\bx,\theta}$ applied to $ 3\times 3$ matrices indicates the 2-dimensional column vector obtained by taking the first two entries (translational components) of the third column (rotational component).
The thermal component of the drift correction term only arises for colonies with asymmetric shapes that induce translational-rotational coupling, and is absent for disclike colonies.  The active component of the drift correction corresponds for a disclike colony to the terms proportional to $ \sigma_F^2 $ and $ \sigma_{\Theta}^2$ in a second order expansion of Eq.~\eqref{Col_force_autocorr}, recalling the complex notation used there.  Note this \Ito\ drift correction will have no effect on the mean-square velocity measured over coarse-grained time scales by essentially the same argument developed in detail for the disclike model in Subsection~\ref{sub:transstat}.   Recalling that the \Ito\ drift correction in Eq.~\eqref{eq:afterito} represents the correlations of the flagellar variables with the colony rotation, calculation of the velocity correlation function evaluated at lags $ \tau $ large compared to the flagellar time scales will produce equal and opposite terms because the relevant correlations are between the velocity at the advanced time $ t + \tau $ and \emph{past} dynamics of $ \Thetacol  $ on $ [t,t+\tau]$ and the velocity  at the retarded time $ t $ and the \emph{future} dynamics of $ \Thetacol$ on $ [t,t+\tau]$.  Therefore we obtain the simple expression~\eqref{eq:coarsespeedgen} for the mean-square velocity inferred over time scales that do not resolve the flagellar dynamics.

Next, to characterize the effective translational diffusion, we cannot simply cite standard results for a rigid object with a fixed speed and prescribed rotational drift and diffusion because of the translational-rotational coupling.  We rather apply the method of moments, computing the differential $ \difd \bXcol (t) \cdot \bXcol (t)$ using \Ito's lemma and then taking the average:
\begin{align} 
\difd \langle \bXcol (t) \cdot \bXcol (t) \rangle 
&= 2 \langle (\bXcol (t))^T  \Rotmat{\Thetacol (t)}  \rangle \left[\Granddraginv 
\begin{pmatrix}
\Fbodymean \cr \Torquemean
\end{pmatrix} \right]_{\bx} \, \difd t \label{eq:methmom}\\
& \qquad \qquad + 4\langle (\bXcol (t))^T \Rotmat{\Thetacol (t) +\pi/2} \rangle \left[k_B T\Granddraginv+\Granddraginv \Actnoisemat \Granddraginv\right]_{\bx,\theta} \, \difd t \nonumber \\
& \qquad \qquad + 2\Tr \left[k_B T \Granddraginv +  \Granddraginv \Actnoisemat \Granddraginv\right]_{\bx,\bx}
\, \difd t \nonumber
\end{align}
We compute similarly, referring also to Eq.~\eqref{eq:thetacolsolve}:
\begin{align*}
\difd \langle (\bXcol (t))^T \Rotmat{\Thetacol (t)}  \rangle 
&= \left[\Granddraginv 
\begin{pmatrix}
\Fbodymean \cr \Torquemean
\end{pmatrix}\right]^T_{\bx} \, \difd t+   2\left[k_B T\Granddraginv+\Granddraginv \Actnoisemat \Granddraginv\right]^T_{\bx,\theta} \Rotmat{-\pi/2} \, \difd t \\
& \qquad \qquad + \meanrot \langle (\bXcol (t))^T  \Rotmat{\Thetacol (t)}\rangle  \Rotmat{\pi/2} 
 \, \difd t \\
& \qquad \qquad - \Diffroteff\langle (\bXcol (t))^T  \Rotmat{\Thetacol (t)} \rangle\, \difd t \\
& \qquad \qquad +2 
\left[k_B T\Granddraginv+\Granddraginv \Actnoisemat \Granddraginv\right]^T_{\bx,\theta}\Rotmat{\pi/2}\, \difd t
\end{align*}
This is just a simple damped harmonic oscillator equation with forcing that cancels out, whose solution can be written in terms of matrix exponentials as
\begin{equation*}
   \langle (\bXcol (t))^T  \Rotmat{\Thetacol (t)}  \rangle
= \int_0^t  
\left[\Granddraginv 
\begin{pmatrix}
\Fbodymean \cr \Torquemean
\end{pmatrix} \right]^T_{\bx} 
 \expe^{(-\Diffroteff \Id +\meanrot \Rotmat{\pi/2}) (t-s)}\, \difd s,
\end{equation*}
where we assume without loss of generality that $ \bXcol (0) = \bzero$.
Upon substitution into Eq.~\eqref{eq:methmom} then integrating, we have:
\begin{align*}
\langle \bXcol (t) \cdot \bXcol (t) \rangle
&= 2 \int_0^t \int_0^{\tp}
\left[\Granddraginv 
\begin{pmatrix}
\Fbodymean \cr \Torquemean
\end{pmatrix} \right]^T_{\bx} 
\expe^{(-\Diffroteff \Id +\meanrot \Rotmat{\pi/2}) (\tp-s)} \\
& \qquad \qquad \qquad \qquad \left\{\left[\Granddraginv 
\begin{pmatrix}
\Fbodymean \cr \Torquemean
\end{pmatrix} \right]_{\bx} 
+ 2\Rotmat{\pi/2}\left[k_B T\Granddraginv+\Granddraginv \Actnoisemat \Granddraginv\right]_{\bx,\theta}\right\} \, \difd s \, \difd \tp \\
& \qquad \qquad +  2\Tr \left[k_B T \Granddraginv +  \Granddraginv \Actnoisemat \Granddraginv\right]_{\bx,\bx} t
\end{align*}
Computing the long-time limit of the double integral:
\begin{align*}
\lim_{t\rightarrow \infty} \frac{1}{t}
    \int_0^t \int_0^{\tp} \expe^{(-\Diffroteff \Id +\meanrot \Rotmat{\pi/2}) (\tp-s)} \, \difd s \, \difd \tp
    &= \lim_{t\rightarrow \infty} \frac{1}{t} \int_0^t \left(\Id - \expe^{(-\Diffroteff \Id +\meanrot \Rotmat{\pi/2}) \tp}\right)
    \left(\Diffroteff \Id -\meanrot \Rotmat{\pi/2}\right)^{-1} \, \difd \tp \\
    &=   \left(\Diffroteff \Id -\meanrot \Rotmat{\pi/2}\right)^{-1} = \frac{\Diffroteff \Id + \meanrot \Rotmat{\pi/2}}{\Diffroteff{}^2 + \meanrot{}^2}
\end{align*}
we obtain:
\begin{align*}
\Difftranseff &= \lim_{t \rightarrow \infty} \frac{\langle \bXcol (t)\cdot \bXcol (t)\rangle }{4t} \\
&= \frac{1}{2} 
\left[\Granddraginv 
\begin{pmatrix}
\Fbodymean \cr \Torquemean
\end{pmatrix} \right]^T_{\bx} 
 \frac{\Diffroteff \Id + \meanrot \Rotmat{\pi/2}}{\Diffroteff{}^2 + \meanrot{}^2}\\
 & \qquad \qquad  \left\{\left[\Granddraginv 
\begin{pmatrix}
\Fbodymean \cr \Torquemean
\end{pmatrix} \right]_{\bx} 
+ 2\Rotmat{\pi/2}\left[k_B T\Granddraginv+\Granddraginv \Actnoisemat \Granddraginv\right]_{\bx,\theta}\right\}
+ \frac{1}{2} \Tr\left[k_B T \Granddraginv +  \Granddraginv \Actnoisemat \Granddraginv\right]_{\bx,\bx} \\
&= \frac{\Diffroteff}{2(\Diffroteff{}^2+\meanrot{}^2)}
\left|\left[\Granddraginv 
\begin{pmatrix}
\Fbodymean \cr \Torquemean
\end{pmatrix} \right]_{\bx}\right|^2
- \frac{\meanrot}{(\meanrot{}^2+\Diffroteff{}^2)}
\left[\Granddraginv 
\begin{pmatrix}
\Fbodymean \cr \Torquemean
\end{pmatrix} \right]_{\bx}\cdot \left[k_B T\Granddraginv+\Granddraginv \Actnoisemat \Granddraginv\right]_{\bx,\theta} \\
& \qquad \qquad + \frac{1}{2} \Tr\left[k_B T \Granddraginv +  \Granddraginv \Actnoisemat \Granddraginv\right]_{\bx,\bx}
\end{align*}
The thermal term $ \frac{1}{2} \Tr[k_B T \Granddraginv]_{\bx,\bx}$ is just the translational diffusivity $ \Difftrans$, and we treat the term $ \left[\Granddraginv \Actnoisemat \Granddraginv\right]_{\bx,\bx} $ as negligible because it arises from randomness decorrelating on the short flagellar time scale.  (If we wished to retain it accurately, we would need to do a fine analysis of the velocity correlations on these flagellar time scales.)  Substitution of the expressions for mean total force, torque, and active noise matrix $ \Actnoisemat$ derived above the yields the expression~\eqref{eq:transdiffgen} in the summary.

\section{Discussion}
\label{sec:discuss}

We have presented in Section~\ref{summarych1} an analytically computable quantitative framework for the effective mobility characteristics of colonial microswimmers typified by choanoflagellate Rosettes.  Extensions to chainlike and other morphologies were given in Section~\ref{sec:gengeo}.  The major assumptions leading to these results is a thin planar geometry for the colony and some asymptotic assumptions about physical parameters, such as smallness of the dynamical fluctuations in the flagella and slowness of the colony's rotational dynamics relative to the flagellar time scales.  
From the analytical formulas, we can characterize the impact of various sources of randomness.  On the one hand, motion of the colony relies on
demographic statistical variations in flagellar placement and/or properties, otherwise the flagellar forces would cancel out.  On the other hand, the dynamical variance of the  angular orientation of the flagella leads to lower colony propulsive forcing and mean instantaneous translational and rotational speed.  This signifies that the colony can be made less mobile if the molecular motors in the flagella cause too much angular forcing noise.  While the approximations we have made to obtain a tractable theoretical expression for the translational diffusivity are clearly not as quantitatively accurate as the other transport statistics we have analyzed, the theoretical expressions do correctly predict wide variability in translational diffusivity across colonies and connect it to the variability in the rotational drift, which scales with colony size as $ N^{-3/2}$.  In terms of the coefficient of variation (ratio of standard deviation to mean), the demographic variability of rotational drift and diffusivity is found to decrease with colony size $N$ as $ N^{-1/2}$ (see also Figures~\ref{long time rot drift N} and~\ref{long time rot diff N}),  while the demographic variability of the root-mean-square speed remains substantial for large colonies (Figure~\ref{speedsize} and accompanying discussion).

Our results in Section~\ref{summarych1} give some  measure, within the context of an idealized model, of experimental observations on the slower swimming speed of \emph{S. rosetta} colonies relative to single cells, and of their tendency to swim in circles through quantification of rotational drift and rotational diffusion~\citep{KoehlSelective}.  Characterization of the rotational effects requires explicit consideration of both dynamic and cell-to-cell variability in the flagellar force and orientation.  We have endeavored in this way to render more  precision to the scaling arguments from~\citet{SolariHydrodynamics2006,colonialmotility}  on how the forces from the constituent flagella of a colony combine to induce the mobility properties of the colony.  Our model does not predict the increase of colony speed with size reported in~\citet{colonialmotility}, but does explain the wide variability in speed across colonies seen in these experiments.

Other than the translational diffusivity, the mobility expressions for the more general colony geometry in Subsection~\ref{sub:gensum} are separately quadratic in the flagellar orientation variables $ \orient{j}$ and flagellar positions $ \flagpos{j}$.  If we neglected variations in the friction coefficients of the colony,  then averages over statistical models of demographic variation could be computed just as was done for disc colonies in Section~\ref{summarych1}.   We leave such analyses for chainlike and other colony configurations for future work.   Our primary purpose here was to conduct a statistical mobility analysis for roughly symmetric disclike colonies in an effort to better understand the observed dynamics of rosette colonies in~\citet{colonialmotility}.  Section~\ref{sec:gengeo} simply shows that the same methodology carries through to other planar morphologies, so long as the noisiness of the cycle-averaged flagellar forces can be taken to be at least somewhat small.  A fully three-dimensional extension of our analytical approach does not seem promising, as the stochastic dynamics of non-commuting rotations would be quite cumbersome to analyze beyond the phenomenological approach in~\citet{colonialmotility}.

Our mathematical modeling framework has some similarities to the approach in the recent work of~\citet{thiffeault_anisotropic_2022} in which the stochastic dynamics of a microswimmer is characterized in terms of the dynamical response to a stochastic model for flagellar forcing, rather than an overall phenomenological ``active Brownian particle'' model for the microswimmer.  In that work as well as ours, the flagellar forcing induces stochastically coupled translation and rotation of the microswimmer.  Though the calculations in~\citet{thiffeault_anisotropic_2022}  were only presented for a single flagellum, their methodology should be capable of analyzing colonial dynamics as well.  Key differences are that our flagellar forcing model in Subsection~\ref{sec:cellmod} has correlated fluctuations in magnitude and direction, which are a bit more amenable to parameterization from the experimental results of~\citet{colonialmotility}, while~\citet{thiffeault_anisotropic_2022} incorporate inertial effects which are found to result in a substantial noise-induced drift in the body frame of anisotropic swimmers.  In our model, which neglects inertia altogether, we do have rather straightforward corrections to the drift in the body frame from fluctuations in the flagellar orientations, as can be seen in the formula~\eqref{eq:msqspeedgen} for the mean-square speed, but the drift correction expression in~\citet{thiffeault_anisotropic_2022} has a more subtle inclusion of microswimmer mass and moment of inertia.  In principle the analysis of general planar colonies in Section~\ref{sec:gengeo} could be extended to include inertia, though the calculations with our flagellar models may become rather complicated.   Useful simplifications might be possible by assuming scale separation between the inertial time scale of the swimming body and the flagellar correlation time scale.  The expression for translational diffusion of the swimmer model in~\citet{thiffeault_anisotropic_2022} corresponds approximately to the first two terms in Eq.~\eqref{eq:transdiffgen}, with the addition of a correction term which is similar in nature to the third term in Eq.~\eqref{eq:transdiffgen} in that it involves a product of the swimmer velocity and the components of the grand diffusion tensor in the body frame, $ k_B T \Granddraginv$, which couple rotation and translation.  Our expression, however, involves the mean rotational rate rather than the rotational diffusion rate in ~\citet{thiffeault_anisotropic_2022}, and an inner product rather of vectors than a product of magnitudes.  No analogue to the remaining terms in Eq.~\eqref{eq:transdiffgen}  appear in the translational diffusion reported in~\citet{thiffeault_anisotropic_2022}

\citet{Fauci_morphology} examined via computational fluid dynamic simulations the role of microvillar collars, and suggested the utility of studying the effects of variability in microvilla lengths on choanoflagellate cell motility.  In principle, one could adopt a similar modeling approach as we have pursued here for demographic stochasticity of microvilli.    As shown in~\citet{Fauci_morphology}, these collars may well have a significant effect on the mobility of a colony.  One might hope the effects of the microvillar colonies could be incorporated in our model by suitably renormalizing the hydrodynamic drag coefficients, but our geometric representation of the cells is likely too crude to give the quantitative accuracy available from the direct simulations in~\citet{Fauci_morphology}.

Here we have focused on Chaonoflagellates and other colonial protozoa where the cellular flagella are known to be far enough apart for the hydrodynamic interactions to be negligible. However,  some protozoa 
form densely populated colonies of flagellated cells.  A prime example is the green algae Volvox, which forms colonies of approximately 50000 cells about 2000 of which are densely packed flagellated cells on its surface used for motion \citep{SheltonVolvox}.  
\cite{BrumleyVolvox} confirm that densely packed flagella and cilia are hydrodynamically interacting and theorize that the interactions can generate metachronal waves. An interesting direction for future development of the work presented here is to extend it to densely packed colonies with coordination, rather than independence, between the flagella.

\appendix
\section{Numerical Simulations} \label{sim_settings}
Unless otherwise specified, all simulations are based on the physical parameters as given in Table~\ref{Tab:Parameter table}.  
In each figure, an individual colony sample is simulated by generating a set of the flagellar displacements $ \{S_i\}_{i=1}^N$ with the indicated distribution and relaxed flagellar orientations $\Theta_i^{(0)}\sim N(0,0.0004)$.  
 $F_i^{(0)}$ is evaluated as a uniformly distributed random variable between 1 and 3 $ \si{\pico\newton}$, based on the range reported in Table \ref{Tab:Parameter table}. 
These flagellar parameters are held fixed for all simulations in a figure where the parameter $\sigmaThe^2$ is varied, effectively simulating different trajectories of a given colony over different dynamical variances of the flagellar orientation.

We use the standard Euler-Marayama method with time step $ \Delta t = 0.1 \gamma_\Theta^{-1} = 0.1\gamma_F^{-1}$. In all simulations, we initialize the flagellar angles $ \Theta_i(0)$ with their stationary distribution $ N(\Theta_i^{(0)},\sigmaThe^2)$ and the flagellar forces with their stationary distribution $ N(F_i^{(0)},\sigma_F^2)$.
The center of mass of the colony starts at $ \mathbf{\mathbf{X}^{(c)}}(0) = \boldsymbol{0}$ and colony orientation is initialized as $\Theta^{(c)}(0)=0$.

From the numerical simulation of the trajectory of a given colony, the various transport statistics are computed as follows.
Rotational drift  is estimated by simply dividing the total trajectory displacement by the simulation time.  Rotational diffusivity is estimated  by taking the sample variance of the angular change over 999 subsequent intervals of length $ \SI{10}{\second}$, leaving out the first interval starting from $ t=0$.  The instantaneous speed of a colony is computed by dividing the total force by the translational drag, sampled at times spaced by $\SI{10}{\second}$.  The time spacings here are chosen to be long relative to the decorrelation time $\SI{0.1}{\second}$ of the flagella.  For the translational diffusivity, we need to consider increments that are long relative to the decorrelation time of the colony rotation, which from Figure~\ref{long time rot diff N} is on the order of $ \SI{10}{\second}$.  Thus, we estimate the translational diffusivity by
averaging the displacement variance over 99 subsequent intervals of length $\SI{100}{\second}$  over a simulation trajectory of $ \SI{10000}{\second}$, leaving out the first interval starting from $t=0$. 

\section{Expectations of Functionals of Gaussian Random Processes}
\label{app:gaussavg}
Here we collect some basic results about certain averages of Gaussian random functions that we need in our calculations in Subsection~\ref{sec:methods}. We note first that if 
$ Y (t)$ is a Gaussian random process with mean $ \muY (t) \equiv \langle Y (t) \rangle$ and correlation function $ \CY (t,t^{\prime}) \equiv \langle (Y (t)-\muY (T)) \otimes (Y(t^{\prime})-\muY(t^{\prime})) \rangle$ and $ \varphi (t)$ is a continuous deterministic function, then 
\begin{equation*}
   \int_0^t \varphi (t^{\prime}) Y (t^{\prime}) \, \difd t^{\prime} 
\end{equation*}
is a Gaussian random variable with mean 
$\int_0^t \muY (t^{\prime}) \, \difd t^{\prime}$  and variance $ \int_0^t \int_0^t \varphi (t^{\prime}) \CY (t^{\prime},t^{\prime\prime}) \varphi (t^{\prime \prime})  \, \difd t^{\prime} \, \difd t^{\prime \prime}$~\citep[Sec. 3.9.3]{GrigoriuStochCalc}.  Similarly, we can compute the covariance between two integrals with possibly different continuous test functions $ \varphi (t)$, $\psi (t)$:
\begin{equation*}
\Cov \left( \int_0^t \varphi (t^{\prime}) Y (t^{\prime}) \, \difd t^{\prime},  \int_0^t \psi (t^{\prime}) Y (t^{\prime}) \, \difd t^{\prime}  \right)=\int_0^t \int_0^t \varphi (t^{\prime}) \CY (t^{\prime},t^{\prime\prime}) \psi (t^{\prime \prime})  \, \difd t^{\prime} \, \difd t^{\prime \prime}.
\end{equation*}
Also, we have the covariance between such integrals with the Gaussian random process itself:
\begin{equation*}
\Cov \left(Y(t),\int_0^t \varphi (t^{\prime}) Y (t^{\prime}) \, \difd t^{\prime}\right)=
\int_0^t \CY (t,t^{\prime}) \varphi (t^{\prime}) \, \difd t^{\prime}.
\end{equation*}

These relationships imply that the various averages over Gaussian random functions are really reducible to averages over Gaussian random variables.  These can be done by using the moment generating function of a Gaussian random vector $ \bZ$ of dimension $d$ with mean $ \bmu$ and covariance matrix $ \Cmat$ with diagonal entries equal to the variances $ \sigma_j^2$:
\begin{equation}
    \langle \expe^{\mathbf{s}\cdot\bZ} \rangle  = \expe^{\mathbf{s}\cdot \bmu+\frac{1}{2} \mathbf{s} \cdot \Cmat \cdot \mathbf{s}} \text{ for } \bs \in \mathbb{C}^d
    \label{eq:jointmgf}
\end{equation}
From this we can derive a few formulas we will use in our calculations:
\begin{align}
    \langle Z_2 \expe^{s_1 Z_1} \rangle &=s_1 C_{1,2}
    \expe^{s_1 \mu_1+\frac{1}{2} s_1^2 \sigma_1^2}, \\ \langle Z_2^2 \expe^{s_1 Z_1} \rangle &=
    \left(\sigma_2^2 + (\mu_2 +s_1 C_{1,2})^2\right) \expe^{s_1 \mu_1+\frac{1}{2} s_1^2 \sigma_1^2}, \\
       \langle Z_2 Z_3 \expe^{s_1 Z_1} \rangle &= \left(C_{2,3}+(\mu_2+s_1 C_{1,2})(\mu_3+s_1 C_{1,3})\right)
    \expe^{s_1 \mu_1+\frac{1}{2} s_1^2  \sigma_1^2}
    \label{eq:zez}
\end{align}
This latter result follows from evaluating the first two partial derivatives of Eq.~\eqref{eq:jointmgf} with respect to $ s_2$ and/or $s_3$ at $ (s_1,s_2,s_3)=(1,0,0)$.

By choosing $ s_1=\mathi$ and $ s_2=0$ in Eq.~\eqref{eq:jointmgf}, and taking real and imaginary parts, we have:
\begin{equation}
    \langle \cos Z_1 \rangle = \cos \mu_1 \expe^{-\frac{1}{2} \sigma_1^2}, \qquad \qquad 
     \langle \sin Z_1 \rangle = \sin \mu_1 \expe^{-\frac{1}{2} \sigma_1^2} \label{eq:trigavg}
\end{equation}

\section{Demographic Variability Calculation}
\label{app:demovar}
Starting from the expression~\eqref{eq:msqspeedsum} and taking averages with respect to the demographic distribution of static flagellar parameters as described in Section~\ref{sec:discmodel}, we obtain:
\begin{align*}
\bExp[\Vcolmsq] &=\gamtrans^{-2}
\left[N \left(\sigma_F^2 + \Fmom{2}\right) 
+ e^{-\sigmaThe^2} \sum_{j=1}^N \sum_{\substack{\jp=1 \\ j \neq j}}^N (\Fmom{1})^2 \expe^{-\varTh^2} \sinc \left(\frac{\flagdispran}{\colrad}\right) \expe^{2\pi \mathi(j - \jp)/N}\right]  \\
&= \gamtrans^{-2}
\left[N \left(\sigma_F^2+ \Fmom{2}\right) 
-N  (\Fmom{1})^2 \expe^{-\varTh^2-\sigmaThe^2} \sinc \left(\frac{\flagdispran}{\colrad}\right)\right] \\
&= \frac{N}{\gamtrans^{2}} \left(\sigma_F^2+ \Fmom{2}
-(\Fmom{1})^2 \expe^{-\varTh^2-\sigmaThe^2} \sinc \left(\frac{\flagdispran}{\colrad}\right)\right)  
\end{align*}

\begin{align*}
&\gamtrans^4\Var[\Vcolmsq]
=  \sum_{j=1}^N \Var[(F_j^{(0)})^2]+
 \expe^{-2\sigmaThe^2} \sum_{\substack{j,\jp=1 \\\jp\neq j}}^N  \expe^{\frac{4 \pi\mathi (j-\jp)}{N}}
\Var \left[F_j^{(0)} F_{\jp}^{(0)} \expe^{\mathi(\Theta^{(0)}_j-\Theta^{(0)}_{\jp})} \expe^{\frac{2\pi\mathi}{N \celllen} (\flagdisp{j} - \flagdisp{\jp})} \right]  \\
&  + 2\expe^{-\sigmaThe^2}\sum_{\substack{j,\jp=1 \\\jp\neq j}}^N \expe^{\frac{2\pi \mathi (j-\jp)}{N}}
\Cov ((F_j^{(0)})^2,F_j^{(0)} F_{\jp}^{(0)} \expe^{\mathi(\Theta^{(0)}_j-\Theta^{(0)}_{\jp})} \expe^{\frac{2\pi\mathi}{N \celllen} (\flagdisp{j} - \flagdisp{\jp})} ) \\
&  +  2\expe^{-\sigmaThe^2}\sum_{\substack{j,\jp=1 \\\jp\neq j}}^N \expe^{\frac{2\pi \mathi (\jp-j)}{N}}
\Cov ((F_j^{(0)})^2,F_j^{(0)} F_{\jp}^{(0)} \expe^{\mathi(\Theta^{(0)}_{\jp}-\Theta^{(0)}_{j})} \expe^{\frac{2\pi\mathi}{N \celllen} (\flagdisp{\jp} - \flagdisp{j})} ) \\
&  + \expe^{-2\sigmaThe^2} \sum_{\substack{j,\jp=1 \\ \jp \neq j}}^N
\Cov\left(
    F_j^{(0)} F_{\jp}^{(0)} \expe^{\mathi(\Theta^{(0)}_j-\Theta^{(0)}_{\jp})} \expe^{\frac{2\pi\mathi}{N \celllen} (\flagdisp{j} - \flagdisp{\jp})} ,
    F_j^{(0)} F_{\jp}^{(0)} \expe^{\mathi(\Theta^{(0)}_{\jp}-\Theta^{(0)}_{j})}\expe^{\frac{2\pi\mathi}{N \celllen} (\flagdisp{\jp} - \flagdisp{j})} \right) \\
& +  \expe^{-2\sigmaThe^2} \sum_{\substack{j,\jp,\jpp=1 \\ j\neq \jp \neq \jpp}}^N 
\expe^{\frac{2 \pi\mathi (2j-\jp-\jpp)}{N}}
\Cov\left(
    F_j^{(0)} F_{\jp}^{(0)} \expe^{\mathi(\Theta^{(0)}_j-\Theta^{(0)}_{\jp})}\expe^{\frac{2\pi\mathi}{N \celllen} (\flagdisp{j} - \flagdisp{\jp})} , F_j^{(0)} F_{\jpp}^{(0)} \expe^{\mathi(\Theta^{(0)}_j-\Theta^{(0)}_{\jpp})}\expe^{\frac{2\pi\mathi}{N \celllen} (\flagdisp{j} - \flagdisp{\jpp})} \right) \\
&  +  \expe^{-2\sigmaThe^2}\sum_{\substack{j,\jp,\jpp=1 \\ j\neq \jp \neq \jpp}}^N 
\expe^{\frac{2 \pi\mathi (j+\jpp-2\jp)}{N}}
\Cov\left(
    F_j^{(0)} F_{\jp}^{(0)} \expe^{\mathi(\Theta^{(0)}_j-\Theta^{(0)}_{\jp})}\expe^{\frac{2\pi\mathi}{N \celllen} (\flagdisp{j} - \flagdisp{\jp})} , F_{\jp}^{(0)} F_{\jpp}^{(0)} \expe^{\mathi(\Theta^{(0)}_{\jpp}-\Theta^{(0)}_{\jp})}\expe^{\frac{2\pi\mathi}{N \celllen} (\flagdisp{\jpp} - \flagdisp{\jp})} \right)
 \\
& +  \expe^{-2\sigmaThe^2} \sum_{\substack{j,\jp,\jpp=1 \\ j\neq \jp \neq \jpp}}^N 
\expe^{\frac{2 \pi\mathi (\jpp-\jp)}{N}}
\Cov\left(
    F_j^{(0)} F_{\jp}^{(0)} \expe^{\mathi(\Theta^{(0)}_j-\Theta^{(0)}_{\jp})}\expe^{\frac{2\pi\mathi}{N \celllen} (\flagdisp{j} - \flagdisp{\jp})} , F_j^{(0)} F_{\jpp}^{(0)} \expe^{\mathi(\Theta^{(0)}_{\jpp}-\Theta^{(0)}_{j})}\expe^{\frac{2\pi\mathi}{N \celllen} (\flagdisp{\jpp} - \flagdisp{j})} \right) \\
& + \expe^{-2\sigmaThe^2}  \sum_{\substack{j,\jp,\jpp=1 \\ j\neq \jp \neq \jpp}}^N 
\expe^{\frac{2 \pi\mathi (j-\jpp)}{N}}
\Cov\left(
    F_j^{(0)} F_{\jp}^{(0)} \expe^{\mathi(\Theta^{(0)}_j-\Theta^{(0)}_{\jp})}\expe^{\frac{2\pi\mathi}{N \celllen} (\flagdisp{j} - \flagdisp{\jp})} , F_{\jp}^{(0)} F_{\jpp}^{(0)} \expe^{\mathi(\Theta^{(0)}_{\jp}-\Theta^{(0)}_{\jpp})}\expe^{\frac{2\pi\mathi}{N \celllen} (\flagdisp{\jp} - \flagdisp{\jpp})} \right)
\end{align*}
\begin{equation}
    \begin{aligned}
\gamtrans^4\Var[\Vcolmsq]&=  N \Var[(F^{(0)})^2] + 
 \twoflagvar \expe^{-2\sigmaThe^2}\left[\sum_{j=1}^N \sum_{\jp=1}^N \expe^{\frac{4 \pi\mathi (j-\jp)}{N}}
- N\right] \\
& \qquad + 4 \expe^{-\sigmaThe^2} \Real \twoflagthrice
\left[\sum_{j=1}^N \sum_{\jp=1}^N \expe^{\frac{2 \pi\mathi (j-\jp)}{N}}
- N\right]  +N (N-1) \expe^{-2\sigmaThe^2} \twoflagmod \\
& \qquad + 2N \expe^{-2\sigmaThe^2} \Real \threeflagcov \left[ \sum_{j=1}^N \sum_{\jp=1}^N \sum_{\jpp=1}^N  \expe^{\frac{2 \pi\mathi (2j-\jp-\jpp)}{N}} -  \sum_{j=1}^{N} \sum_{\jp=1}^N \expe^{\frac{4 \pi\mathi (j-\jp)}{N}} \right. \\
& \qquad \qquad \qquad \qquad \qquad \qquad \left.- 2 \sum_{j=1}^{N} \sum_{\jp=1}^N \expe^{\frac{2 \pi\mathi (j-\jp)}{N}} + 2N\right] \\
& \qquad + 2N \expe^{-2\sigmaThe^2} \Real \threeflagcovconj\left[ \sum_{j=1}^N \sum_{\jp=1}^N \sum_{\jpp=1}^N  \expe^{\frac{2 \pi\mathi (\jpp-\jp)}{N}} - \sum_{j=1}^N \sum_{\jp=1}^N 1 - \sum_{j=1}^N \sum_{\jp=1}^{N}  \expe^{\frac{2 \pi\mathi (j-\jp)}{N}} \right.\\
& \qquad \qquad \qquad \qquad \qquad \qquad \left. -\sum_{j=1}^N \sum_{\jpp=1}^N \expe^{\frac{2 \pi\mathi (\jpp-j)}{N}} + 2N\right] \\
&= N \Var[(F^{(0)})^2] + 
 \twoflagvar \expe^{-2\sigmaThe^2}\left[\left|\sum_{j=1}^N  \expe^{\frac{4 \pi\mathi j }{N}}\right|^2
- N\right] + 4 \expe^{-\sigmaThe^2} \Real \twoflagthrice\left[\left|\sum_{j=1}^N  \expe^{\frac{2 \pi\mathi j }{N}}\right|^2-N\right] \\
& \qquad \qquad +(N^2-N) \twoflagmod \expe^{-2\sigmaThe^2}\\
& + 2 \Real \threeflagcov \expe^{-2\sigmaThe^2}\left[ \sum_{j=1}^N\expe^{\frac{4 \pi\mathi j}{N}} 
\left(\sum_{\jp=1}^N \expe^{-\frac{2 \pi\mathi\jp}{N}}\right)^2 - \left|\sum_{j=1}^N  \expe^{\frac{4 \pi\mathi j }{N}}\right|^2   - 2 \left|\sum_{j=1}^N  \expe^{\frac{2 \pi\mathi j }{N}}\right|^2 + 2N\right] \\
& + 2 \Real \threeflagcovconj\expe^{-2\sigmaThe^2}\left[ \sum_{j=1}^N \left|\sum_{\jp=1}^N  \expe^{\frac{2 \pi\mathi \jp}{N}}\right|^2 -  N^2 - 2\left| \sum_{\jp=1}^{N} \expe^{\frac{2 \pi\mathi \jp}{N}} \right|^2+ 2N\right] \\
&= N \Var[(F^{(0)})^2] +  0- N \twoflagvar \expe^{-2\sigmaThe^2}
-4N \expe^{-\sigmaThe^2} \Real \twoflagthrice
+(N^2-N) \twoflagmod \expe^{-2\sigmaThe^2} \\
& \qquad \qquad 
+ 2 \Real \threeflagcov \expe^{-2\sigmaThe^2} \left[ 0 - 0 - 0 + 2N\right] 
+ 2 \Real \threeflagcovconj \expe^{-2\sigmaThe^2} \left[0 -  N^2 - 0+ 2N\right] \\
&= N (\Var[(F^{(0)})^2]- \twoflagvar \expe^{-2\sigmaThe^2}) -4N \expe^{-\sigmaThe^2} \Real \twoflagthrice
 +(N^2-N) \expe^{-2\sigmaThe^2} \twoflagmod  \\
 & \qquad  \qquad + 4N \expe^{-2\sigmaThe^2} \Real \threeflagcov + (4N-2N^2) \expe^{-2\sigmaThe^2}\Real \threeflagcovconj
\label{eq:varvelcalc}
    \end{aligned}
\end{equation}
The statistical components are evaluated  as:
\begin{align*}
\twoflagvar &\equiv
    \Var \left[F_j^{(0)} F_{\jp}^{(0)} \expe^{\mathi(\Theta^{(0)}_j-\Theta^{(0)}_{\jp})} \expe^{\frac{2\pi\mathi}{N \celllen} (\flagdisp{j} - \flagdisp{\jp})} \right] \text{ for } j \neq \jp\\
&    = \bExp\left[ (F_j^{(0)})^2 (F_{\jp}^{(0)})^2
    \expe^{2\mathi(\Theta^{(0)}_j-\Theta^{(0)}_{\jp})}\expe^{\frac{4\pi\mathi}{N \celllen} (\flagdisp{j} - \flagdisp{\jp})} \right]
    -  \bExp \left[ F_j^{(0)} F_{\jp}^{(0)}\expe^{\mathi(\Theta^{(0)}_j-\Theta^{(0)}_{\jp})}\expe^{\frac{2\pi\mathi}{N \celllen} (\flagdisp{j} - \flagdisp{\jp})} \right]^2 \\
    &= \bExp[(F^{(0)})^2]^2 \expe^{-4\varTh^2}
    \sinc^2(2\flagdispran/\colrad)
    -\bExp[F^{(0)}]^4 \expe^{-2 \varTh^2}
    \sinc^4(\flagdispran/\colrad) \\
    \twoflagmod &\equiv\Cov\left(
    F_j^{(0)} F_{\jp}^{(0)} \expe^{\mathi(\Theta^{(0)}_j-\Theta^{(0)}_{\jp})} \expe^{\frac{2\pi\mathi}{N \celllen} (\flagdisp{j} - \flagdisp{\jp})} ,
    F_j^{(0)} F_{\jp}^{(0)} \expe^{\mathi(\Theta^{(0)}_{\jp}-\Theta^{(0)}_{j})}\expe^{\frac{2\pi\mathi}{N \celllen} (\flagdisp{\jp} - \flagdisp{j})} \right) \text{ for } j \neq \jp \\
    &= \bExp[(F^{(0)})^2]^2-\bExp[F^{(0)}]^4 \expe^{-2\varTh^2} \sinc^4 (\flagdispran/\colrad)\\
    \twoflagthrice &\equiv 
    \Cov ((F_j^{(0)})^2,F_j^{(0)} F_{\jp}^{(0)} \expe^{\mathi(\Theta^{(0)}_j-\Theta^{(0)}_{\jp})} \expe^{\frac{2\pi\mathi}{N \celllen} (\flagdisp{j} - \flagdisp{\jp})}) \text{ for } j \neq \jp \\ &= \bExp[F^{(0)}] \expe^{-\varTh^2} \sinc^2(\flagdispran/\colrad)
    \left(\bExp[(F^{(0)})^3]-\bExp[(F^{(0)})^2]\bExp[F^{(0)}]\right), \\
    \threeflagcov &\equiv \Cov\left(
    F_j^{(0)} F_{\jp}^{(0)} \expe^{\mathi(\Theta^{(0)}_j-\Theta^{(0)}_{\jp})}\expe^{\frac{2\pi\mathi}{N \celllen} (\flagdisp{j} - \flagdisp{\jp})} , F_j^{(0)} F_{\jpp}^{(0)} \expe^{\mathi(\Theta^{(0)}_j-\Theta^{(0)}_{\jpp})}\expe^{\frac{2\pi\mathi}{N \celllen} (\flagdisp{j} - \flagdisp{\jpp})} \right) 
  \text{ for } j\neq \jp \neq \jpp \\
    &= \Var \left[F_j^{(0)} 
     \expe^{\mathi \Theta^{(0)}_j} \expe^{\frac{2\pi\mathi}{N \celllen} \flagdisp{j}} \right]\bExp[F^{(0)}]^2 \expe^{-\varTh^2} \sinc^2 (\flagdispran/\colrad) \\
    &=  \bExp[F^{(0)}]^2 \expe^{-\varTh^2} \sinc^2 (\flagdispran/\colrad) \left(\bExp[(F^{(0)})^2] \expe^{-2\varTh^2} \sinc(2\flagdispran/\colrad)
   -\bExp[F^{(0)}]^2\expe^{-\varTh^2}\sinc^2(\flagdispran/\colrad)\right), \\
    \threeflagcovconj &=
\Cov\left(
    F_j^{(0)} F_{\jp}^{(0)} \expe^{\mathi(\Theta^{(0)}_j-\Theta^{(0)}_{\jp})}\expe^{\frac{2\pi\mathi}{N \celllen} (\flagdisp{j} - \flagdisp{\jp})} , F_j^{(0)} F_{\jpp}^{(0)} \expe^{\mathi(\Theta^{(0)}_{\jpp}-\Theta^{(0)}_j)}\expe^{\frac{2\pi\mathi}{N \celllen} (\flagdisp{\jpp} - \flagdisp{j})} \right)  \text{ for } j\neq \jp \neq \jpp   \\
    &=  \bExp[F^{(0)}]^2 \expe^{-\varTh^2} \sinc^2 (\flagdispran/\colrad)
   \left(\bExp[(F^{(0)})^2]-\bExp[F^{(0)}]^2\expe^{-\varTh^2} \sinc^2 (\flagdispran/\colrad)\right)  
\end{align*}
Substituting these expressions into Eq.~\eqref{eq:varvelcalc}, we obtain 
\begin{align*}
\gamtrans^4     \Var (\Vcolmsq) 
     & = N^2 \expe^{-2\sigmaThe^2} \left(\bExp[(F^{(0)})^2]- \bExp[F^{(0)}]^2 \expe^{-\varTh^2} \sinc^2 (\flagdispran/\colrad) \right)^2 \\
     & \qquad \qquad -4 \expe^{-\sigmaThe^2} \expe^{-\varTh^2} \sinc^2(\flagdispran/\colrad) N \bExp[(F^{(0)})^3] \bExp[F^{(0)}] \\
     & \qquad \qquad 
     +N \bExp[(F^{(0}))^4] - (1+ \expe^{-2\sigmaThe^2}+ \expe^{-2\sigmaThe^2-4\varTh^2}
    \sinc^2(2\flagdispran/\colrad)) N \bExp[(F^{(0)})^2]^2 \\
& \qquad \qquad    -6\expe^{-2\sigmaThe^2-2 \varTh^2}
    \sinc^4(\flagdispran/\colrad)N\bExp[F^{(0)}]^4 \\
& \qquad \qquad    +4\sinc^2(\flagdispran/\colrad)
    (\expe^{-\sigmaThe^2-\varTh^2}+ \expe^{-2\sigmaThe^2-3\varTh^2}\sinc(2\flagdispran/\colrad)+\expe^{-2\sigmaThe^2-\varTh^2}) \\
    & \qquad \qquad \qquad \qquad \times N\bExp[F^{(0)}]^2  \bExp[(F^{(0)})^2] \\
&=N^2 \expe^{-2\sigmaThe^2} \left(\Fmom{2}- \expe^{-\varTh^2} \sinc^2 (\flagdispran/\colrad) (\Fmom{1})^2 \right)^2 \\
     & \qquad \qquad +4(1- \expe^{-\sigmaThe^2} \expe^{-\varTh^2} \sinc^2(\flagdispran/\colrad)) N \bExp[(F^{(0)})^3] \bExp[F^{(0)}] \\
     & \qquad \qquad 
     +N \Fcum{4} + (2- \expe^{-2\sigmaThe^2}- \expe^{-2\sigmaThe^2-4\varTh^2}
    \sinc^2(2\flagdispran/\colrad)) N \bExp[(F^{(0)})^2]^2 \\
& \qquad \qquad    +6(1-\expe^{-2\sigmaThe^2-2 \varTh^2}
    \sinc^4(\flagdispran/\colrad))N\bExp[F^{(0)}]^4 \\
& \qquad \qquad    +4\left[\sinc^2(\flagdispran/\colrad)
    (\expe^{-\sigmaThe^2-\varTh^2}+ \expe^{-2\sigmaThe^2-3\varTh^2}\sinc(2\flagdispran/\colrad)+\expe^{-2\sigmaThe^2-\varTh^2})-3\right] \\ & \qquad \qquad \qquad \times N\bExp[F^{(0)}]^2  \bExp[(F^{(0)})^2] \\
 &=   N^2 \expe^{-2\sigmaThe^2} \left(\Fmom{2}- \expe^{-\varTh^2} \sinc^2 (\flagdispran/\colrad) (\Fmom{1})^2 \right)^2 \\
     & \qquad \qquad +4(1- \expe^{-\sigmaThe^2} \expe^{-\varTh^2} \sinc^2(\flagdispran/\colrad)) N \Fcum{3} \bExp[F^{(0)}] \\
     & \qquad \qquad 
     +N \Fcum{4} + (2- \expe^{-2\sigmaThe^2}- \expe^{-2\sigmaThe^2-4\varTh^2}
    \sinc^2(2\flagdispran/\colrad))  N \bExp[(F^{(0)})^2]^2 \\
& \qquad \qquad    +(-2+8\expe^{-\sigmaThe^2-\varTh^2}\sinc^2(\flagdispran/\colrad)-6\expe^{-2\sigmaThe^2-2 \varTh^2}
    \sinc^4(\flagdispran/\colrad))N\bExp[F^{(0)}]^4 \\
& \qquad \qquad    +4\sinc^2(\flagdispran/\colrad)
    (-2\expe^{-\sigmaThe^2-\varTh^2}+ \expe^{-2\sigmaThe^2-3\varTh^2}\sinc(2\flagdispran/\colrad)+\expe^{-2\sigmaThe^2-\varTh^2}) \\
    & \qquad \qquad \qquad \times N\bExp[F^{(0)}]^2  \bExp[(F^{(0)})^2] \\
\end{align*}

\bibliography{ref}
\bibliographystyle{unsrtnat}

\end{document}